\documentclass[a4paper,11pt]{article}
\pdfoutput=1 
\usepackage[T1]{fontenc} 
\usepackage[utf8]{inputenc}
\usepackage{caption,amsmath,amssymb,amsfonts,graphicx,multirow, enumitem}
\usepackage{pstricks,pdflscape,morefloats,algorithmicx,algpseudocode,slashed}
\usepackage[section]{algorithm}
\usepackage{booktabs}
\usepackage{color}
\usepackage{hyperref}
\usepackage{enumitem,mathtools}
\makeatletter
\def\cl@chapter{\@elt {theorem}}
\makeatother
\usepackage{cleveref} 
\usepackage{subfigure}
\usepackage[titletoc]{appendix}
\usepackage{cprotect}
\usepackage{makecell}
\usepackage{listings}
\newcommand{\shat}{\hat s}
\newcommand{\pt}{p_{T}}
\newcommand{\pperp}{p_T}
\newcommand{\dY}{\Delta {\rm Y}}
\newcommand{\sigeff}{\sigma_{\rm eff}}
\newcommand{\muf}{\mu_{\rm F}}
\newcommand{\mur}{\mu_{\rm R}}

\newcommand{\bi}[1]{
	\textbf{\emph{#1}}
}

\usepackage{xspace}
\newcommand{\angantyr}{Angantyr\xspace}
\newcommand{\pythia}{P\protect\scalebox{0.8}{YTHIA}\xspace}
\newcommand{\pytppp}{P\protect\scalebox{0.8}{YTHIA}8\xspace}

\def\beq{\begin{equation}}
\def\eeq{\end{equation}}
\def\bear{\begin{eqnarray}}
\def\eear{\end{eqnarray}}

\begin{document}
\vspace*{1.0cm}

\vspace{1cm}

\begin{center}
	{\Large \textbf{Double parton scattering in four-jet  production in proton-proton collisions at the LHC}\\
	\vspace{.7cm} \large
          Oleh Fedkevych$^{a,b}$\footnote{\texttt{fedkevyc@uni-muenster.de, oleh.fedkevych@ge.infn.it}} and
	Anna Kulesza$^{a}$\footnote{\texttt{anna.kulesza@uni-muenster.de}}\\
	\vspace{.3cm}
	\textit{\normalsize
		$^a$ Institute for Theoretical Physics, WWU M\"unster, D-48149 M\"unster, Germany\\
		$^b$ Dipartimento di Fisica, Universit\`{a} di Genova and INFN, Sezione 
di Genova, Via Dodecaneso 33, 16146, Genoa, Italy 		}\\
	}
\end{center}   

\vspace*{2cm}
\begin{abstract}
We study the contribution from double parton scattering (DPS) to four-jet 
production at the LHC, both at the leading order accuracy and after incorporating the effects of QCD radiation.  Apart from DPS, we also include and discuss the contribution from single parton scattering (SPS). We find that the QCD radiation impacts theoretical predictions significantly, with DPS contributions more affected than the SPS.  We also examine a number 
of observables in regard to their effectiveness for discrimination between DPS and SPS events and propose sets of kinematical cuts to improve the prospects of measuring DPS in four-jet production.
\end{abstract}

\clearpage
\tableofcontents
\setcounter{footnote}{0}

\section{Introduction}
\label{s:intro}

A composite nature of hadrons leads to a complicated structure of the underlying event in hadronic collisions. 
In particular it gives rise to a possibility of several interactions per one collision, a phenomenon referred to as  \textit{multiple parton interactions} (MPI). Rapidly increasing fluxes of partons in hadrons at small momentum fractions $x$ make their occurence more frequent at higher collision energies or, alternatively, at lower invariant masses of the measured system.  A particular subset of MPI with two hard interactions per single hadron-hadron collision is called \textit{double parton scattering} (DPS). It is a simplest possible MPI system. Provided the final state carries enough transverse momenta, it also is a relatively clean system experimentally. DPS processes have been first observed by the AFS \cite{Akesson:1986iv} and the UA2 \cite{Alitti:1991rd} collaborations. After these pioneering works a series of measurements were performed at the Tevatron collider 
\cite{Abe:1993rv, Abe:1997xk, Abe:1997bp,   Abazov:2009gc,  Abazov:2014fha, Abazov:2015nnn, Abazov:2014qba, Abazov:2015fbl} 
and at the LHC \cite{Aaboud:2016fzt, Aaij:2012dz, Aad:2013bjm, Chatrchyan:2013xxa, Aad:2014kba, Chatrchyan:2013qza, Aaij:2015wpa, Aaboud:2016dea, Aaij:2016bqq, Khachatryan:2016ydm,  Sirunyan:2017hlu, Aaboud:2018tiq, CMS:2021ijt}.

Among different DPS production channels 
the four-jet DPS production takes a special place due to high abundance of the multi-jet events in hadron-hadron collisions and correspondingly, a 
large DPS cross section. The four-jet DPS production was measured in multiple experiments at various colliders: ISR~\cite{Akesson:1986iv}, SPS~\cite{Alitti:1991rd}, Tevatron~\cite{Abe:1993rv} and the LHC \cite{Chatrchyan:2013qza, Aaboud:2016dea, CMS:2021ijt}.  The theoretical efforts to describe (four-) jet production from multiple scatterings have a long history~\cite{Landshoff:1978fq, Paver:1982yp, Humpert:1983pw, Humpert:1984ay, Ametller:1985tp, Mangano:1988sq}. In more recent years studies focused on, among other, including modelling of radiation effects~\cite{Berger:2009cm} and LHC's potential to gain new information on the properties of two-parton distribution functions in the transverse plane~\cite{Domdey:2009bg}. The importance of four-jet production in the context of DPS studies, especially in the back-to-back regime, was further highligted in~\cite{Blok:2010ge, Blok:2011bu, Blok:2013bpa}. In this series of papers, concurrently to other efforts at that time,  a theoretical DPS framework that accounts for perturbative splittings of a single parton into two was developed. It was later 
followed by a proposal to implement the new approach in the \pythia~\cite{Sjostrand:2014zea, Sjostrand:2006za} event generator through modifying its MPI model with the help of a dedicated tune ~\cite{Blok:2015rka}. Furthermore, the DPS observables and kinematics of four-jet production ~\cite{Maciula:2014pla, Maciula:2015vza} as well as calculations in the high-energy factorization approach~\cite{Kutak:2016mik, Kutak:2016ukc} were explored. In parallel, there has been an enormous progress in theoretical understanding and decription of DPS on a more fundamental level~\cite{Blok:2011bu, Blok:2013bpa, Diehl:2011tt, Diehl:2011yj, Gaunt:2011xd, Gaunt:2012dd,  Ryskin:2011kk, Ryskin:2012qx, Manohar:2012pe,Diehl:2017kgu}, see also recent review in~\cite{Diehl:2017wew}. Concurrently  to  the aforementioned advances in DPS description, development of Monte Carlo DPS and MPI models progressed significantly~\cite{Sjostrand:1985vv, Sjostrand:1987su, Sjostrand:2004pf, Corke:2009tk, Corke:2011yy, Sjostrand:2017cdm, Bahr:2008dy, Gieseke:2016fpz, Martin:2012nm}.

The DPS four-jet production unavoidably occurs simultaneously with the four-jet production in \textit{single parton scattering} (SPS). Theoretical 
predictions for this process are known at the next-to-leading order (NLO) 
accuracy~\cite{Bern:2011ep, Badger:2012pf}. The di-jet production, which is a major building block of double scattering resulting in four jets, has been studied up to the next-to-next-to-leading order (NNLO) in perturbation theory~\cite{Currie:2017eqf}. Although a significant progress has been made recently towards DPS calculations at NLO~\cite{Diehl:2017kgu, Diehl:2019rdh} no DPS process has been described at this accuracy yet.

In this work, we revisit the theoretical predictions for four-jet DPS production in $pp$ collisions at the LHC. We first review the LO calculations at the partonic level, both for DPS and SPS, and discuss selected differential quantities as well as their uncertainties. We also investigate the impact of longitudinal parton correlations present in double parton distribution functions, as defined below.  The LO analysis is then extended by studying the effect of initial and final state radiation on both the SPS and DPS total cross sections and distributions. The radiation is simulated with the parton shower algorithm as implemented in the \pythia event 
generator. 

In this sense, the work presented here goes beyond parton level LO calculations in the collinear factorization~\cite{Maciula:2014pla, Maciula:2015vza}, or adding to it one (undetected) real emission, as done in~\cite{Berger:2009cm}.  Our studies can be also seen as parallel to the effort of including higher-order effects in the DPS cross sections using the framework of high-energy factorization of~\cite{Kutak:2016mik, Kutak:2016ukc}. The developments presented here are also different from the option of  second hard scattering built in \pythia. While \pythia constructs two-parton parton distribution functions from single parton distributions in a very specific way which cannot be changed by a user, and requires additional 
adjustments to obtain correct normalization of DPS cross sections, our approach allows to specify how double parton distributions are modelled.
Besides, in earlier versions of \pythia (before 8.240)  the DPS events are dependent on the ordering of the hard interactions constituting the DPS 
event. Furthermore, our approach offers another way to study effects of showering on DPS predictions in nuclear collisions~\cite{inpreparation}, different from the \angantyr model of \pythia~\cite{Bierlich:2018xfw}\footnote{As it was recently shown in \cite{Fedkevych:2019ofc} the \angantyr model can be used to simulate four-jet DPS production in pA collisions.}.

The paper is structured as follows: in Chapter 2 we briefly review the DPS framework used in our studies. This is followed in Chapter 3 by the description of the implementation and details of various technical aspects of the simulation. This chapter also contains our numerical results and their discussion. We conclude and present an outlook for future work in \mbox{Chapter 5}.

\section{Phenomenology of double parton scattering }
\label{s:theory}

We begin with a description of the framework in which four-jet DPS production is studied. Its origins go back to the work of Paver and Treleani~ \cite{Paver:1982yp} and  Mekhfi~\cite{Mekhfi:1983az}.  The results of  \cite{Paver:1982yp},  \cite{Mekhfi:1983az} were later generalized and extended to the  case of $n$-hard interactions  \cite{Diehl:2011tt, Diehl:2011yj}.

The total DPS cross section is given by the following expression
\begin{eqnarray}
	\sigma^{\rm DPS}_{AB}  &=&
	\frac{1}{1 + \delta_{AB}}
	\sum\limits_{i, j, k, l}
	\int \, dx_1\, dx_2\, dx_3\, dx_4 \ 
	\hat{\sigma}_{i \, k \rightarrow A} \,
	\hat{\sigma}_{j \, l \rightarrow B} \times \nonumber\\
	&\times&
	\Gamma_{i j / h_A}(x_1, x_2, \bi b, Q_1, Q_2) \,
	\Gamma_{k l / h_B}(x_3, x_4, \bi b, Q_1, Q_2),
	\label{eq:dps_cross_secton_non_factorized}
\end{eqnarray}
where $A$ and $B$ denote final states in $i \, k \rightarrow A$  and $j \, l \rightarrow B$ processes 
and objects \small{$\Gamma_{ij/h}$} are called  \textit{generalized two parton distribution functions} which  
can be in a first approximation thought of as a probability to find two partons $i$ and $j$ with longitudinal momentum fractions $x_1$ and $x_2$ separated by distance $|\bi b|$ in a transverse plane of a hadron $h$.

In the following, we assume factorization of $\Gamma_{ij / h}$ into a product of longitudinal and transverse 
dependent pieces
\begin{equation}
	\Gamma_{ij / h}(x_1, x_2, \bi b, Q_1, Q_2) \approx D_{i j / h}(x_1, x_2, 
Q_1, Q_2) \, F(\bi b).
	\label{eq:factorization}
\end{equation}
Using Eq.~\ref{eq:factorization} one can write a total DPS cross section for pp collisions as
\begin{eqnarray}
	\sigma^{\rm DPS}_{AB} &=&
	\frac{1}{1 + \delta_{AB}}	
	\frac{1}{\sigma_{\rm eff}}
	\sum\limits_{i, j, k, l}
	\int \, dx_1\, dx_2\, dx_3\, dx_4 \ 
	\hat{\sigma}_{i \, k \rightarrow A} \,
	\hat{\sigma}_{j \, l \rightarrow B} \times \nonumber\\
	&\times& 
	D_{i j}(x_1, x_2, Q_1, Q_2) \,
	D_{k l}(x_3, x_4, Q_1, Q_2),
	\label{eq:dps_GS_formula}
\end{eqnarray}
where we have dropped the subscript $h$ in order to simplify notation. The prefactor $1 /\left(1 + \delta_{AB}\right)$ 
in Eq.~\ref{eq:dps_GS_formula} was  introduced to reflect the fact that in case of production of two indistinguishable finals states $A$ and $B$ 
one has to divide a total DPS cross section by 2. In the following we refer to the distributions $D_{ij}$ in Eq.~\ref{eq:dps_GS_formula} as to
\textit{double parton distribution functions} (dPDFs).
The quantity $\sigma_{\rm eff}$ in Eq.~\ref{eq:dps_GS_formula} is given by 
\begin{eqnarray}
	\sigma_{\rm eff} = \frac{1}{\int d^2b \, F^2(\bi b)},
	\label{eq:sig_eff_def}
\end{eqnarray}
and can be interpreted as an effective interaction area. It should be noted, however, that  the factorization of longitudinal and transverse pieces is an assumption driven by a practical need of modelling of two-parton parton distribution functions for phenomenology purposes. 
The generalized distributions evolve differently in the position and in momentum space~\cite{Diehl:2011tt, Diehl:2011yj}, and consequently at ${\bi b}=0$ and  ${\bi b} \neq 0$~\cite{Gaunt:2011xd}. This is inconsistent 
with the factorization assumption, which should hold for all values of  ${\bi b}$. 
As discussed in \textit{e.g.}~\cite{Diehl:2017kgu}, the cornerstone theoretical expression for the DPS cross section is given by Eq.~(\ref{eq:dps_cross_secton_non_factorized}), with \small{$\Gamma_{ij/h}$}  evolving according to a homogenous double DGLAP equation~\cite{Diehl:2011tt, Diehl:2011yj}.

Assuming no partonic correlations in $x$-space one can substitute 
\begin{eqnarray}
	D_{ij}(x_1, x_2, Q_1, Q_2) \approx f_{i}(x_1, Q_1) f_{j}(x_2, Q_2)
\end{eqnarray}
in Eq. \ref{eq:dps_GS_formula} which gives us  the ``pocket formula of DPS''
\begin{eqnarray}
	\sigma^{\rm DPS}_{AB} = \frac{1}{1 + \delta_{AB}} 
	\sum\limits_{i, j, k, l}
	\frac{\sigma_{i \, k \rightarrow A} \, \sigma_{j \, l \rightarrow B}}{\sigma_{\rm eff}}.
	\label{eq:dps_pocket_formula}
\end{eqnarray}
Such factorization violates momentum and number (flavour) dPDF sum rules proposed by Gaunt and Stirling \cite{Gaunt:2009re}.
One can avoid unphysical contributions by multiplying a factorized product of PDFs by an appropriate cutoff function, for example
\begin{eqnarray}
	D_{ij}(x_1, x_2, Q_1, Q_2) \approx f_{i}(x_1, Q_1) \, f_{j}(x_2, Q_2) \, 
\theta(1 - x_1 - x_2),
	\label{eq:factorization_dPDF_naive}
\end{eqnarray}
where $\theta(1 - x_1 - x_2)$ excludes unphysical region where $x_1 + x_2 
> 1$. However, one has to keep in mind that 
Eq.~\ref{eq:factorization_dPDF_naive} still violates momentum and number sum rules for dPDFs. 

Notwithstanding all the above concerns, Eqs.~\ref{eq:dps_GS_formula} and  
\ref{eq:factorization_dPDF_naive} are still commonly used for a phenomenological modelling of DPS. In this paper we refer to the dPDFs approximated according to  Eq. \ref{eq:factorization_dPDF_naive} as to ``naive'' dPDFs. We also should notice that ``naive'' dPDFs, as in \mbox{Eq. \ref{eq:factorization_dPDF_naive}}, do not allow to reduce \mbox{Eq. \ref{eq:dps_GS_formula}}
to the ``pocket formula'' Eq. \ref{eq:dps_pocket_formula}.  Instead,  by substituting 	Eq. \ref{eq:factorization_dPDF_naive} into 
\mbox{Eq. \ref{eq:dps_GS_formula}}, we get
\begin{eqnarray}
	\sigma^{\rm DPS}_{AB}  &=&
	\frac{1}{1 + \delta_{AB}}	
	\frac{1}{\sigma_{\rm eff}}
	\sum\limits_{i, j, k, l}
	\int \, dx_1\, dx_2\, dx_3\, dx_4 \ 
	f_{i}(x_1, Q_1) \, f_{k}(x_3, Q_1) \, \hat{\sigma}_{ik \rightarrow A} \times \nonumber\\
	&\times& 
	f_{j}(x_2, Q_2) \, f_{l}(x_4, Q_2) \, \hat{\sigma}_{jl \rightarrow B} \, 
\theta(1 - x_1 - x_2) \, \theta(1 - x_3 - x_4).
	\label{eq:dps_naive_formula}
\end{eqnarray} 
In our analysis  we   use two different models of  dPDFs. Namely, we use ``naive'' dPDFs, as in \mbox{Eq.~\ref{eq:factorization_dPDF_naive}}, constructed 
out of MSTW2008~\cite{Martin:2009iq} or CT14~\cite{Dulat:2015mca} leading 
order (LO) PDFs.  In order to estimate the impact of partonic correlations in $x$-space we compare between results obtained according to  Eq. \ref{eq:dps_GS_formula} and supplemented with either ``naive'' dPDFs or with GS09 dPDFs \cite{Gaunt:2009re}. The latter are  LO dPDFs, $D_{i j / h}(x_1, x_2, Q_1, Q_2)$ which evolve according to the double DGLAP equation and  obey the momentum and number sum rules. Their initial parametrization is predominantly based on MSTW2008LO as input single PDFs. 

At the root of the difference in the evolution of generalized distributions in the position and in momentum space lies perturbative splitting of one parton into two partons~\cite{Diehl:2011tt, Diehl:2011yj}, a mechanism 
which contribution to double parton distributions needs to be accounted for. The corresponding $1 \to 2$ term in the \small{$\Gamma_{ij/h}$} has a 
$1/\bi b^2$ behaviour at small $\bi b$, and renders the cross section in Eq.~\ref{eq:dps_cross_secton_non_factorized} UV-divergent. The UV divergence is an artefact of using the DPS description outside of its region of validity, where the SPS picture is more suitable. 
A consistent scheme which treats the problem of UV divergencies, as well as a closely related problem of double counting between DPS and SPS, was proposed in~\cite{Diehl:2017kgu}.  Earlier approaches and discussion of the problems can be found in~\cite{Blok:2011bu, Blok:2013bpa, Gaunt:2012dd, Ryskin:2011kk,  Ryskin:2012qx,  Manohar:2012pe}.  According to the scheme of ~\cite{Diehl:2017kgu}, both DPS cross section and SPS cross section 
(at the same order in perturbation theory as DPS) contribute to the total 
cross section, while the double counting and the UV divergencies  are removed by suitably designed subtraction terms. However, in the case of four-jet production it is not possible to apply this prescription since the corresponding SPS calculation at NNLO is technically out of reach. Instead, we rely on the observation made in~\cite{Diehl:2017kgu} that the SPS contribution and the associated subtraction terms lose their relevance if the considered system probes sufficiently low values of $x$ in two-parton distribution functions (see also~\cite{Blok:2013bpa}). The four-jet production at relatively low $\pt$ can be then seen as a promising setup to study DPS, in addition to \textit{e.g.} same-sign W-pair production~\cite{Diehl:2017kgu, Kulesza:1999zh, Gaunt:2010pi, Cao:2017bcb, Cabouat:2019gtm}.

Observables used in DPS phenomenology can be often characterized as ``transverse'' and ``longitudinal'', depending on the direction of momenta of the final state particles which are predominantly probed. Both types of  observables exploit the fact that the individual partonic collision in a DPS event must obey momentum conservation, \textit{i.e.} make use of correlations between final state particles either in DPS or SPS topologies.  In addition, the rapidity-based observables not only exploit differences between SPS and DPS topologies~\cite{Kom:2011bd, Kom:2011nu}, but also are directly sensitive to the correlations among the incoming momenta fractions for the parton coming from the same hadron. It has been observed in \textit{e.g.}~\cite{Kom:2011bd} that the DPS observables of the transverse type are particularly sensitive to real radiation effects. In particular, in many cases the DPS-sensitive variables based upon $p_T$ imbalance and angular correlations  between produced jets are trivial at the partonic level. Consequently, adding the real radiation results is needed in order to obtain a realistic description of these observables.

\section{Predictions for four-jet production in pp collisions}
\label{s:fourjet}

The numerical calculations of  DPS production at the parton-level are performed using an in-house built Monte Carlo programme, which calculates LO 
matrix elements for  $2 \to 2 \, \otimes \, 2 \to2$ scattering. In order to add initial and final state radiation (ISR and FSR) effects to our parton-level simulations 
we use the \pythia event generator \cite{Sjostrand:2006za, Sjostrand:2014zea} with modifications necessary  to read and ``shower'' output of our DPS code \cite{TSjostrand:private_com}. In the first step we output the DPS events using modified (``double'') Les Houches Event (LHE) file records 
\cite{Alwall:2006yp}, see Appendix~\ref{s:double_lhe} for more technical details. The double LHE files are then supplied   to \pythia for showering using  \verb|SecondHard:generate = on| and \verb|PartonLevel:MPI = off| settings. While generating partonic events, we use the original Lund 
Monte Carlo algorithm proposed by \mbox{Bengtsson} \cite{Bengtsson:1982jr} which allows to generate colour charges of the initial and final state partons within the leading colour approximation\cprotect\footnote{The same algorithm is used in the \pythia event generator.}.  The generation  of 
the colour charges is required to take into account colour coherence effects~\cite{Ellis:1986bv}, \cite{Marchesini:1987cf}.

For each individual $2 \to 2$ scattering, all 8 subprocesses involving combinations of (anti-)quarks and/or gluons are taken into account. The scales of the two partonic collisions are treated  as independent and chosen 
equal to the value of the transverse momentum partons in the two di-jets. 
In each case, the central value of the factorization scale is set to be equal to renormalization scale,  
$\mu_{{\rm DPS}, i}=\mu_{\rm F, i}=\mu_{\rm R, i}= H_{T,i}/2= |p_{T, {i}}|$, where $p_{i}$ is the momentum of one of the partons in the collision $i$ and  $i \in \{1,2\}$ with ${1,2}$ indicating the two scatterings. The SPS predictions are obtained using the MadGraph package~\cite{Alwall:2014hca}. In order to account for the NLO effects in the normalization of the SPS total cross section,we apply an effective K-factor~\cite{Bern:2011ep, Badger:2012pf} to the LO predictions, similarly to approach of 
Ref.~\cite{Maciula:2015vza, Kutak:2016ukc}. The scale of the SPS process is chosen as  $\mu_{\rm SPS}=\muf=\mur= H_T/2= \frac{1}{2} \sum^4_{i = 1} p_{T, i}$, in agreement with~\cite{Bern:2011ep, Badger:2012pf}. In the analysis at the partonic level  we produce  four partons and apply to them  a jet separation criterium of $R_{ij}=\sqrt{(y_i-y_j)^2+(\phi_i-\phi_j)^2}>0.4$, unless otherwise stated. In the parton shower analysis  final state particles are clustered by means of the  FastJet~\cite{Cacciari:2011ma} package with the  anti-$k_t$ jet clustering algorithm \cite{Cacciari:2008gp} and  jet radius taken to be $R=0.4$. Whenever comparing the SPS predictions to the showered DPS results, the SPS results are also showered using the same version of \pythia and the same method of 
clustering of partons into jets is applied. Unless otherwise stated, we consider a rapidity coverage of $-4.7<y_j<4.7$  for all jets $j$. Since the generation of DPS events with jets carrying high transverse momentum is 
suppressed due to low fluxes of incoming parton pairs and, correspondingly, high-$\pperp$ jets originate predominantly from SPS, we apply an upper 
cut on $p_{T,j}$ for all jets.

A number of checks at various stages of the calculation have been performed. The MadGraph set-up for the computation of SPS cross sections have been cross-checked against the outcome of the ALPGEN code~\cite{Mangano:2002ea} for the process $pp \to jj gg$. The cross section for results for the individual $2 \to 2$ cross sections, \textit{i.e.} the building blocks of the DPS cross sections, have been also checked against MadGraph. The procedure to assign colour to initial and final state partons in LHE files 
was checked by comparing \pythia-showered results against results based on  MadGraph LHE events, also showered with \pythia. Our results for the ratios of the DPS to SPS+DPS  total cross sections agree with results of LO partonic results of Ref.~\cite{Maciula:2015vza} within a few percent, depending on the choice of the value of $R_{ij}$. We have also checked that with our set-up for SPS calculations we can reasonably well reproduce the LO and LO matrix element matched to parton shower (ME+PS) four-jet cross sections from~\cite{Bern:2011ep}.

\subsection{Parton-level analysis}

\begin{figure}
\includegraphics[width=0.45\linewidth]{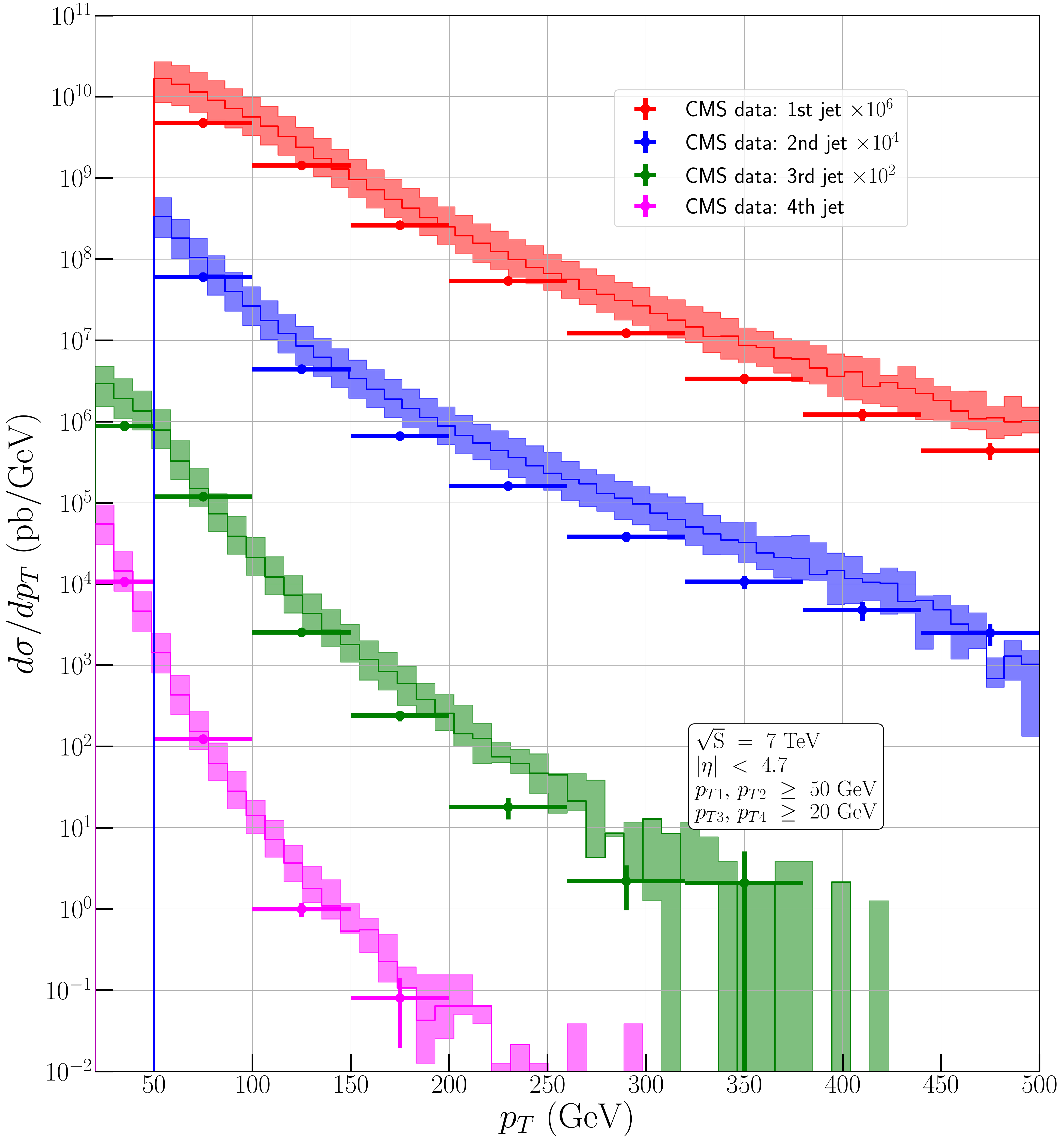}
\includegraphics[width=0.45\linewidth]{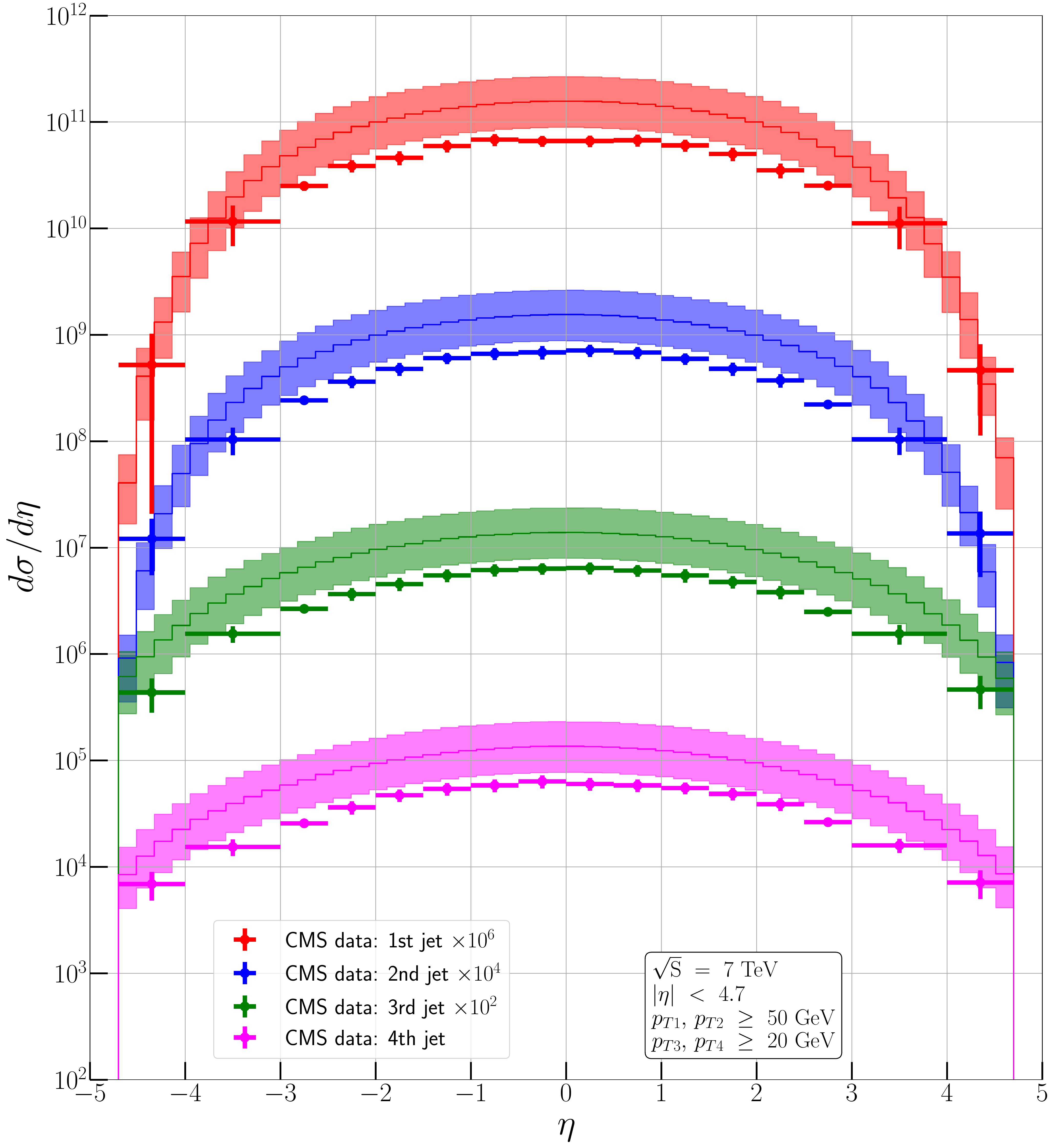}
\caption{Comparison of the parton-level SPS+DPS predictions for four-jet production with CMS data~\cite{Chatrchyan:2013qza} for 7 TeV collision, see text for more explanation.  }
\label{fig:spsdps_vs_cms}
\end{figure}

We begin our studies by investigating the DPS and SPS four-jet production 
at the partonic level. In order to judge the reliability of the predictions, in Fig.~\ref{fig:spsdps_vs_cms} we check how well they fare against CMS measurement of four jet production at \mbox{$\sqrt S=7$ TeV}~\cite{Chatrchyan:2013qza}, which uses relatively low cuts on jets transverse momentum: \mbox{$\pperp > 50$ GeV} for the two most leading jets and $\pperp 
> 20$ GeV for the third and fourth jet. In accordance with experimental analysis, we use here $R_{jj}=0.5$.

The uncertainty bands for DPS and SPS events are estimated independently and combined together afterwards.
For the SPS events we  start by generating central value predictions at LO using MSTW2008 LO PDFs.
Since NLO corrections typically lead to the change in the normalization of the cross section predictions, similarly to the approach of  \cite{Maciula:2015vza}, we multiply the LO predictions by a K-factor of $0.5$.
Such value of the  K-factor follows from the fixed-order four-jet NLO in \cite{Badger:2012pf} and has been obtained for the MSTW PDFs. 
After producing central value predictions we obtain the conservative estimate of uncertainties due to the scale variation by performing simultaneous changes of factorization and renormalization scales up and down by a factor of 2 in the LO calculations. Then the SPS envelope is constructed by choosing the maximal up and down uncertainty band out of all possible options.

The theoretical predictions are obtained as a sum of the SPS result (including the K-factor of 0.5) and the DPS result, with their corresponding error bands. 
While we are not in position to calculate the DPS prediction at NLO, nor have additional information on the NLO effects~\footnote{Arguments against applying an effective K-factor to double di-jet production constituting 
DPS can be found in \textit{e.g.}~\cite{Kutak:2016ukc}.}, we take the spread provided by the scale uncertainty and variation in 
the parameter of $\sigeff = 15\pm5$ mb for two different PDF sets, MSTW2008LO and CT14 LO, used in the ``naive'' dPDFs construction,  Eq.~\ref{eq:factorization_dPDF_naive}. These effects are then combined using the envelope method~\cite{Dittmaier:2011ti, Kramer:2012bx}. Finally, the central values of the DPS and SPS predictions as well as the uncertainty estimates are combined. We observe an overall agreement within uncertainty with 
the data for both each jet's $\pperp$  and $y$, see Fig.~\ref{fig:spsdps_vs_cms}, with tendency for theoretical results to overestimate the data.

In Table~\ref{tab:dps_4j_dps_vs_sps}, we compare central values of the total cross cross sections  for the LHC collision energy of 7 and 13 TeV and two sets of cuts on jets' $p_{T}$: a stricter  $35\ {\rm GeV} < p_{T, j} < 100\ {\rm GeV}$ and a looser one $20\ {\rm GeV} < p_{T, j}  < 100\ {\rm GeV}$.  The upper cut on jet's $\pperp$ is inspired by the well-known fact that DPS is enhanced at lower partonic energies, and by the earlier literature~\cite{Maciula:2015vza}. In the DPS computations we  use ``naive'' dPDFs constructed out of MSTW2008 LO PDFs. Correspondingly, 
the SPS results are also obtained with MSTW2008 PDFs. Apart from the least DPS-favourable case of $\sqrt S =7$ TeV and $35\ {\rm GeV} <  p_{T, j} < 100\ {\rm GeV}$, in all other cases the upper cut on maximal  $p_{T,j}$ leads to higher total DPS cross sections than SPS cross sections.   As 
expected, the DPS cross sections increase greatly when the minimal $p_{T,j}$ cut is lowered, and the ratio of  DPS to SPS cross sections improves. 
Due to growing parton fluxes at low fractions of momenta $x$, an increase 
in the ratio is also observed at higher LHC energies.

\begin{center}
\begin{tabular}{ | c | c | c | c |}
\hline
\thead{Cuts and collision energy}	  			 & \thead{$\sigma_{\rm SPS}$ for \\ 
          										   $pp \xrightarrow{\rm SPS} 4j$ process} 	& \thead{$\sigma_{\rm DPS}$ for \\
          															 						  $pp \xrightarrow{\rm DPS} 4j$ process} 	
          															 						& \thead{
          															 						$\frac{\sigma_{\rm DPS}}
          															 						{\sigma_{\rm DPS} + \sigma_{\rm SPS}}$ }\\\hline		
\thead{$\sqrt{S} = 7$ TeV, ${|y| < 4.7}$,\\ 
       ${p_T \in [35, 100]}$ GeV}				& 	   	76.15	 							& 43.55  	& 36 $\%$\\\hline
       
\thead{$\sqrt{S} = 7$ TeV, ${|y| < 4.7}$,\\ 
       ${p_T \in [20, 100]}$ GeV}				& 		2062.79	 							& 3759.59	& 65 $\%$ \\\hline

\thead{$\sqrt{S} = 13$ TeV, ${|y| < 4.7}$,\\ 
       ${p_T \in [35, 100]}$ GeV}				& 	  	316.78								& 333.83		& 51 $\%$ \\\hline

\thead{$\sqrt{S} = 13$ TeV, ${|y| < 4.7}$,\\ 
       ${p_T \in [20, 100]}$ GeV}				&  		7319.50								& 22062.80	& 75 $\%$\\\hline

\end{tabular}
\captionof{table}{LO DPS and SPS cross sections (in nanobarns) for $pp \rightarrow 4j$ for two sets of cuts on jets'  $\pperp$ and two LHC collision energies.}
\label{tab:dps_4j_dps_vs_sps}
\end{center}

\begin{figure}
\includegraphics[width=0.5\linewidth]{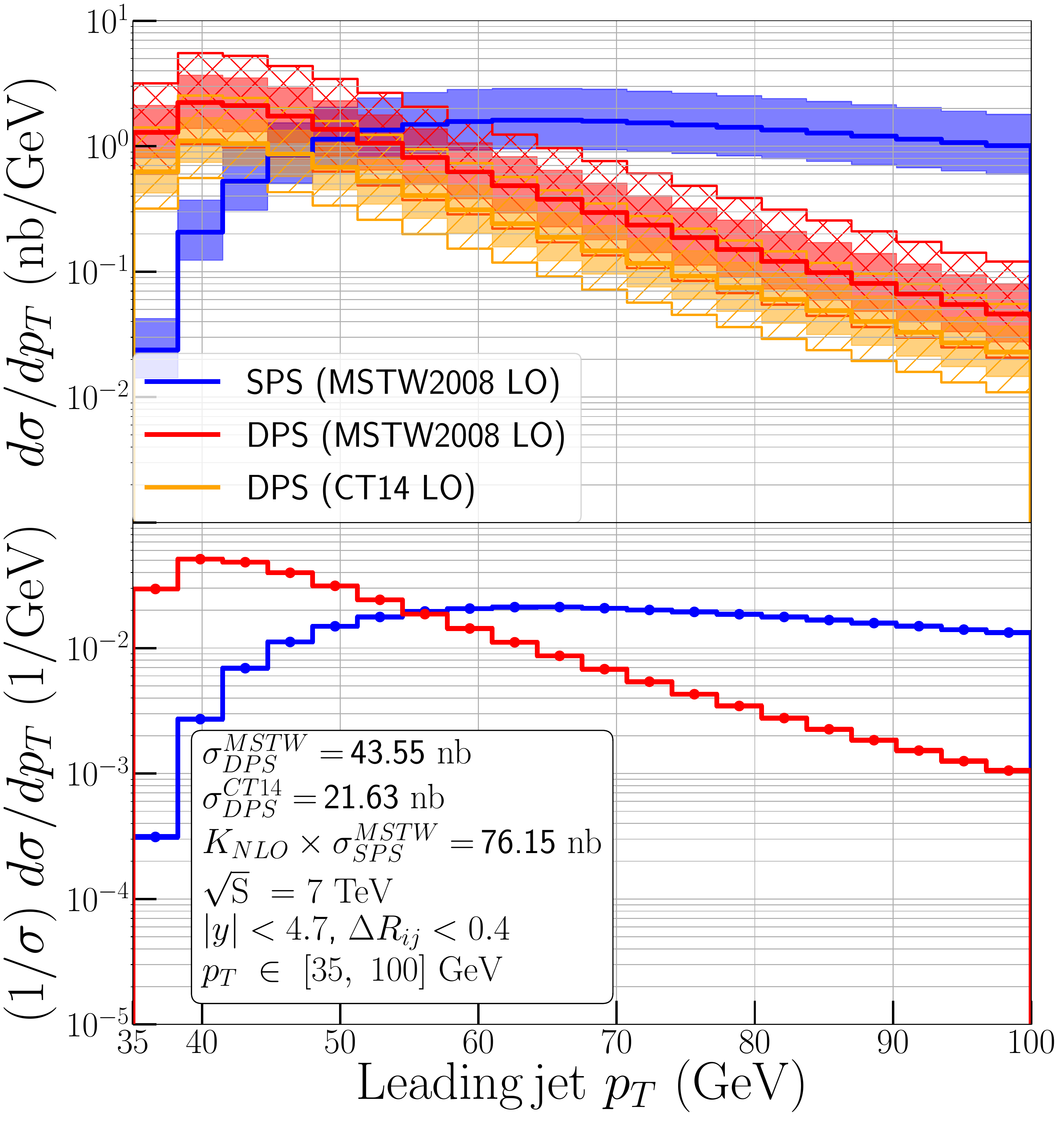}
\includegraphics[width=0.5\linewidth]{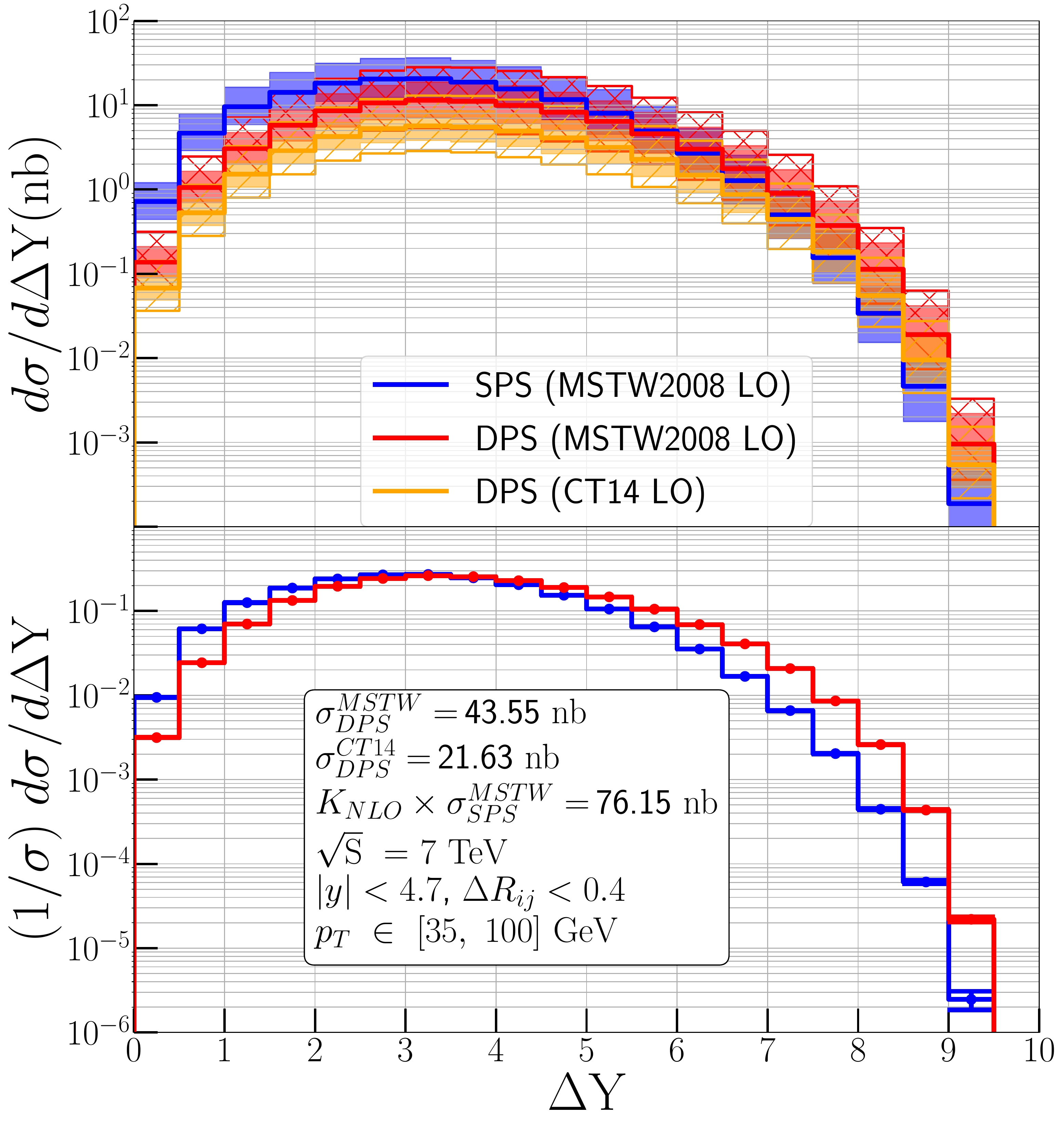}
\caption{Comparison of SPS and DPS leading jet $p_T$ distributions  (left)  and  ${\rm \Delta Y} = {\rm max}|y_i - y_j|$ distributions for four-jet events with \mbox{$p_T \in [35, 100]$ GeV} at $\sqrt{S} = 7 \, {\rm 
TeV}$. Upper panels show absolute values, while lower panels show shapes of the distributions.}
\label{fig:sps_vs_dps_var1}
\end{figure}

\begin{figure}
\includegraphics[width=0.5\linewidth]{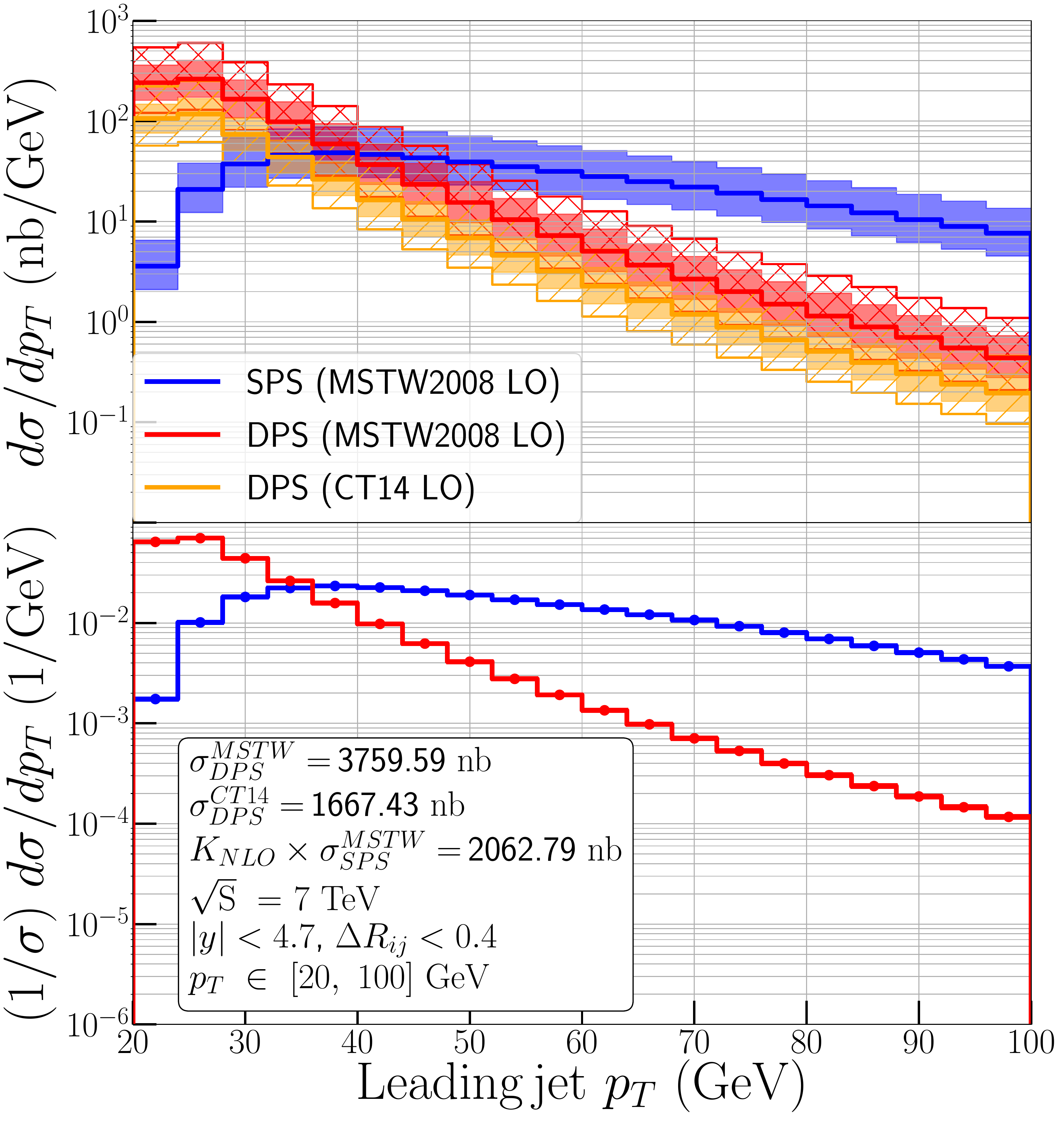}
\includegraphics[width=0.5\linewidth]{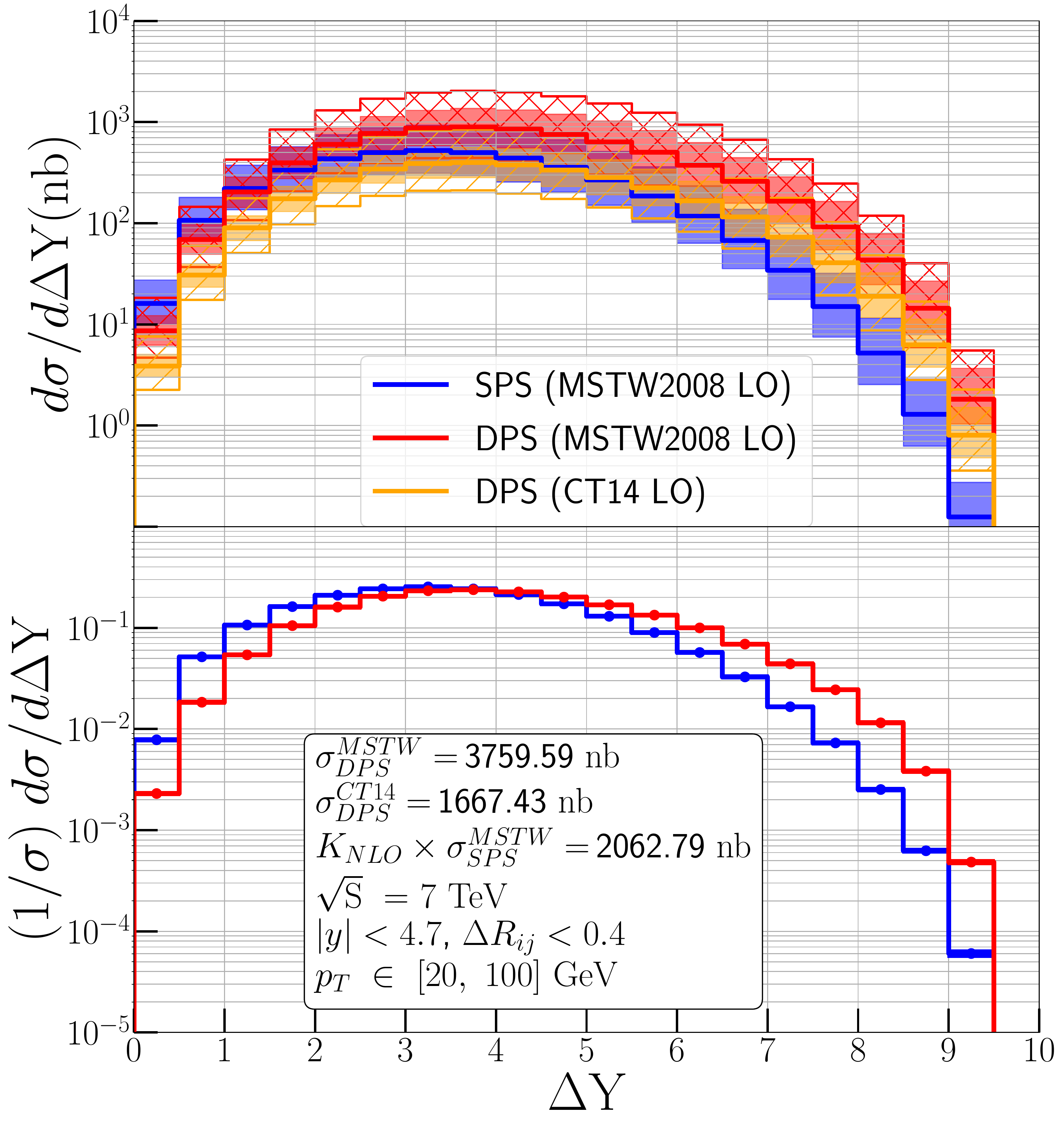}
\caption{Same as in Fig.~\ref{fig:sps_vs_dps_var1} but with \mbox{$p_T \in [20, 100]$ GeV}.}
\label{fig:sps_vs_dps_var2}
\end{figure}

\begin{figure}
\includegraphics[width=0.5\linewidth]{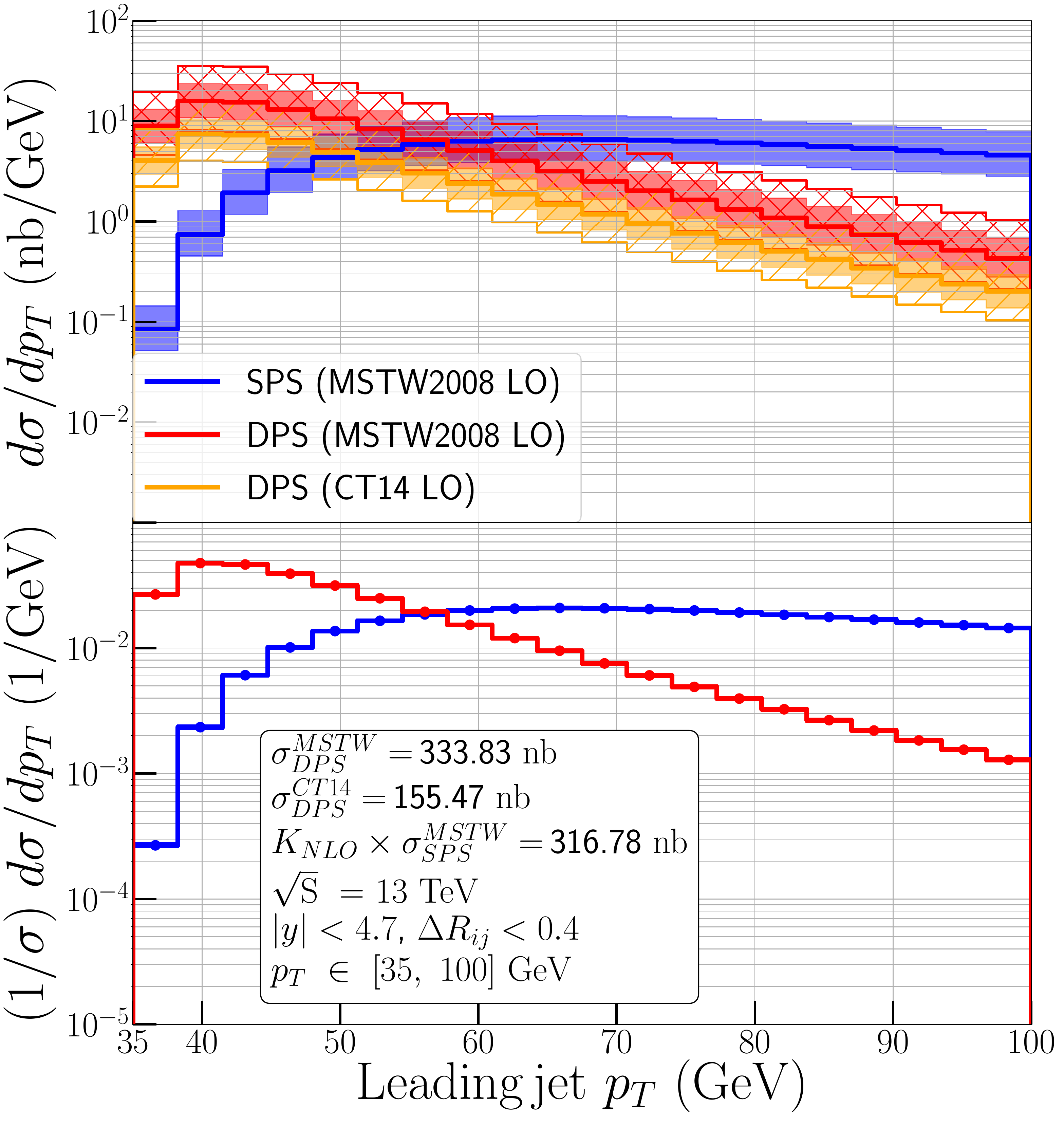}
\includegraphics[width=0.5\linewidth]{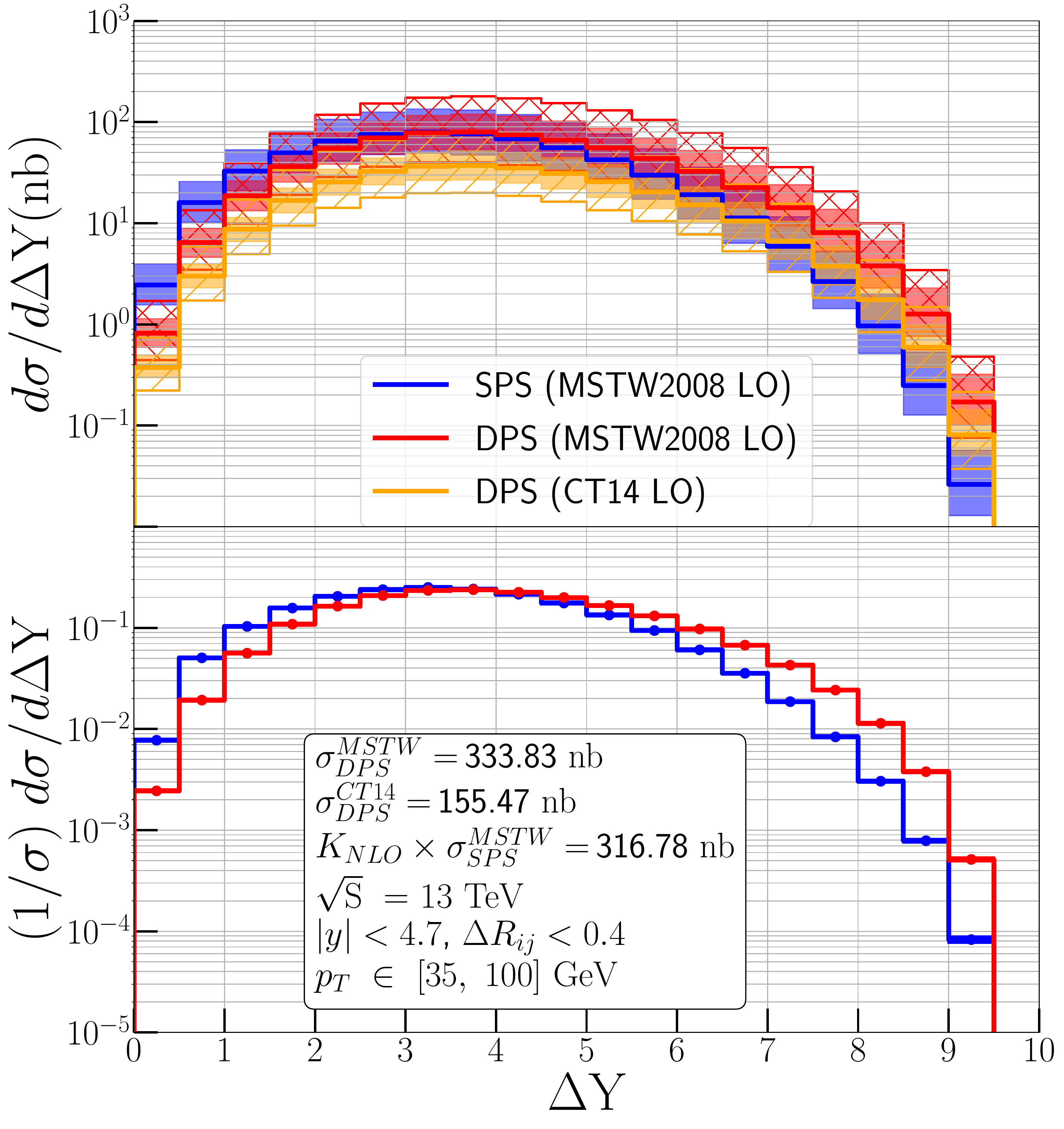}
\caption{Same as in Fig.~\ref{fig:sps_vs_dps_var1} but at $\sqrt{S} = 13 \, {\rm TeV}$. }
\label{fig:sps_vs_dps_var3}
\end{figure}

\begin{figure}
\includegraphics[width=0.5\linewidth]{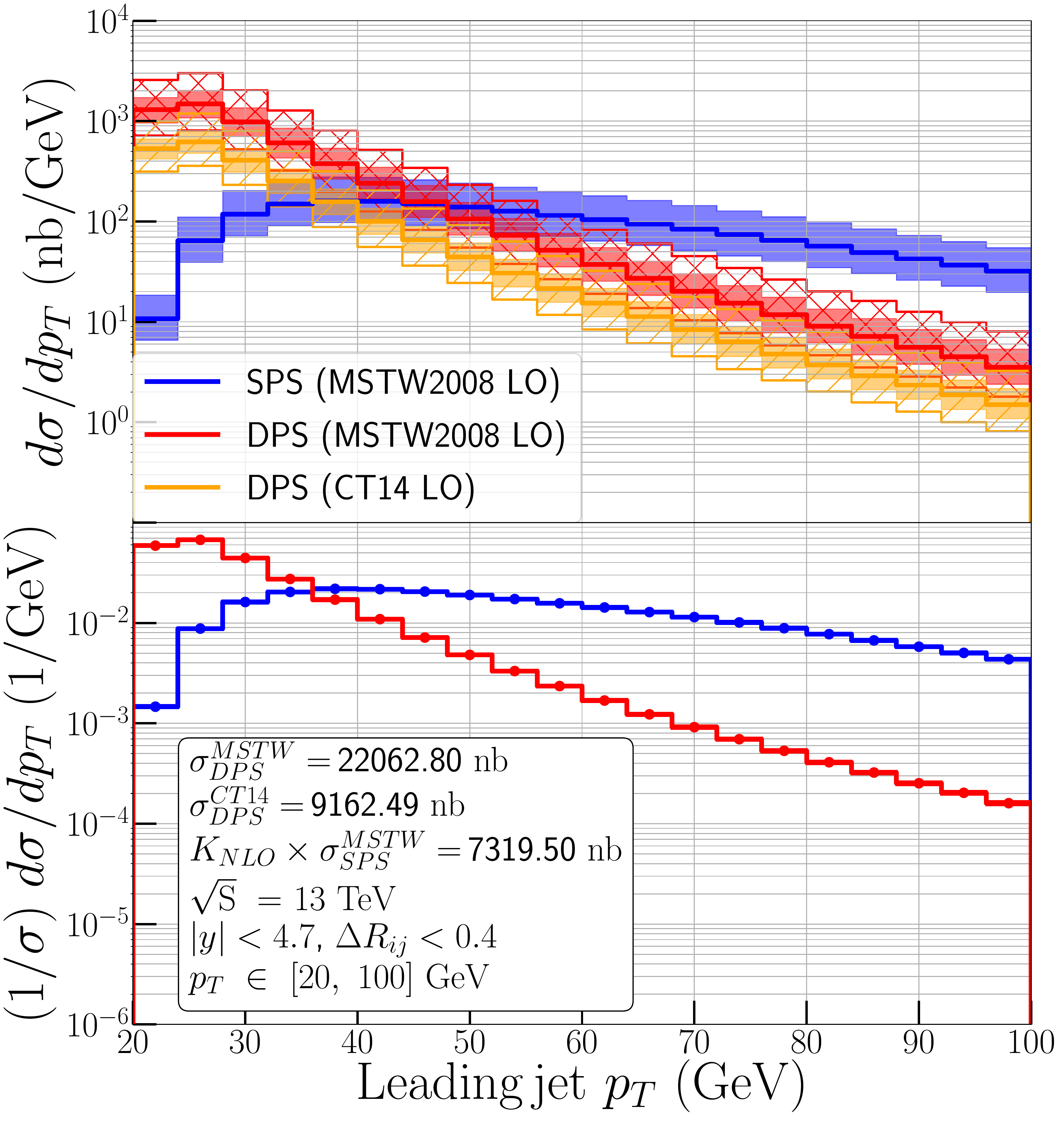}
\includegraphics[width=0.5\linewidth]{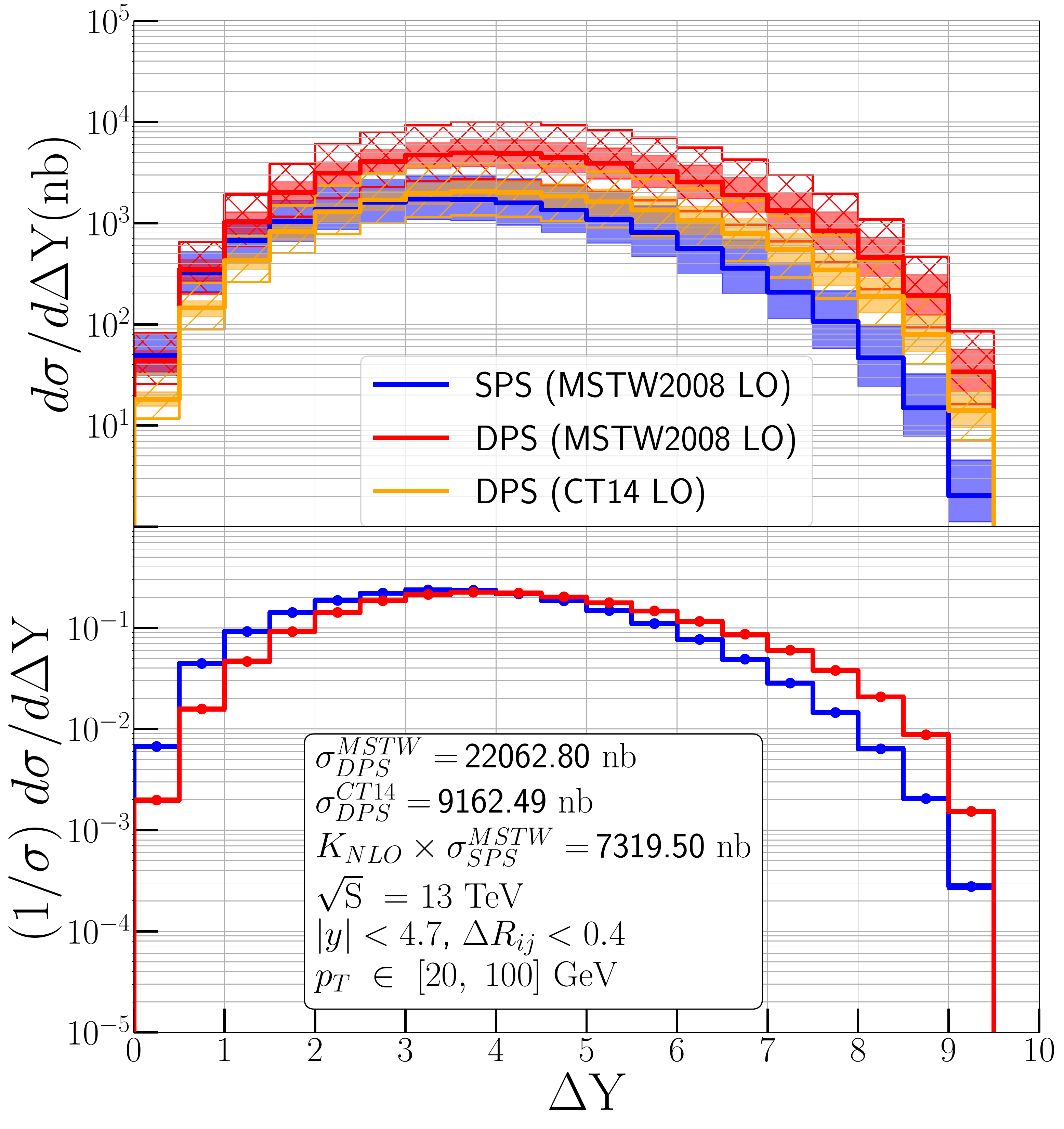}
\caption{Same as in Fig.~\ref{fig:sps_vs_dps_var2} but at $\sqrt{S} = 13 \, {\rm TeV}$.}
\label{fig:sps_vs_dps_var4}
\end{figure}

Table~\ref{tab:dps_4j_dps_vs_sps} provides indication on how the central values of the total cross sections for 7 and 13 TeV collisions behave under the chosen sets of cuts. An estimate of the scale uncertainties on the 
central values is given in Figs.~\ref{fig:sps_vs_dps_var1}-\ref{fig:sps_vs_dps_var4}. 
There we show predictions for the leading jet $\pperp$ and the rapidity difference $\dY \equiv {\rm max} |y_j- y_k|$  distributions~\cite{Kom:2011bd} for the same values of LHC energy and the same sets of cuts on jets' $p_{T,j}$ as in Table~\ref{tab:dps_4j_dps_vs_sps}.
Note that, unlike distributions in Fig.~\ref{fig:spsdps_vs_cms}, we do not combine DPS and SPS predictions together. Moreover, we plot DPS distributions generated with MSTW2008 and CT14 PDFs separately, such that the effect due to the choice of the PDF set remains clearly visible.
Shaded areas in all plots show the size of scale variation error, obtained by varying the central renormalization and factorization scales by a factor of  2 and $1/2$ simultaneously.  Hatched areas correspond to the additional variation of the $\sigeff$ parameter in the $15\pm 5$ mb range, corresponding to the measurement in four-jet final state by the ATLAS collaboration~\cite{Aaboud:2016dea}.  Histograms in  Figs.~\ref{fig:sps_vs_dps_var1}-\ref{fig:sps_vs_dps_var4}  confirm  naive expectations that lowering the cut on minimal $\pperp$ of jets leads to an increase in the DPS scattering vs. SPS scattering at low $\pperp$ of the leading jet, as well as at high $\dY$~\footnote{However, the results at high $\dY$ should be taken with caution. It is known that description of forward-backward jets with large rapidity separation, also known as Mueller-Navalet jets \cite{Mueller:1986ey}, requires accounting for ladder emission of gluons using the BFKL formalism~\cite{Fadin:1975cb, Kuraev:1976ge, Kuraev:1977fs, Balitsky:1978ic}. The DPS production of jets with large rapidity separation was studied in \cite{Maciula:2014pla}.}.  The shapes of the distributions, 
shown in lower panels of  Figs.~\ref{fig:sps_vs_dps_var1}-\ref{fig:sps_vs_dps_var4}, also become more steep as the minimal $\pperp$-cut decreases. 
Similar qualitative changes are observed as the energy of the hadronic collision increases.

In the remaining part of this paper we will present the DPS predictions obtained only with dPDFs built on the basis of the MSTW2008LO PDFs. Firstly,  the only publicly existing dPDFs package, GS09~\cite{Gaunt:2009re}, is built on the basis of the MSTW2008 LO parametrization. Secondly, Figs.~\ref{fig:sps_vs_dps_var1}-\ref{fig:sps_vs_dps_var4} clearly demonstrate that although the DPS predictions come with a large theoretical uncertainty, qualitative behaviour of the results is similar, independently of the LO PDFs used.  Furthermore, the SPS predictions we refer to~\cite{Bern:2011ep, Badger:2012pf}  are obtained using the MSTW2008 PDFs. Finally, most 
of our analysis presented in the following is concerned with shapes of the distributions, where the PDF effects mostly cancel out.%

\begin{figure}[h]
\center{\includegraphics[width=0.7\linewidth]{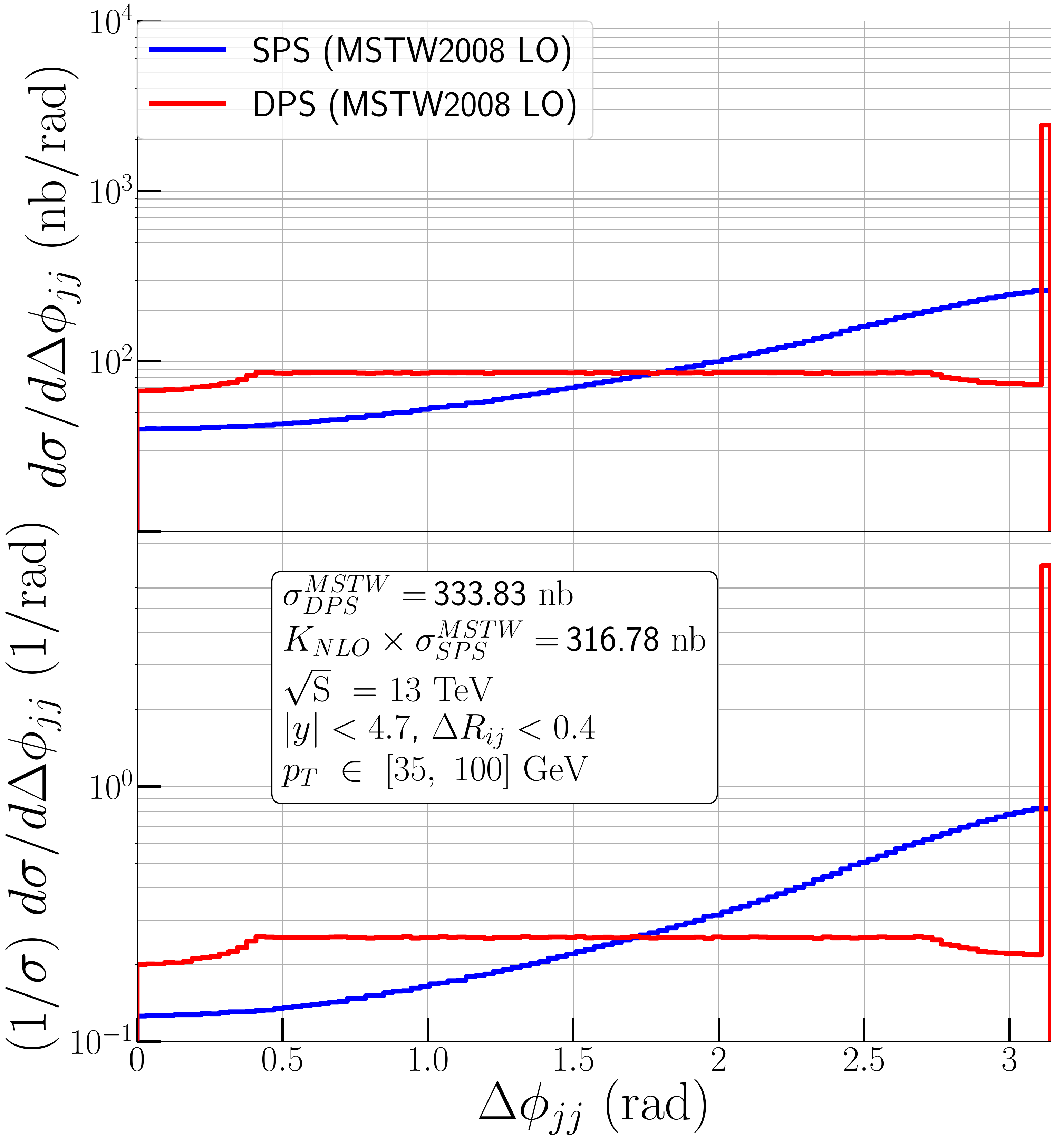}}
\caption{DPS and SPS distributions in azimuthal difference between two most remote in rapidity jets (upper panel) and their normalized shapes (lower panel) at $\sqrt S=13$ TeV.}
\label{fig:sps_vs_dps_var7}
\end{figure}

One of the observables which is considered to have a good discriminating power between DPS and SPS is the azimuthal difference between the two most remote rapidity jets, $\Delta \phi_{jj}$~\cite{Maciula:2015vza}. We show this distribution for  \mbox{$\sqrt S=13$ TeV} 
and $35\ {\rm GeV} < p_{T, j} < 100\ {\rm GeV}$. Indeed, as Fig.~\ref{fig:sps_vs_dps_var7} demonstrates, 
while away from $\Delta \phi_{jj}=\pi$ the DPS events are expected to be almost equally distributed, the SPS events prefer back-to-back configurations for the two most distant jet in rapidity, leading to a strong depletion of the cross section at  small $\Delta \phi_{jj}$. The small dents in $\Delta \phi_{jj}$ DPS distribution are due to jet clustering, since the jet separation obviously cuts out a certain number of events where the 
partons have small angular separation.

\begin{figure}[h]
\begin{minipage}[h!]{1.0\linewidth}
\center{\includegraphics[width=0.7\linewidth]{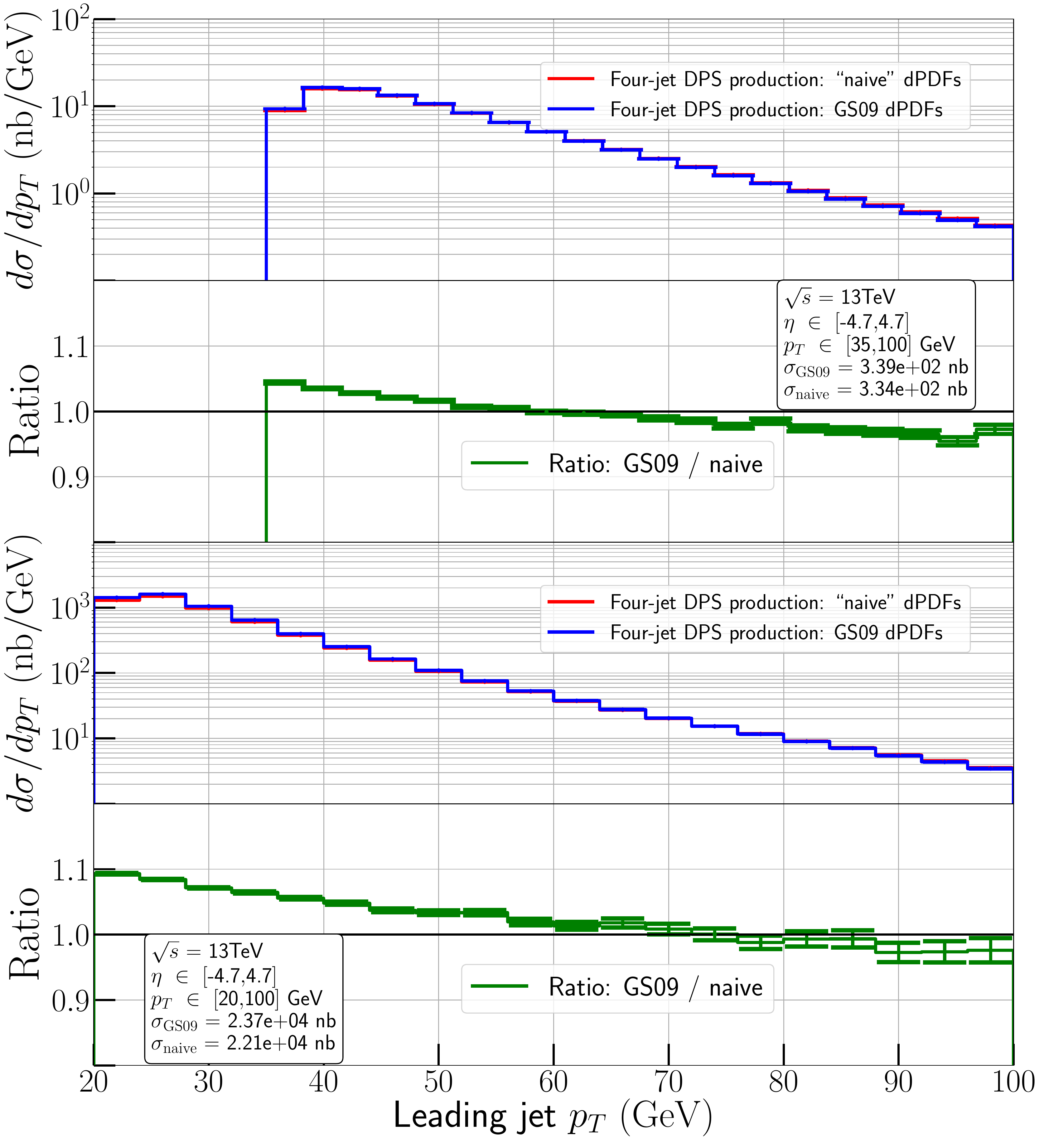}}
\end{minipage}
\caption{Leading jet $p_T$ distribution in four-jets DPS production at $\sqrt S=13$ TeV, calculated using  ``naive'' dPDFs based on MSTW2008LO and GS09 dPDFS. Upper two panels show results obtained with the \mbox{$35\ 
{\rm GeV} < p_{T, j} < 100\ {\rm GeV}$} cut, lower two panels are for 
the \mbox{$20\ {\rm GeV} < p_{T, j}  < 100\ {\rm GeV}$} cut.}
\label{fig:gs09_pT_13TeV}
\end{figure} 

\begin{figure}[h!]
\begin{minipage}[h!]{1.0\linewidth}
\center{\includegraphics[width=0.7\linewidth]{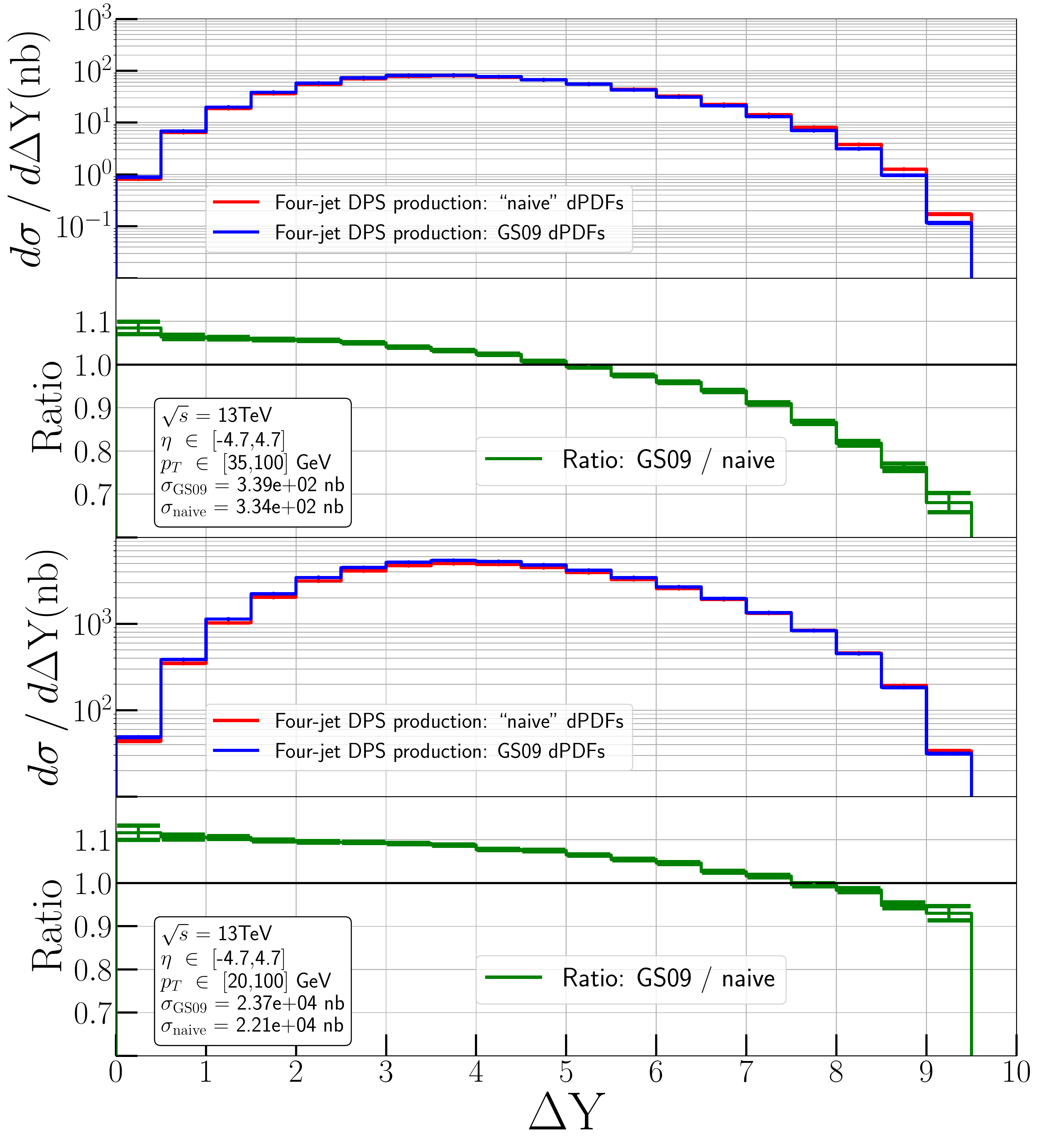}}
\end{minipage}
\caption{Maximal difference in jets' rapidity distribution in four-jets DPS production at $\sqrt S=13$ TeV, calculated using  ``naive'' dPDFs based on MSTW2008LO and GS09 dPDFS. Upper two panels show results obtained with the  $35\ {\rm GeV} < p_{T, j} < 100\ {\rm GeV}$ cut, lower two panels are for the $20\ {\rm GeV} < p_{T, j}  < 100\ {\rm GeV}$ cut.  }
\label{fig:gs09_dY_13TeV}
\end{figure} 

Our results given in 
Figs.~\ref{fig:sps_vs_dps_var1}-\ref{fig:sps_vs_dps_var4} agree with the results of  \cite{Maciula:2015vza} at qualitative level. We shall note, however, that the qualitative comparison between $\Delta \phi_{jj}$ distributions in \cite{Maciula:2015vza} and our results in Fig.~\ref{fig:sps_vs_dps_var7} shows some differences at small and large values of $\Delta \phi_{jj}$. More precisely, 
the $\Delta \phi_{jj}$ distribution given in \cite{Maciula:2015vza} does not have a peak at $\Delta \phi_{jj} = \pi$ as well as curvatures at small and large values of  $\Delta \phi_{jj}$ caused by the  $R_{ij}=\sqrt{(y_i-y_j)^2+(\phi_i-\phi_j)^2}>0.4$ separation cut.

In a next step we investigate the impact of longitudinal correlations within the pairs of partons taking part in the double scattering, as implemented in the GS09 package, on the predictions. To this end, in Figs. ~\ref{fig:gs09_pT_13TeV} and \ref{fig:gs09_dY_13TeV} we compare the leading jet $\pperp$ distributions at $\sqrt S=13$ TeV obtained using the GS09 dPDFs and using ``naive'' dPDFs for the same two sets of cuts on  $p_{T,j}$ 
as above. For the whole range of the $\pperp$ considered, as well as for the practically accessible values of $\dY$, the difference between the two distributions is not more than 10\%.  At  very high values of $\dY$, where large $x$ values are probed, this difference grows bigger. However, given that the effect is overall small, we will not consider it in further 
studies and from now on only use the ``naive'' dPDF modelling.

\subsection{Impact of QCD radiation}

We now move on to discussion of the impact of the initial and final state 
radiation on the DPS and SPS four-jet predictions. We use LHE files with events generated at $\sqrt{S} = 13 \, {\rm TeV}$ where all final state partons have \mbox{$\pperp \in [35, 150]$ GeV} and $R_{ij}>0.4$.
Our DPS predictions rely on the
assumption of factorization of generalized double parton distribution functions into longitudinal and transverse dependent parts as in Eq.~\ref{eq:factorization}. As a consequence of   Eq.~\ref{eq:factorization}, the effective interaction area $\sigma_{\rm eff}$ turns into a constant parameter defined as in Eq.~\ref{eq:sig_eff_def}\footnote{Similar approach is also can be used to describe triple parton interactions, see \textit{e.g.} \cite{dEnterria:2016ids}. }. 
In our simulations we choose the actual value of $\sigma_{\rm eff} = 15$ mb to get an agreement with the four-jet measurements of $\sigma_{\rm eff}$ performed by ATLAS collaboration \cite{Aaboud:2016dea}.  As explained above, the DPS predictions are obtained using the ``naive'' dPDFs based on the MSTW2008LO set. Since the effect of adding radiation is expected to be universal, \textit{i.e.} 
independent on the form of the dPDFs, this is sufficient for the discussion to follow. The LHE files with the DPS events are generated by our in-house C++ code, while the SPS events are generated with the MadGraph event 
generator~\cite{Alwall:2014hca}. In the following, the SPS events are not 
longer multiplied by a $\rm K$-factor.

In order to add radiation effects to our simulations we supply SPS and DPS LHE files to the \pythia event generator. To shower the SPS events we follow a standard procedure where we first  create LHE files using the MadGraph package and supply them to  \pythia to add ISR and FSR to our simulations. The DPS case, however, is more involved: since the standard LHE event record contains information only about one hard interaction, one has 
to adapt the LHE files to contain  also information about the second hard 
interaction in a DPS process, see Appendix \ref{s:double_lhe}. The resulted ``double'' LHE  files are passed to the \pythia code which was modified for this purpose \cite{TSjostrand:private_com}. In this way, the information on two individual hard scatterings propagates through the shower.

The DPS events are then showered in the   interleaved way in accordance with the MPI model of \pytppp \cite{Corke:2010yf}. The approach of \cite{Corke:2010yf} has several advantages comparing to the simplified case where two hard processes constituting  a DPS event  are showered independently from each other. The DPS showering  performed in accordance with \cite{Corke:2010yf} implies that ISR and FSR cascades satisfy constraints imposed by energy and momentum conservation at each evolution step. In particular, in this way it is ensured that the kinematic constraint on Bjorken $x$-es given by Eq.~\ref{eq:factorization_dPDF_naive} is preserved by the ISR cascades at each step of the ISR evolution. Another advantage of the approach of  \cite{Corke:2010yf} is the fact that the ISR cascade which first reaches the proton reduces amount  of energy and momentum available for the second ISR cascade. Finally, we shall note that the interleaved ISR evolution accounts for the changes in the parton content of the beam remnants which introduces non-trivial correlations in flavour of interacting partons \cite{Sjostrand:2004pf}. Therefore, our simulations provide a better treatment of the beam remnants comparing to the case where one ``showers'' two hard processes completely independently.\footnote{
We shall note that the \pythia approach to the ISR evolution of the  DPS (MPI)  events does not take into account cases when two ISR cascades merge into a single ISR cascade  at certain step of backward evolution (so called ``$1v2$'' events). Whereas the first results on implementation of such effects in to the parton-shower framework for the same-sign $WW$ production were recently reported \cite{Cabouat:2019gtm}, the implementation of the aforementioned effects  for the four-jet DPS production is highly non-trivial \cite{Diehl:2017kgu} and is beyond the scope of this work.}

Next, we discuss our results for the 4-jet production.
Apart from showing the absolute value of the differential cross sections, we also study their shapes. In all plots in this section, we show an estimate of the statistical error on the (nominal or normalized) cross section in each bin.

It is well-known that if the same $\pperp$ cut is applied on the two observed jets the higher-order calculations for di-jet production become unstable due to restricted phase space available for soft gluon radiation~\cite{Frixione:1997ks}. Although the four-jet production is not affected by this problem, the contributions to the DPS production which originate from double di-jet production might be. For that reason, apart from applying 
symmetric cuts of  $p_{T, j}> 35$ GeV on all four jets, we additionally introduce asymmetric cuts by requesting  $p_{T, 1} > 55$ GeV for the leading jet and $p_{T, j} > 35$ GeV for the remaining jets.~\footnote{However, 
one should note that in order to make sure no instability of that sort affects the DPS calculations entirely, fully asymmetric cuts, \textit{i.e.} with different minimal $\pperp$ for each jet, would be required. Given that the cuts need to be sufficiently separated, a measurement of DPS based on events selected in that way might be very difficult due to low statistics.}

\begin{figure}
\includegraphics[width=0.49\linewidth]{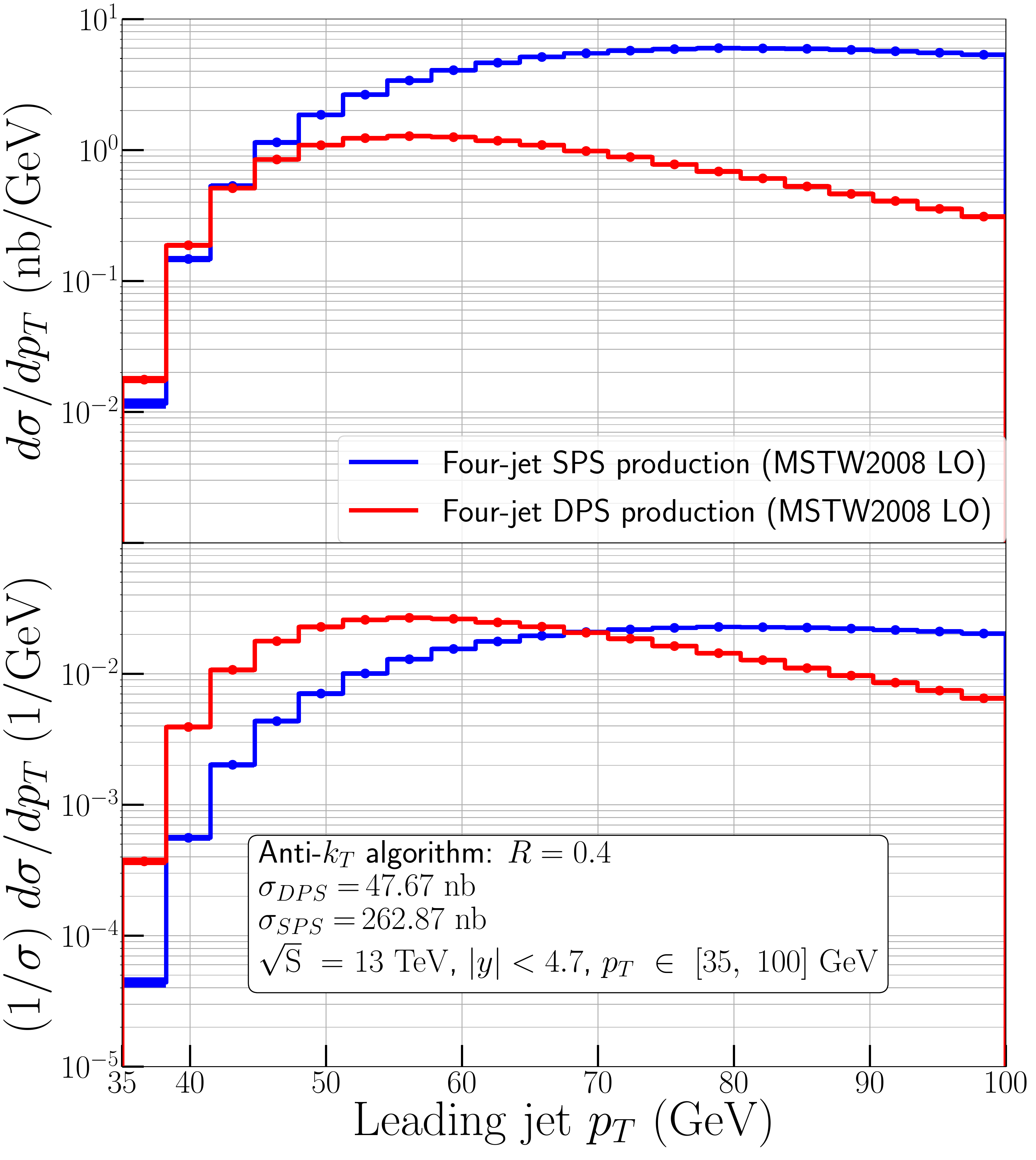}
\includegraphics[width=0.49\linewidth]{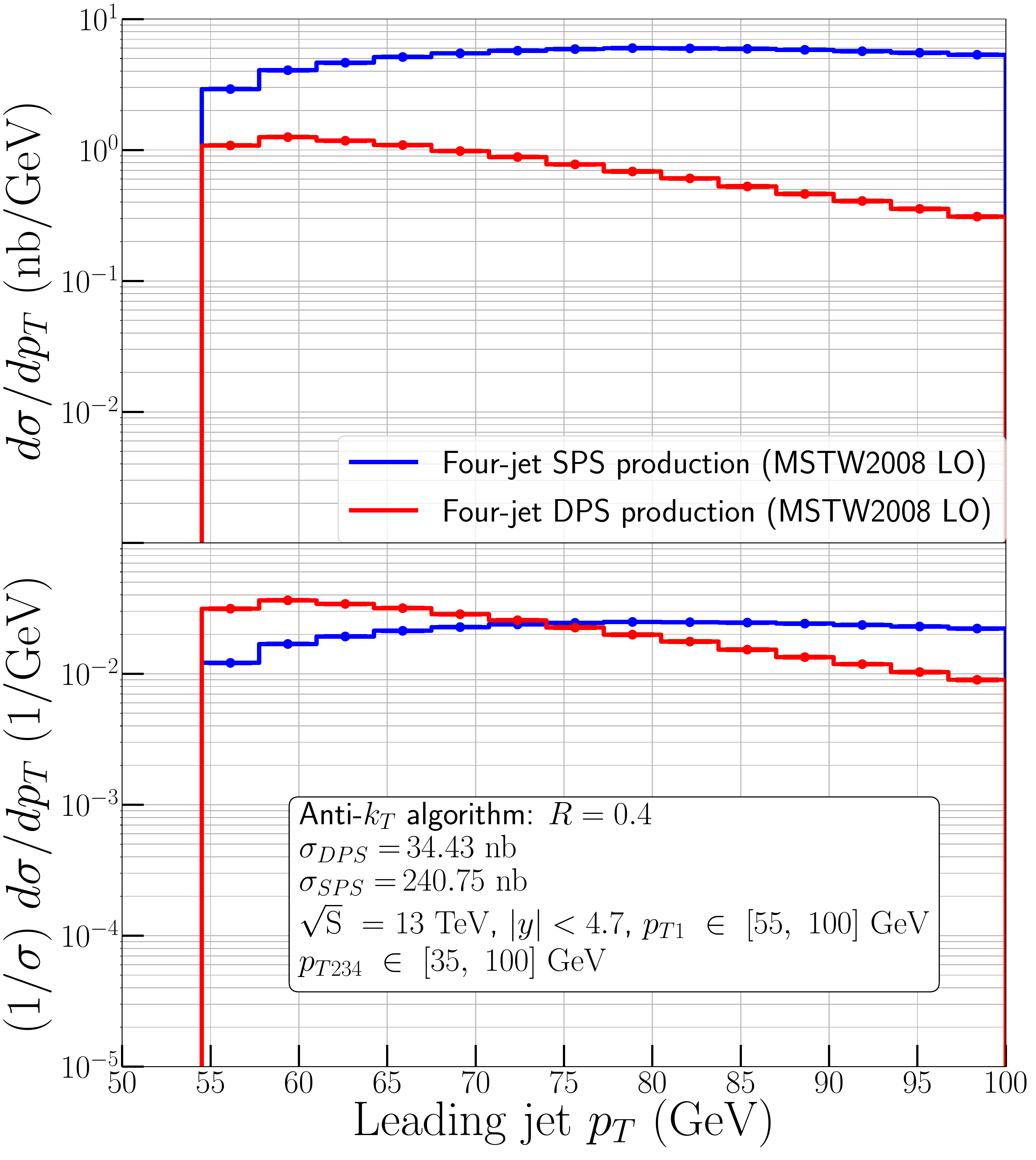}
\caption{DPS and SPS leading jet distributions for four-jet production at 
$\sqrt S =13$ TeV with symmetric cuts $p_{T, j}  \in [35, 100]$ GeV (left) and asymmetric cuts $p_{T, 1}  \in [55, 100]$ GeV, $p_{T, 2,3,4}  \in 
[35, 100]$ GeV (right) after accounting for effects of radiation. Lower panels show normalized distributions.}
\label{fig:sps_vs_dps_ps1}
\end{figure}

\begin{figure}
\includegraphics[width=0.49\linewidth]{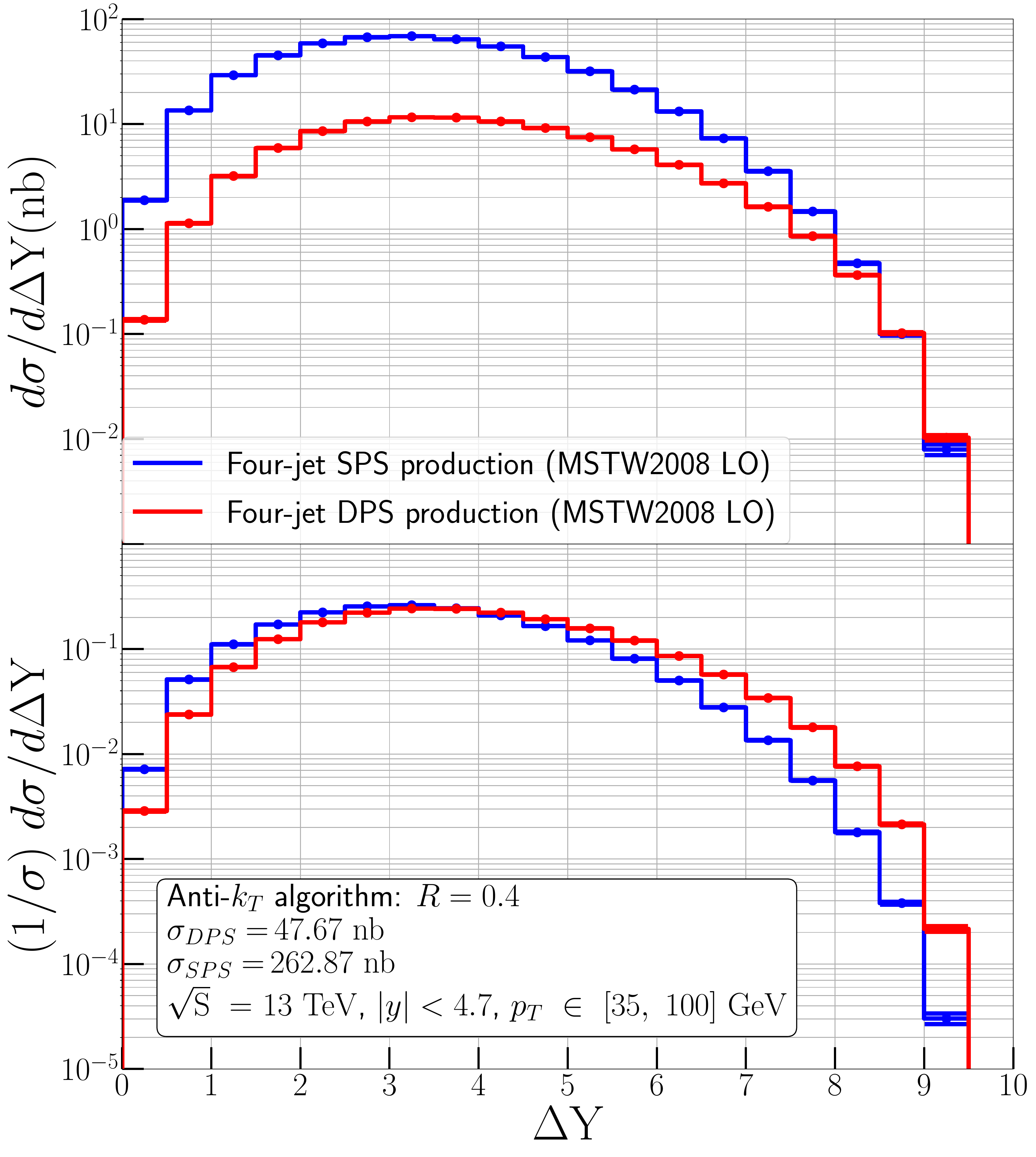}
\includegraphics[width=0.49\linewidth]{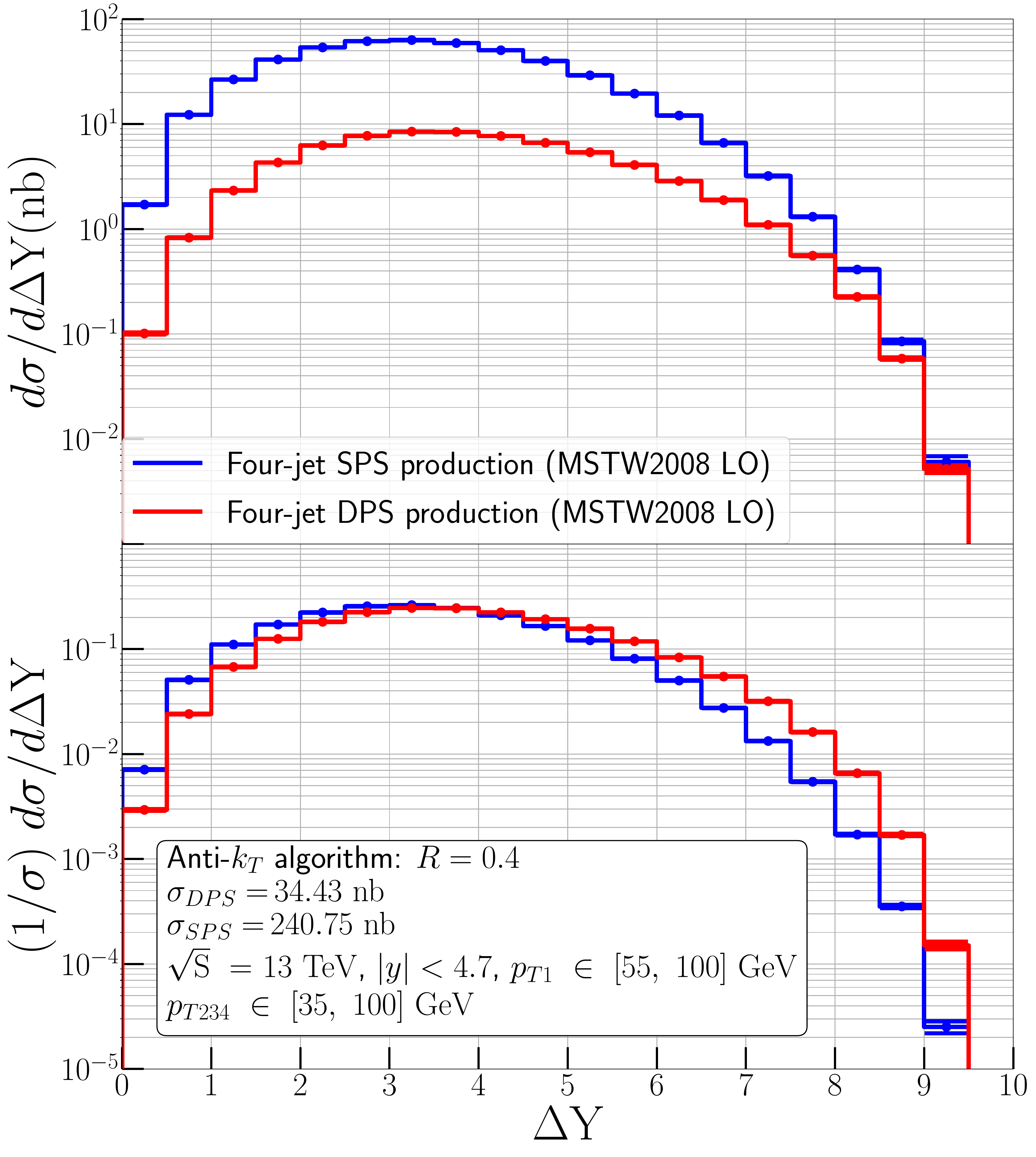}
\caption{DPS and SPS maximal rapidity difference $\Delta {\rm Y} = {\rm 
max}|y_i - y_j|$ distributions for four-jet production at $\sqrt S =13$ 
TeV with symmetric cuts $p_{T, j}  \in [35, 100]$ GeV (left) and asymmetric cuts $p_{T, 1}  \in [55, 100]$ GeV, $p_{T, 2,3,4}  \in [35, 100]$ GeV (right) after accounting for effects of radiation. Lower panels show normalized distributions.}
\label{fig:sps_vs_dps_ps2}
\end{figure}

In Figs.~\ref{fig:sps_vs_dps_ps1} and  ~\ref{fig:sps_vs_dps_ps2} we present distributions in the leading jet $\pperp$ and $\dY$ studied in the previous section, now including the radiation effects. 
We find that adding radiation decreases the total cross section for both the SPS production and DPS production, see also the first row of Table~\ref{tab:dps_4j_impact_of_ps_1e7}.  The exact value of the reduction factor 
depends on the set-up of the calculations, in particular the chosen set of kinematical cuts, but in general the DPS predictions are more impacted by the radiation. As can be seen from the leading jet $\pperp$ distribution discussed in the previous section, most of the partonic DPS events happen to have jets with very low $\pperp$, just passing the cut. This makes 
the DPS distribution much more vulnerable to the effects of radiation, meaning the adjustment of jet's $\pperp$ due to radiation can cause a substantial proportion of the events not to pass the cuts. The same is not true for SPS, where the peak of the  leading jet's  $\pperp$ distribution is 
at much higher values of $\pperp$. Having studied origin of jets passing the selection cuts, it also appears the SPS events have a higher partonic 
center of mass energy $\sqrt {\shat}$,  and correspondingly higher values 
of Bjorken's $x$ carried by the partons participating in a collision, in comparison with individual DPS collisions. As a consequence, radiation in 
the SPS events more often gives rise to an extra jet passing the cut. In this way some SPS events which at partonic level would have not passed the cut are accepted now.  The same effect is much suppressed for the DPS due to lower $x$  values of  incoming partons. The normalized distributions, on the contrary, seem to preserve their main features first observed at the partonic level, \textit{cf.} Fig.~\ref{fig:sps_vs_dps_ps1}. For example,  the DPS leading jet's  $\pperp$  distribution, though it gets flattened out,  remains more peaked at small $\pperp$  than the SPS one, or the $\dY$ DPS distribution stays flatter at very high $\dY$.  It is also worthy noticing, that showering alters the leading jet $\pperp$ distributions more than the $\dY$ distribution, in accordance with the observation of higher impact of radiation on the observables of the ``transverse'' type. Introducing the asymmetric set of cuts does not lead to significant changes in the behaviour of the distributions.

The whole analysis shows clearly that accounting for radiation can dramatically change the size of the four-jet cross sections calculated under assumptions of certain sets of cuts, and conclusions derived from analysis of the partonic level do not hold after a more realistic simulation of the production process is employed. It is interesting to note that a similar dampening effect of radiation on DPS predictions have been observed in the $k_T$ factorization framework~\cite{Kutak:2016mik, Kutak:2016ukc}. However, a comparison between Fig.~\ref{fig:sps_vs_dps_var3} and Figs.~\ref{fig:sps_vs_dps_ps1}, ~\ref{fig:sps_vs_dps_ps2} indicates that the radiation effects can also influence the differences between shapes of DPS and SPS distributions. Hence, even if only the information on normalized distributions is used for distinguishing between SPS and DPS, the impact of radiation should be taken into account.

\begin{figure}
\includegraphics[width=0.49\linewidth]{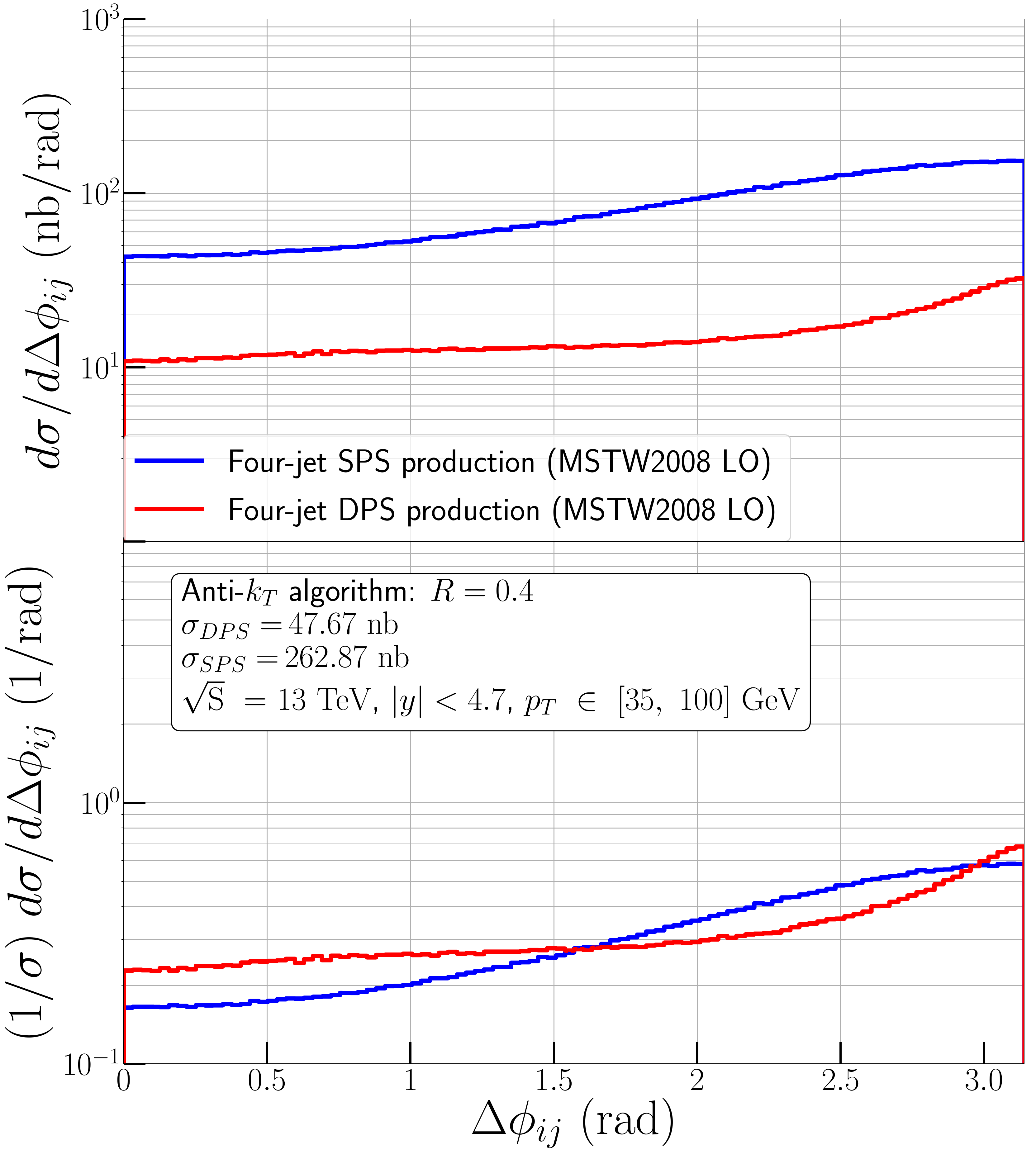}
\includegraphics[width=0.49\linewidth]{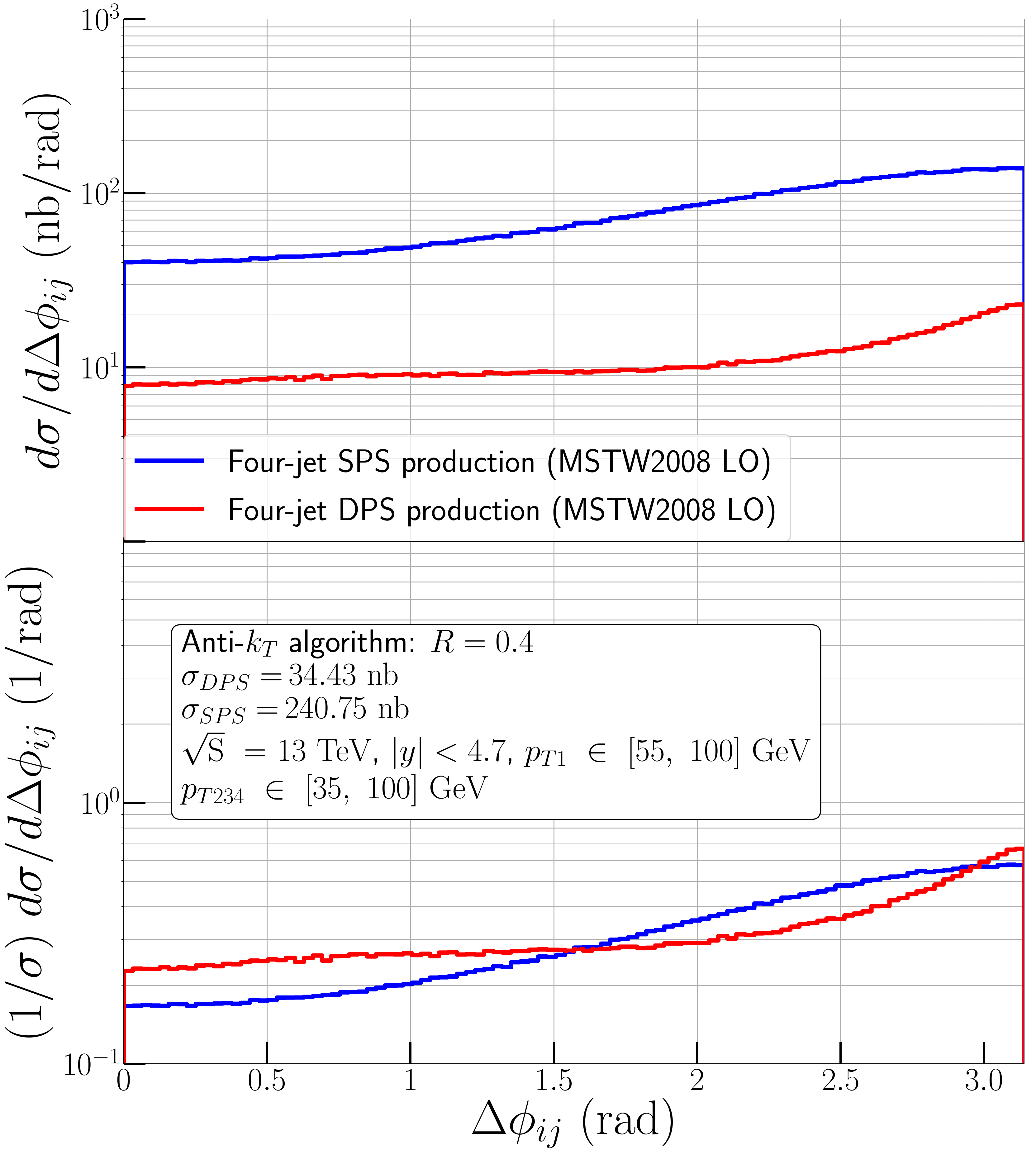}
\caption{DPS and SPS  distributions in azimuthal difference between two most remote jets in rapidity $\Delta \phi_{jj}$ for four-jet 
production at \mbox{$\sqrt S =13$ TeV} with symmetric cuts $p_{T, j}  \in [35, 100]$ GeV (left) and asymmetric cuts \mbox{$p_{T, 1}  \in [55, 100]$ GeV}, $p_{T, 2,3,4}  \in [35, 100]$ GeV (right) after accounting for effects of radiation. Lower panels show normalized distributions.
 }
\label{fig:sps_vs_dps_ps3}
\end{figure}

Next we study the impact of radiation on the differential cross section in $\Delta \phi_{jj}$, discussed at the partonic level in the previous section, see Fig.~\ref{fig:sps_vs_dps_ps3}.  While in that case DPS distribution was higher than the SPS for smaller angular differences $\Delta \phi_{jj}$, adding radiation turns the SPS production mechanism  into a dominant one everywhere. As expected, the typical peak in the DPS distributions calculated at LO due to back-to-back configurations, observed here at maximal value of  $\Delta \phi_{jj}$, vanishes once adtional radiation is introduced. Furthermore, the radiation also changes the shape of the DPS distribution from flat to rising at higher $\Delta \phi_{jj}$, making it almost indistinguishable from the shape of the SPS distribution.  This indicates that the $\Delta \phi_{jj}$ observable does not deliver an efficient discrimination between DPS and SPS mechanisms, in this setup of cuts. 
 However, as we show later, the discrimination power of various observables can be improved on if additional cuts are introduced.

In the following we are going to use different combinations of commonly used DPS-sensitive observables constructed to increase the fraction of DPS 
events. The general idea behind construction of such observables relies on different kinematics of jets produced via DPS and SPS mechanisms. Since 
the DPS events are essentially constructed out of two di-jet events, we expect them to  give rise to two di-jet topologies well separated in rapidity and azimuthal angle $\phi$ and also well balanced in $p_T$.

\begin{figure}
\includegraphics[width=0.49\linewidth]{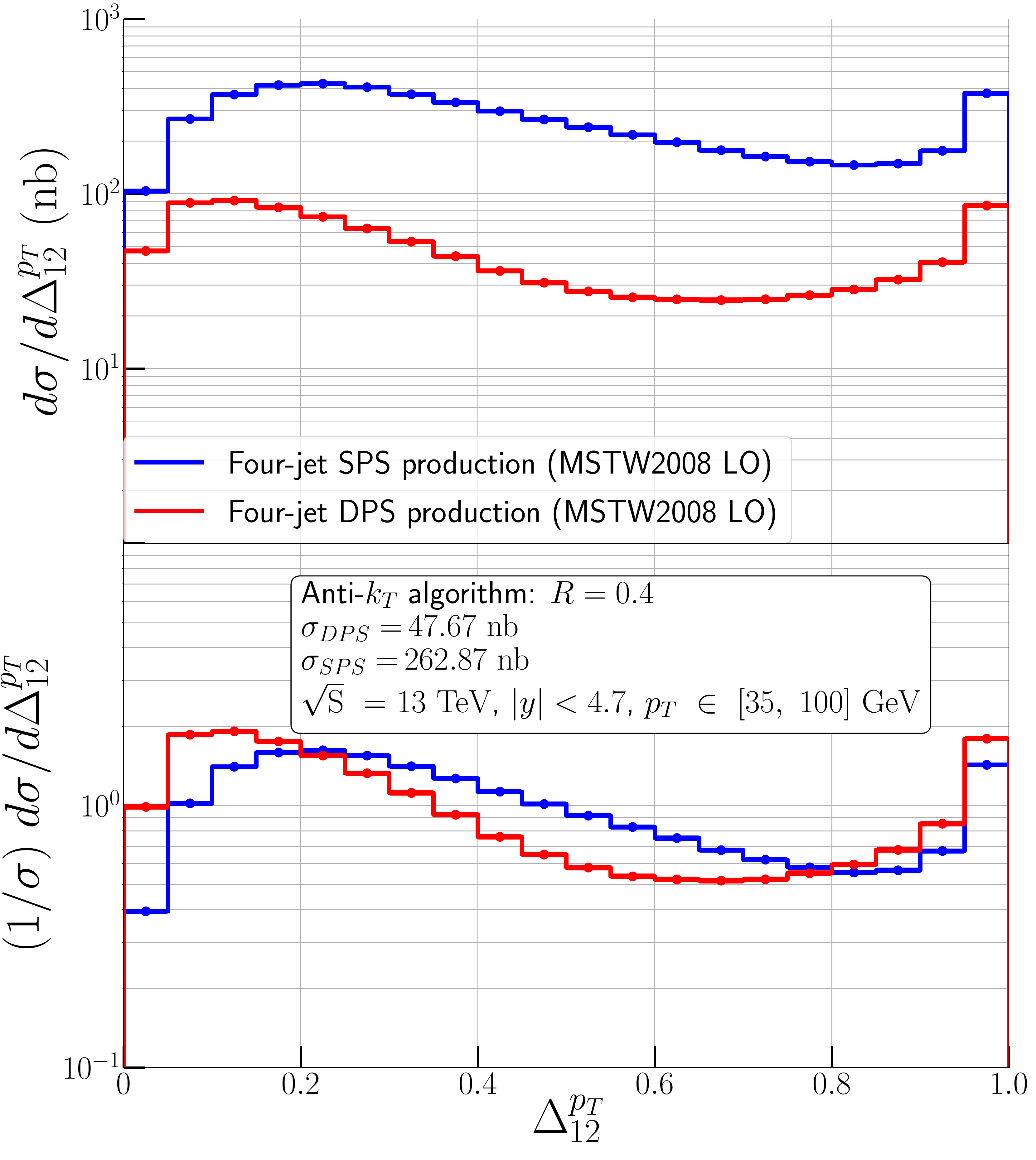}
\includegraphics[width=0.49\linewidth]{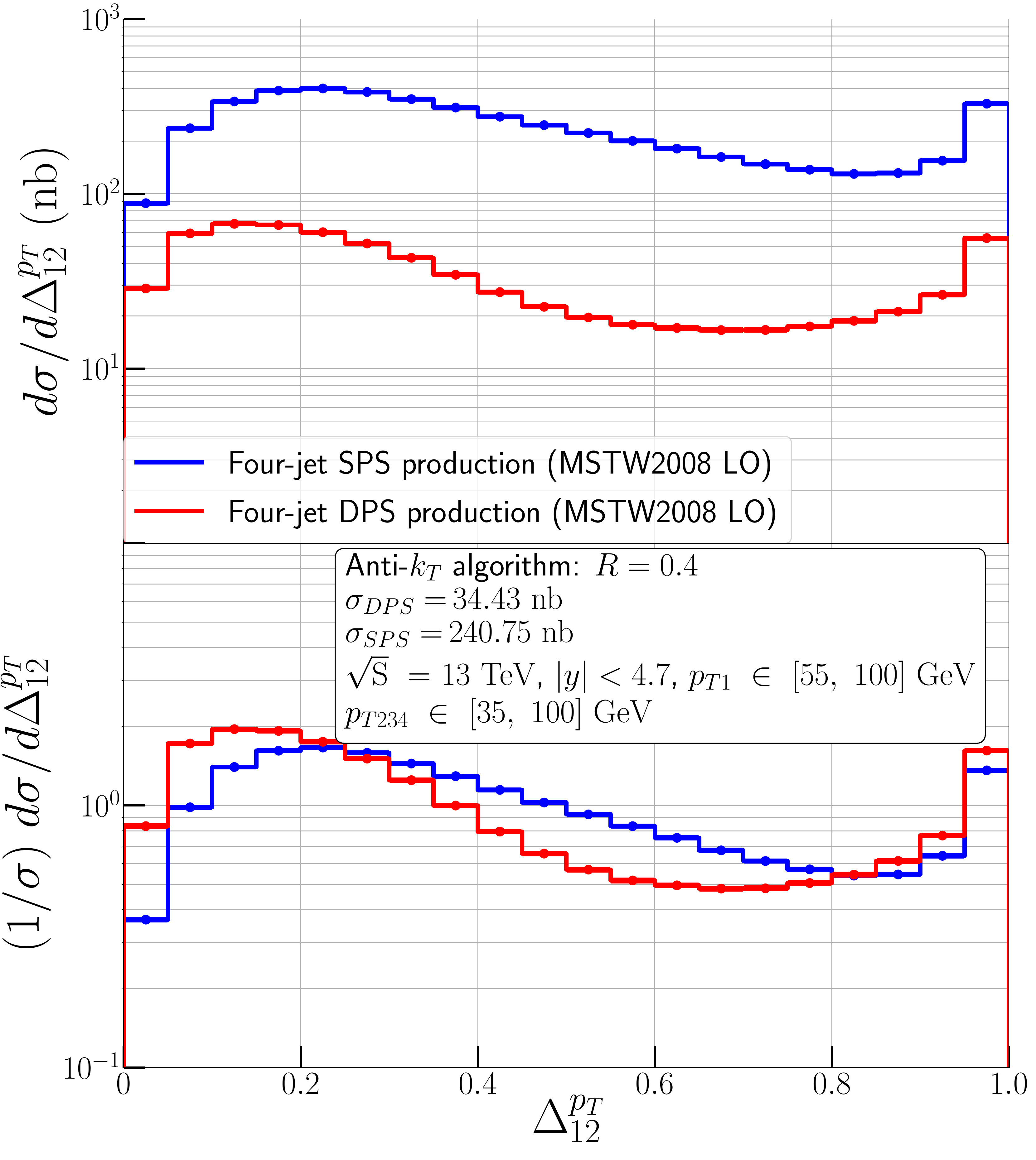}

\caption{DPS and SPS  distributions in transverse momentum imbalance between two hardest jets, $\Delta_{12}^{p_T}$, for four-jet production 
at \mbox{$\sqrt S =13$ TeV} with symmetric cuts $p_{T, j}  \in [35, 100]$ GeV (left) 
and asymmetric cuts \mbox{$p_{T, 1}  \in [55, 100]$ GeV}, $p_{T, 2,3,4}  \in [35, 100]$ GeV (right) after accounting for effects of radiation. Lower panels show normalized distributions.}
\label{fig:sps_vs_dps_ps4}
\end{figure}

\begin{figure}
\includegraphics[width=0.49\linewidth]{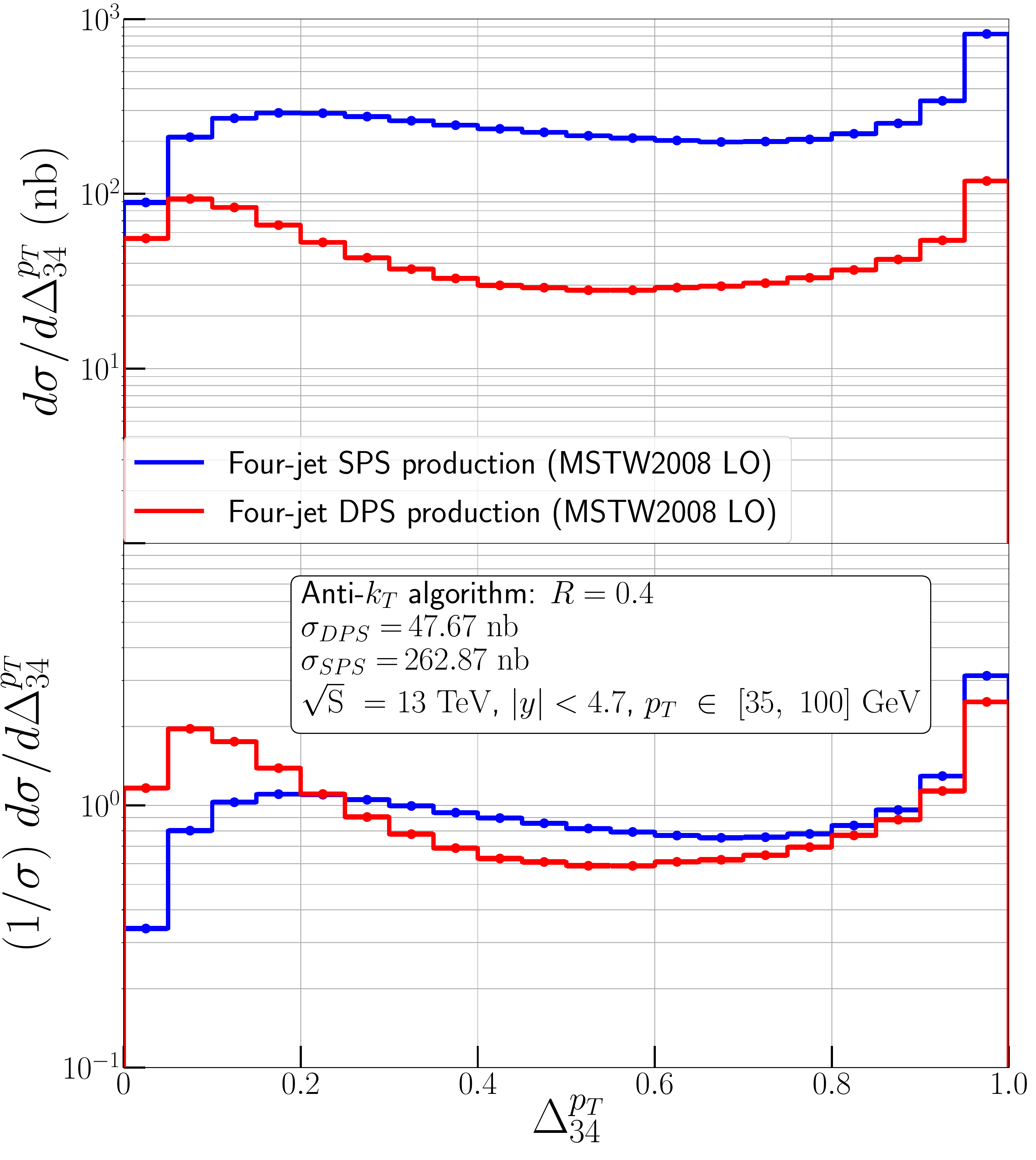}
\includegraphics[width=0.49\linewidth]{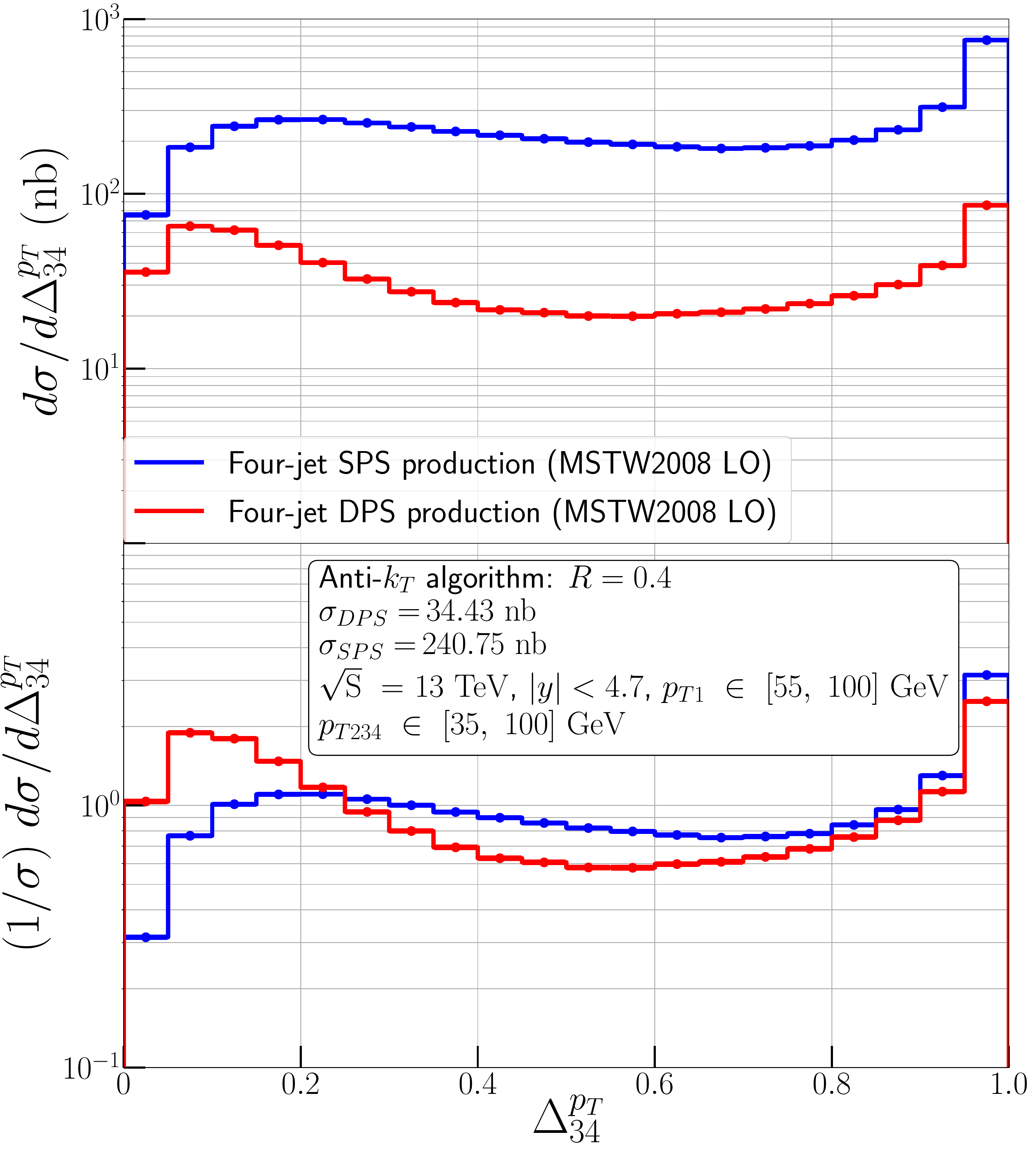}

\caption{DPS and SPS  distributions in transverse momentum imbalance between two softest jets, $\Delta_{34}^{p_T}$, for four-jet 
production at \mbox{$\sqrt S =13$ TeV} with symmetric cuts $p_{T, j}  \in [35, 100]$ GeV (left) and asymmetric cuts 
\mbox{$p_{T, 1}  \in [55, 100]$ GeV}, \mbox{$p_{T, 2,3,4}  \in [35, 100]$ 
GeV} (right) after accounting for effects of radiation. Lower panels show 
normalized distributions.}
\label{fig:sps_vs_dps_ps5}
\end{figure}

\begin{figure}
\includegraphics[width=0.49\linewidth]{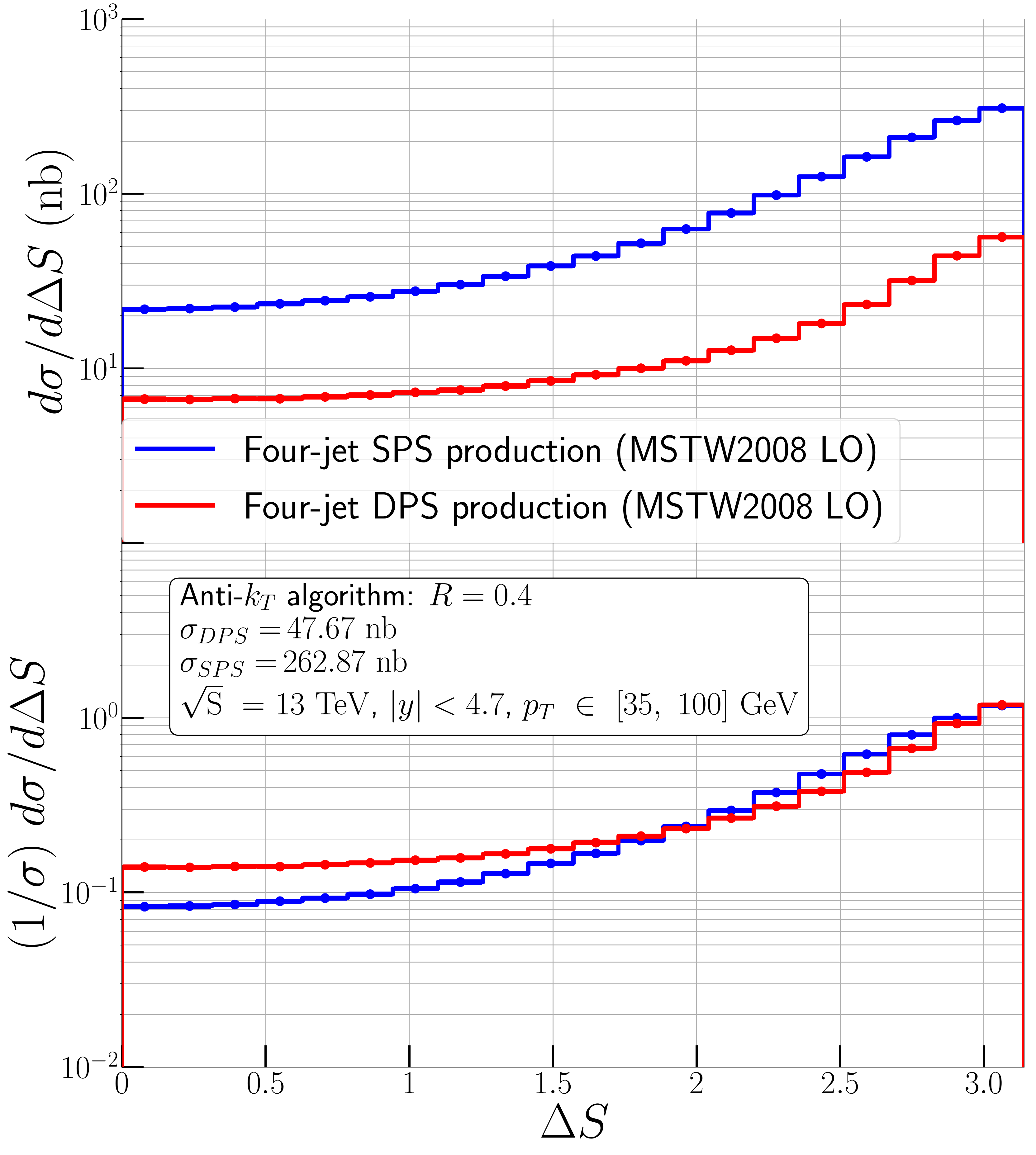}
\includegraphics[width=0.49\linewidth]{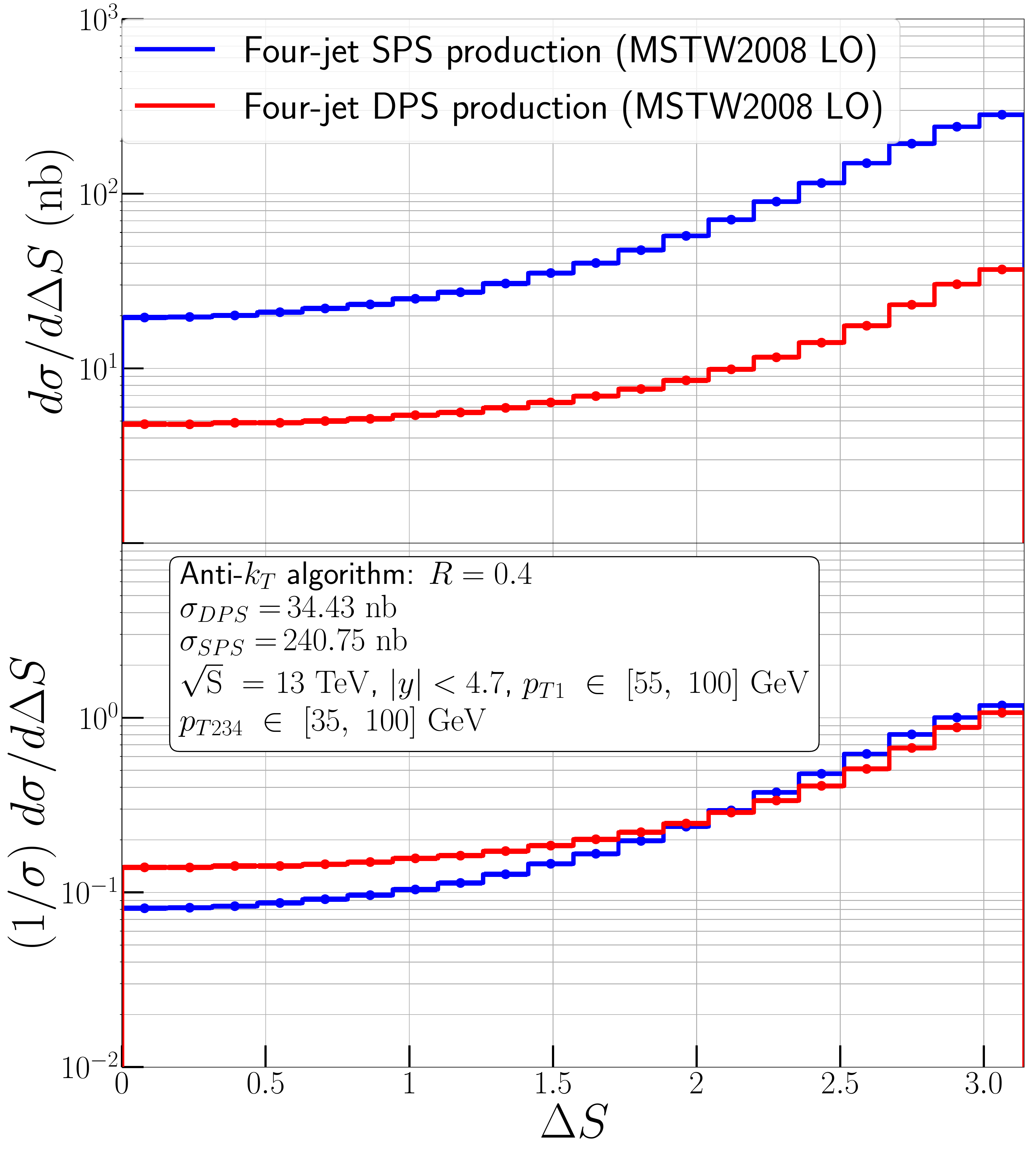}

\caption{DPS and SPS  distributions in $\Delta S$ for four-jet production 
at $\sqrt S =13$ TeV with symmetric cuts $p_{T, j}  \in [35, 100]$ GeV (left) and asymmetric cuts \mbox{$p_{T, 1}  \in [55, 100]$ GeV}, $p_{T, 2,3,4}  \in [35, 100]$ GeV (right) after accounting for effects of radiation. Lower panels show normalized distributions.}
\label{fig:sps_vs_dps_ps6}
\end{figure}

\begin{figure}
\includegraphics[width=0.49\linewidth]{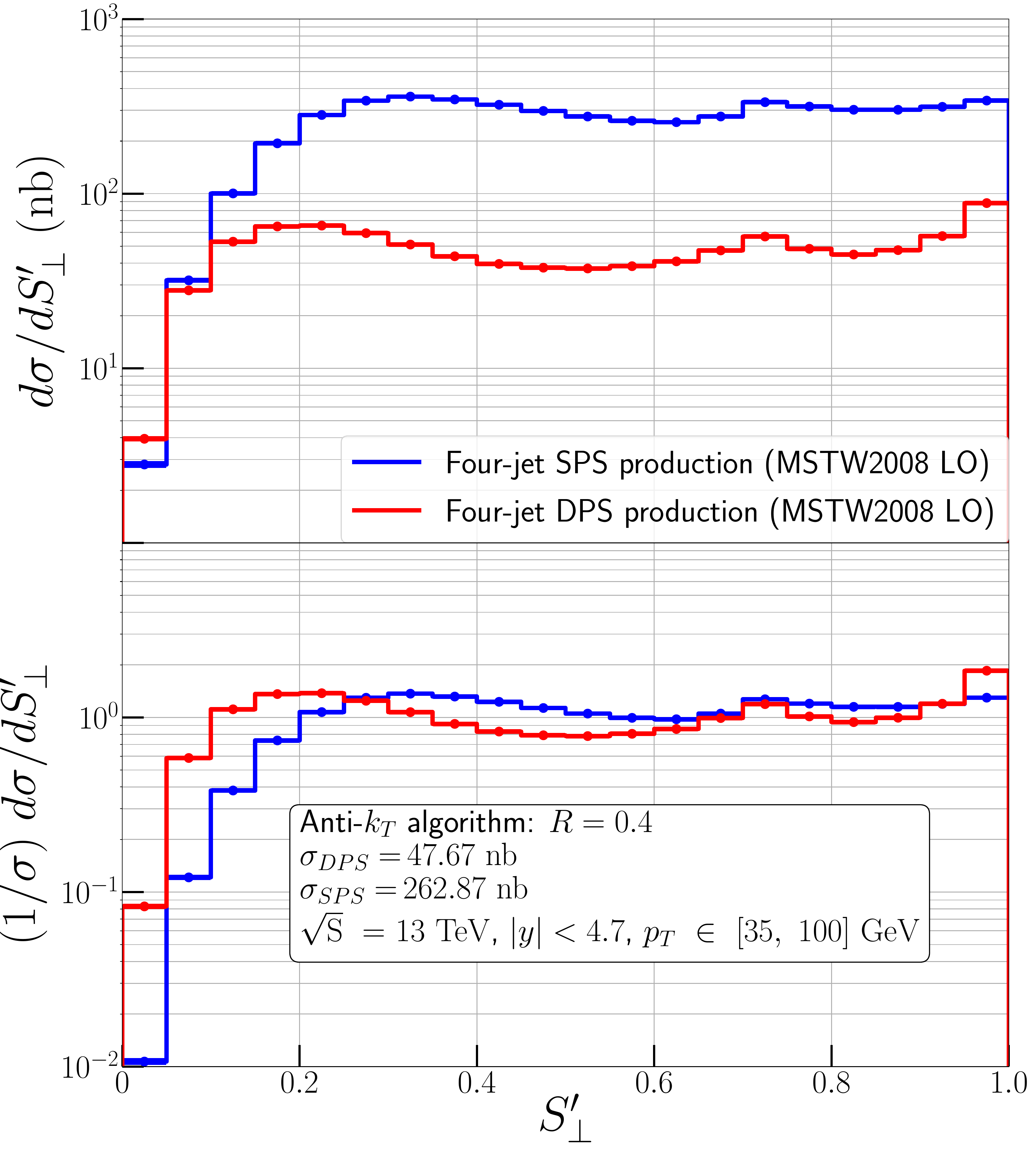}
\includegraphics[width=0.49\linewidth]{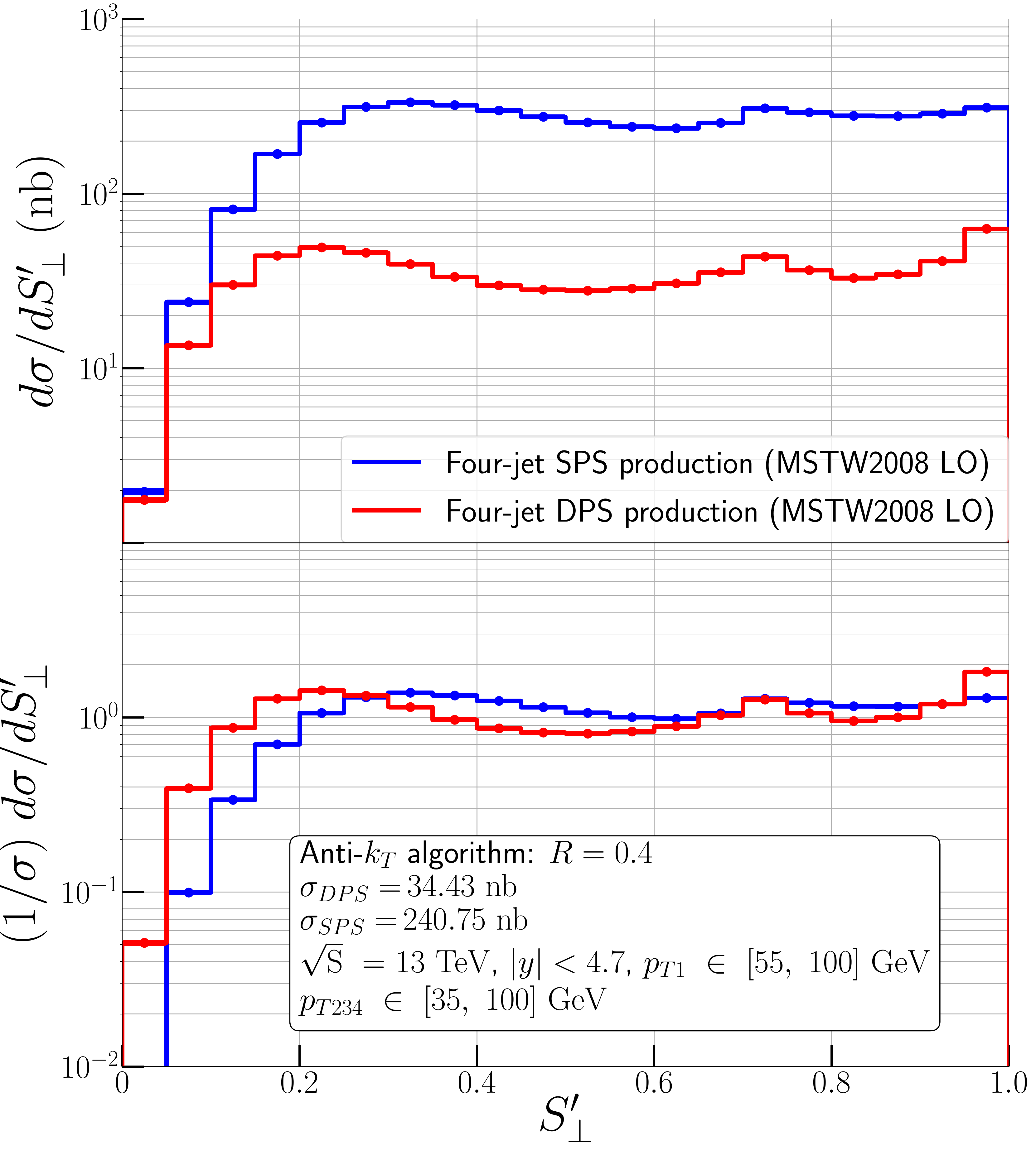}

\caption{DPS and SPS  distributions in $S'_\perp$ for four-jet 
production at \mbox{$\sqrt S =13$ TeV} with symmetric cuts $p_{T, j}  \in [35, 100]$ GeV (left) 
and asymmetric cuts $p_{T, 1}  \in [55, 100]$ GeV, $p_{T, 2,3,4}  \in [35, 100]$ GeV (right) after accounting for effects of radiation. 
Lower panels show normalized distributions.}
\label{fig:sps_vs_dps_ps7}
\end{figure}

Using the aforementioned consideration as a guiding line we can study discriminating power of different DPS sensitive observables.
In Figs.~\ref{fig:sps_vs_dps_ps4}, ~\ref{fig:sps_vs_dps_ps5}, ~\ref{fig:sps_vs_dps_ps6} and ~\ref{fig:sps_vs_dps_ps7} we show the absolute and normalized DPS and SPS differential cross sections in other commonly studied 
observables in this context, such as the momentum imbalances 
\begin{eqnarray}
	\Delta_{12}^{p_T} &\equiv& \frac{|\vec{p}_{T1} + \vec{p}_{T2}|}{|\vec{p}_{T1}| + |\vec{p}_{T2}|},
	\label{eq:DpT12}\\
	\Delta_{34}^{p_T} &\equiv& \frac{|\vec{p}_{T3} + \vec{p}_{T4}|}{|\vec{p}_{T3}| + |\vec{p}_{T4}|},
	\label{eq:DpT34}
\end{eqnarray}
as in four-jet DPS ATLAS \cite{Aaboud:2016dea} measurements,  and the closely related observables
\begin{eqnarray}
	\Delta S &\equiv& \arccos
					\left(
			 			\frac{ \left(\vec{p}_{T1} + \vec{p}_{T2}\right) \cdot \left(\vec{p}_{T3} + \vec{p}_{T4}\right)   }
                    		     { |\vec{p}_{T1} + \vec{p}_{T2}| |\vec{p}_{T3} + \vec{p}_{T4}|} 
                    	\right),
	\label{eq:dS}\\                     
	S'_\perp &\equiv& \frac{1}{\sqrt{2}} \sqrt{  \left( \Delta_{ij}^{p_T} \right)^2   + \left( \Delta_{kl}^{p_T} \right)^2 },
	\label{eq:dS_Berger}
\end{eqnarray}
defined in \cite{Blok:2015rka} and \cite{Berger:2009cm} respectively.

\begin{figure}
\includegraphics[width=0.49\linewidth]{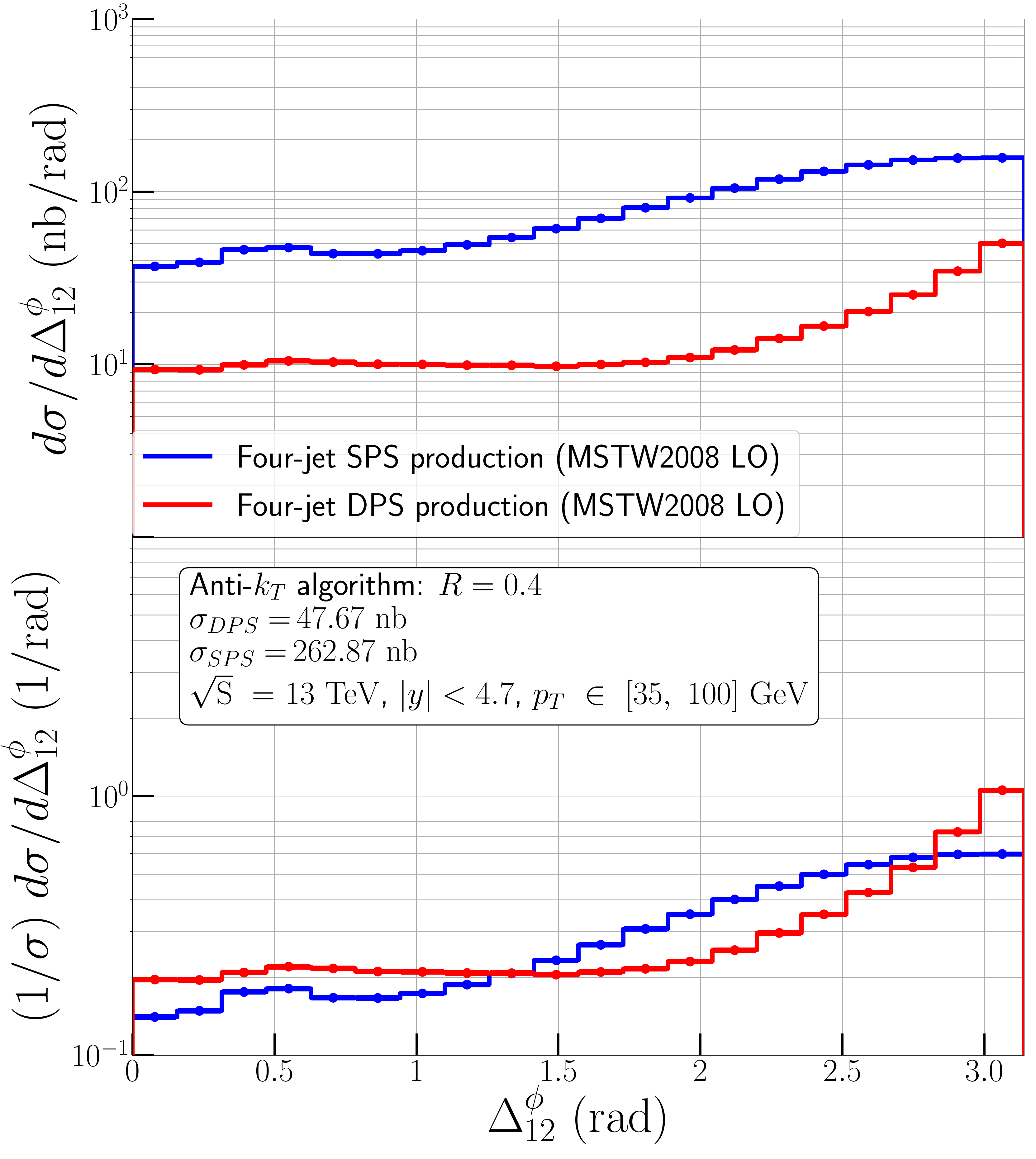}
\includegraphics[width=0.49\linewidth]{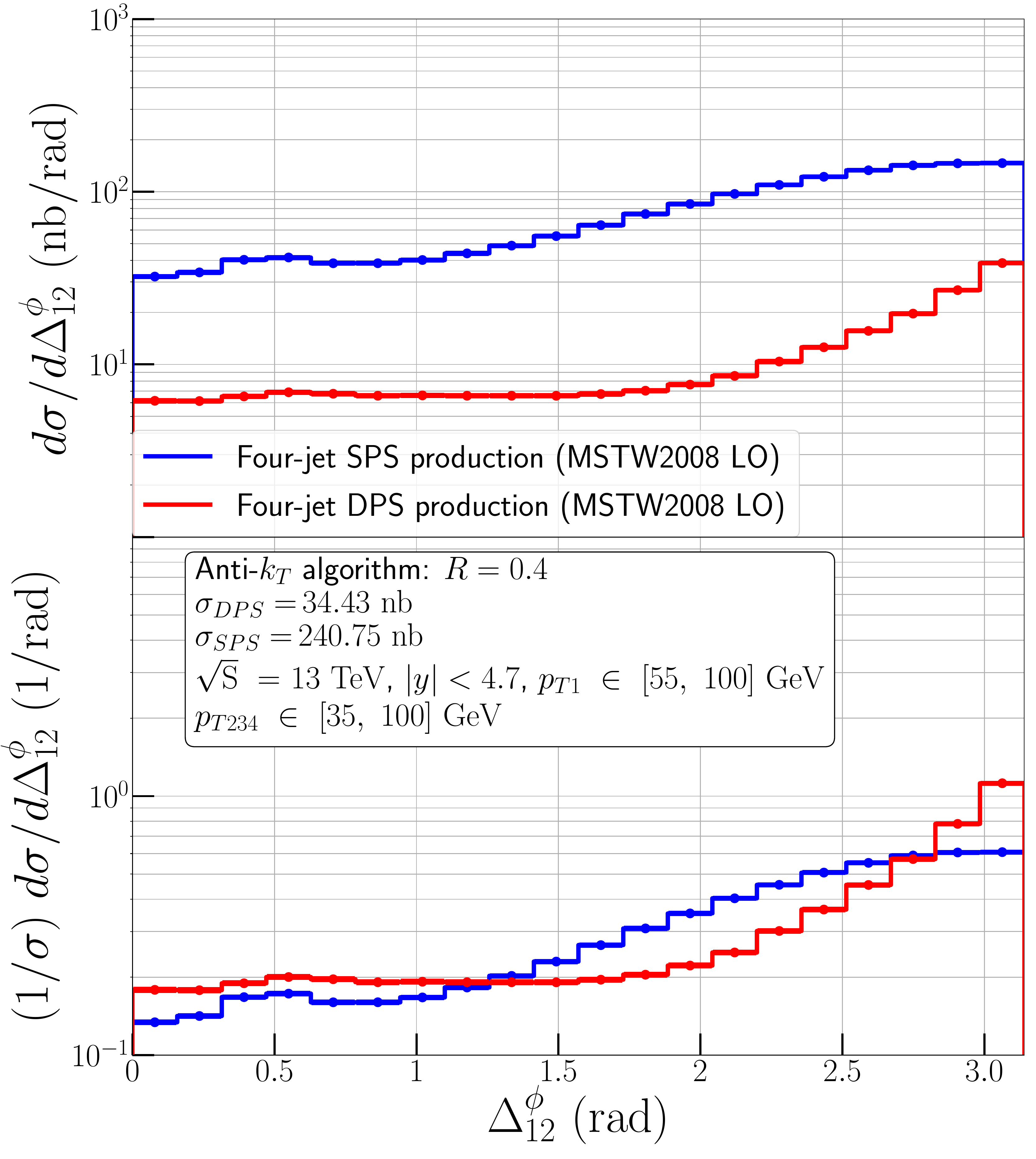}

\caption{DPS and SPS  distributions in azimuthal difference between the two hardest jets, $\Delta_{12}^\phi$ for four-jet production at $\sqrt S =13$ TeV with symmetric cuts $p_{T, j}  \in [35, 100]$ GeV (left) and asymmetric cuts $p_{T, 1}  \in [55, 100]$ GeV, $p_{T, 2,3,4}  \in [35, 100]$ GeV (right) after accounting for effects of radiation. Lower panels show normalized distributions.}
\label{fig:sps_vs_dps_ps8}
\end{figure}

\begin{figure}
\includegraphics[width=0.49\linewidth]{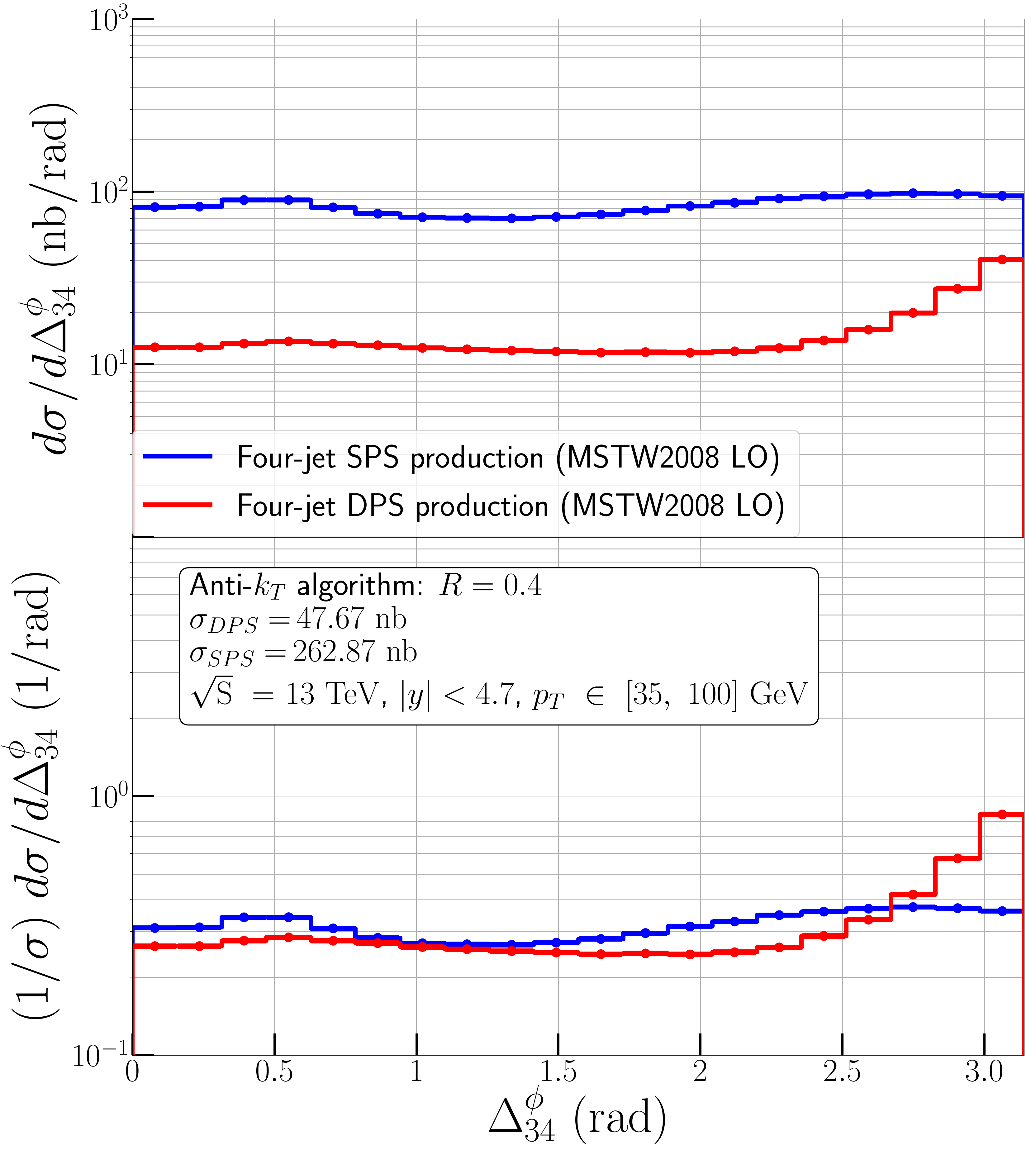}
\includegraphics[width=0.49\linewidth]{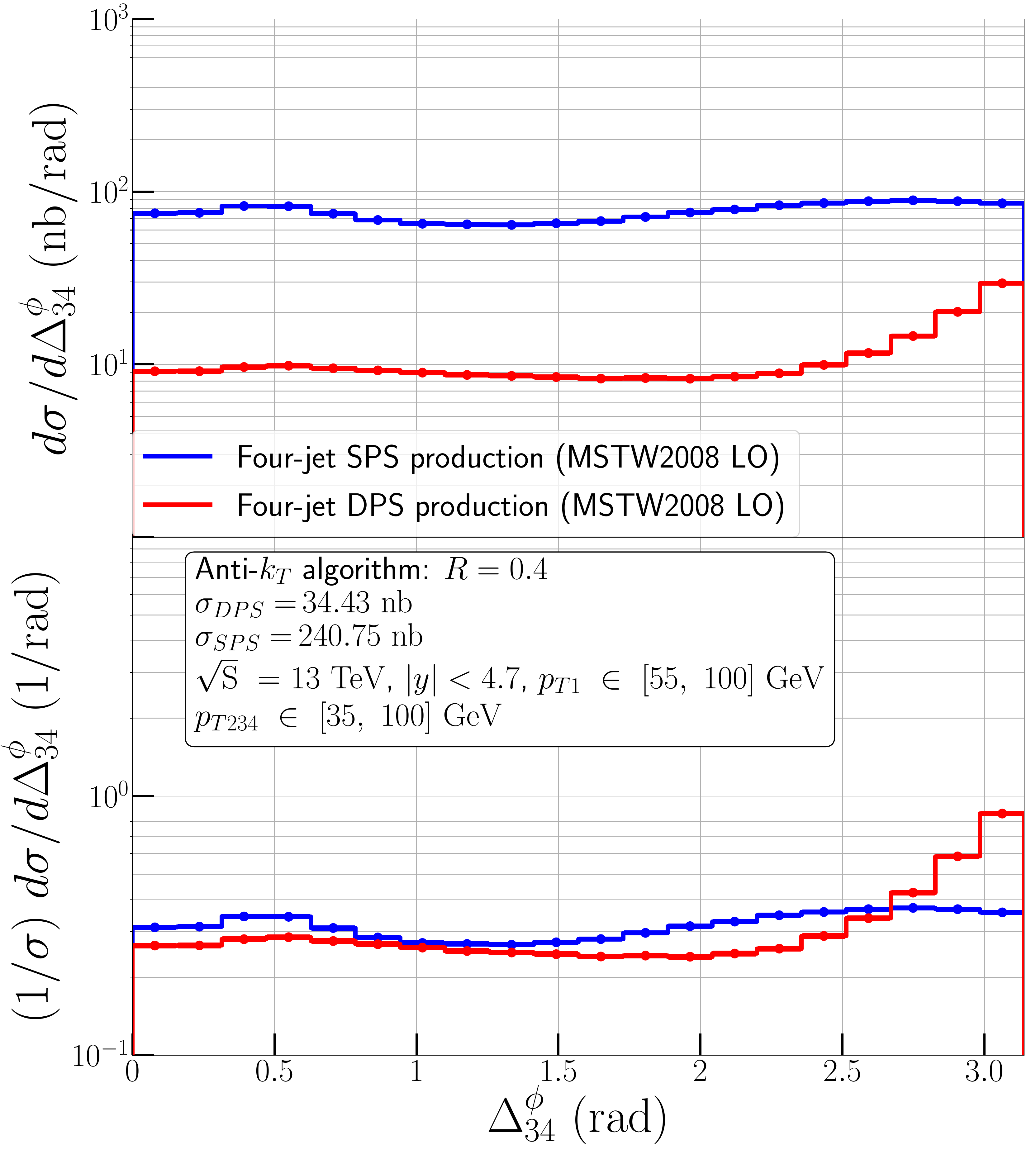}

\caption{DPS and SPS  distributions in azimuthal difference between the two softest jets, $\Delta_{34}^\phi$ , for four-jet production at $\sqrt S 
=13$ TeV with symmetric cuts $p_{T, j}  \in [35, 100]$ GeV (left) and asymmetric cuts $p_{T, 1}  \in [55, 100]$ GeV, $p_{T, 2,3,4}  \in [35, 100]$ GeV (right) after accounting for effects of radiation. Lower panels show normalized distributions.}
\label{fig:sps_vs_dps_ps9}
\end{figure}

\begin{figure}
\includegraphics[width=0.49\linewidth]{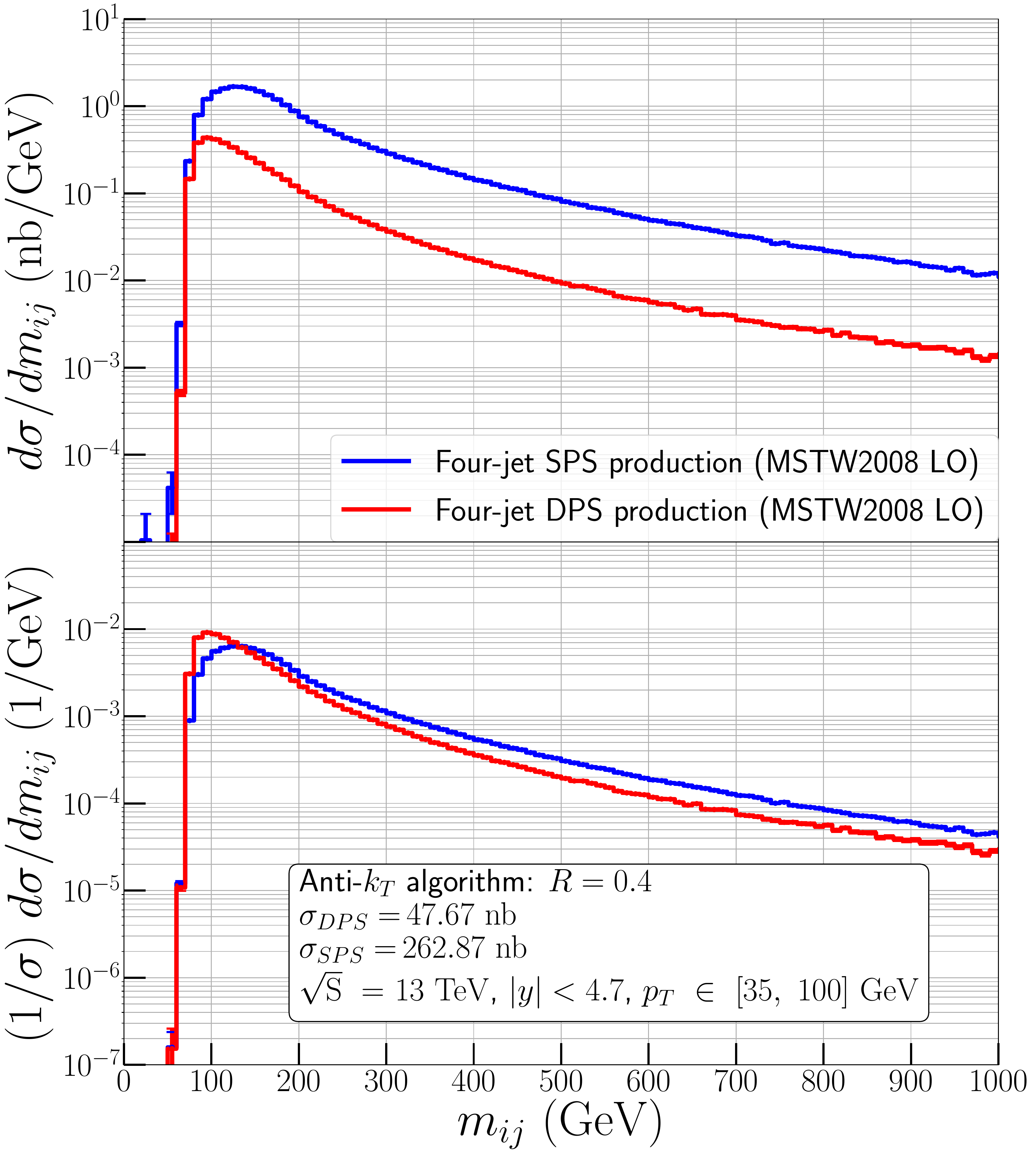}
\includegraphics[width=0.49\linewidth]{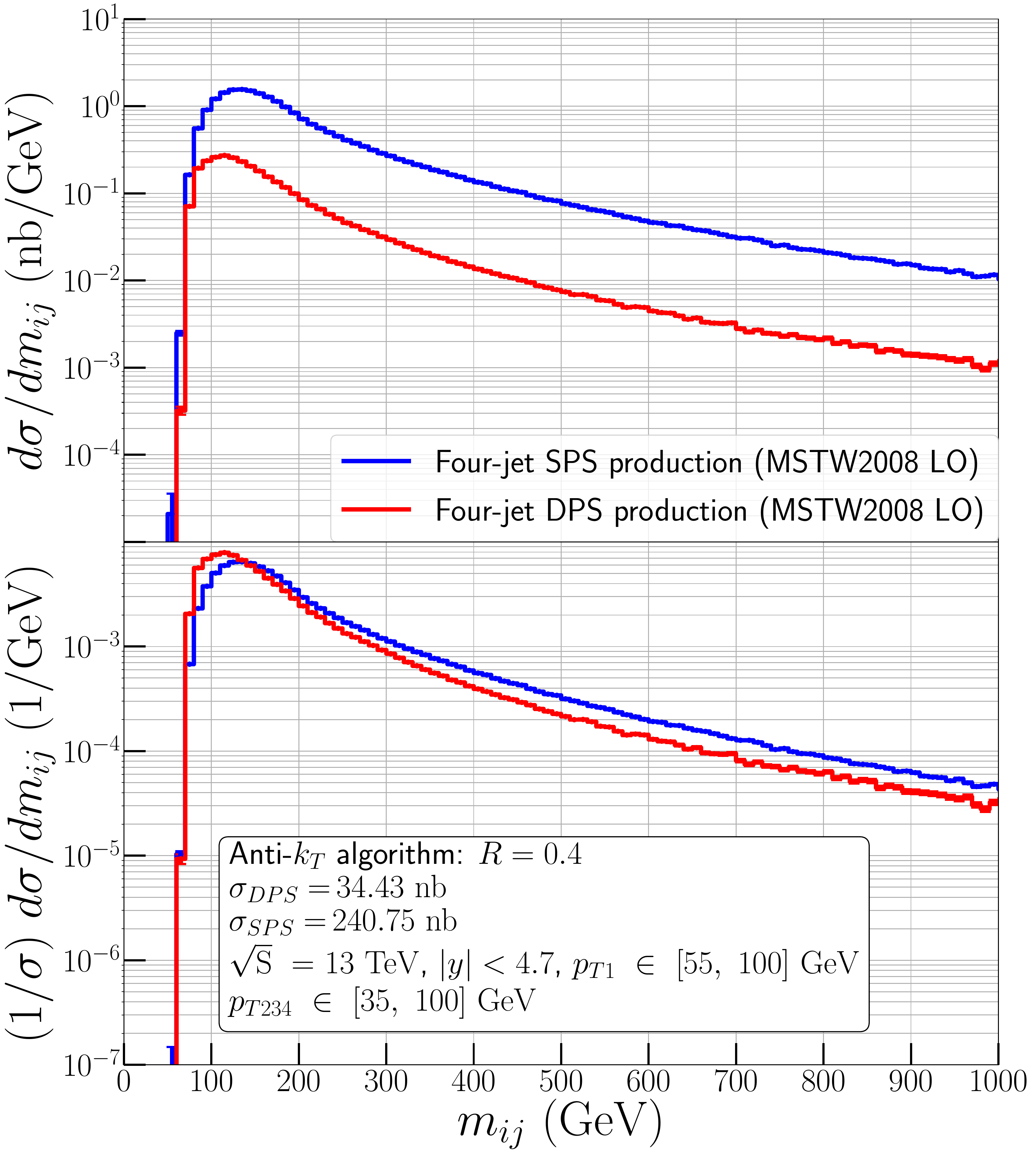}
\caption{DPS and SPS  distributions in the invariant mass of the two jets 
$i$, $j$ with smallest $\Delta_{ij}^{\pperp}$ imbalance, $m_{ij}$, for four-jet production at \mbox{$\sqrt S =13$ TeV} with symmetric cuts $p_{T, j}  \in [35, 100]$ GeV (left) 
and asymmetric cuts \mbox{$p_{T, 1}  \in [55, 100]$ GeV}, $p_{T, 2,3,4}  \in [35, 100]$ GeV (right) after accounting for effects of radiation. Lower panels show normalized distributions.}
\label{fig:sps_vs_dps_ps10}
\end{figure}

We also study the azimuthal differences
\begin{eqnarray}
	\Delta \phi_{12} &\equiv& |\phi_1 - \phi_2|,
 	\label{eq:DPhi12}\\
  	\Delta \phi_{34} &\equiv& |\phi_3 - \phi_4|,
 	\label{eq:DPhi34}
\end{eqnarray}
as in \cite{Aaboud:2016dea}, and the invariant mass $m_{ij}$ of the two jets, $i$ and $j$, with lowest $\Delta^{p_T}_{ij}$, see Figs.~\ref{fig:sps_vs_dps_ps8}, ~\ref{fig:sps_vs_dps_ps9}, and ~\ref{fig:sps_vs_dps_ps10}, respectively. 
We also shall note that in Eqs.~\ref{eq:DpT12} - \ref{eq:dS} and Eqs.~\ref{eq:DPhi12} - \ref{eq:DPhi34} we assume that $|\vec{p}_{T1}| > |\vec{p}_{T2}| > |\vec{p}_{T3}| > |\vec{p}_{T4}|$ whereas in Eq.~\ref{eq:dS_Berger} we consider all non-equivalent combinations of jets $i$, $j$, $k$, $l$. 
For all observables, there is no value at which the DPS production would prevail over the SPS. We also find that some shapes of the DPS distributions at the parton level do not survive after showering. The case in point 
is the $S'_\perp$ DPS distribution, showing distinct peaks at the minimal 
and maximal value of $S'_\perp$ at the parton level~\cite{Berger:2009cm} which get washed out by the radiation effects, cf. Fig.~\ref{fig:sps_vs_dps_ps7}. On the positive side, for some observables the shape of the DPS and SPS distributions can differ substantially, as shown in the lower parts of Figs.~\ref{fig:sps_vs_dps_ps1}-\ref{fig:sps_vs_dps_ps10}. In particular, compared to SPS, we observe narrower distributions for DPS at small 
values of $\Delta_{12}^{\pperp}$ and $\Delta_{34}^{\pperp}$. These regions correspond to back-to-back configurations of the two di-jets. The observed enhancements are then expected from theoretical considerations~\cite{Blok:2010ge, Blok:2011bu}, if we identify the two leading jet pair and the subleading jet pair as coming from two separate hard collisions, which can happen rather often. We also note that in the back-to-back region the 
contributions to the cross section from the perturbative splitting mechanism are less relevant~\cite{Blok:2013bpa}, meaning our predictions should 
not be much affected if these contributions are taken into account in the 
simulation. Similarly to the leading jet $\pperp$ and $\Delta{\rm Y}$ distributions, also for the other observables shown here we do not see a significant differences between results obtained with our symmetric and asymmetric cuts.

\begin{figure}
\includegraphics[width=0.49\linewidth]{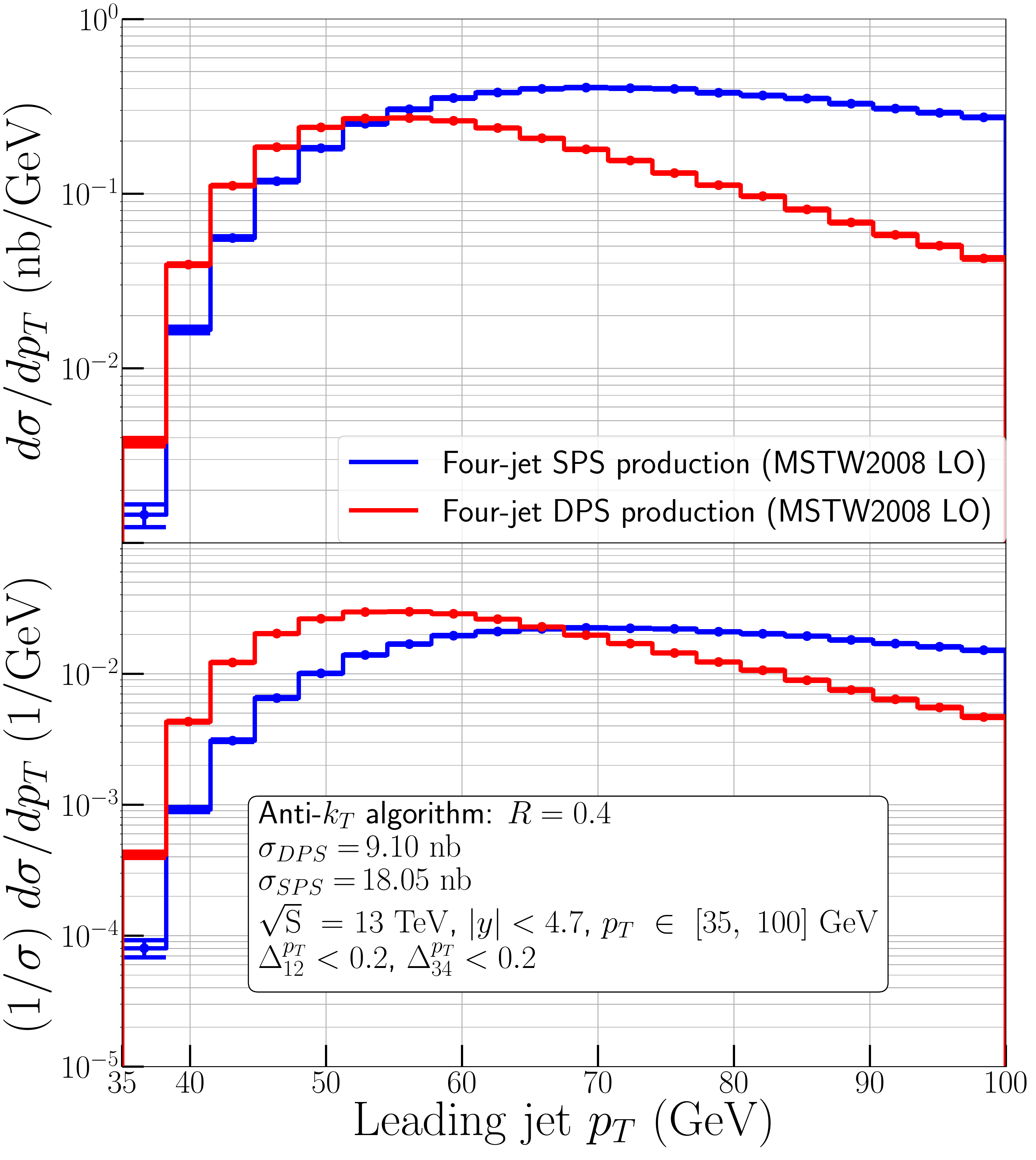}
\includegraphics[width=0.49\linewidth]{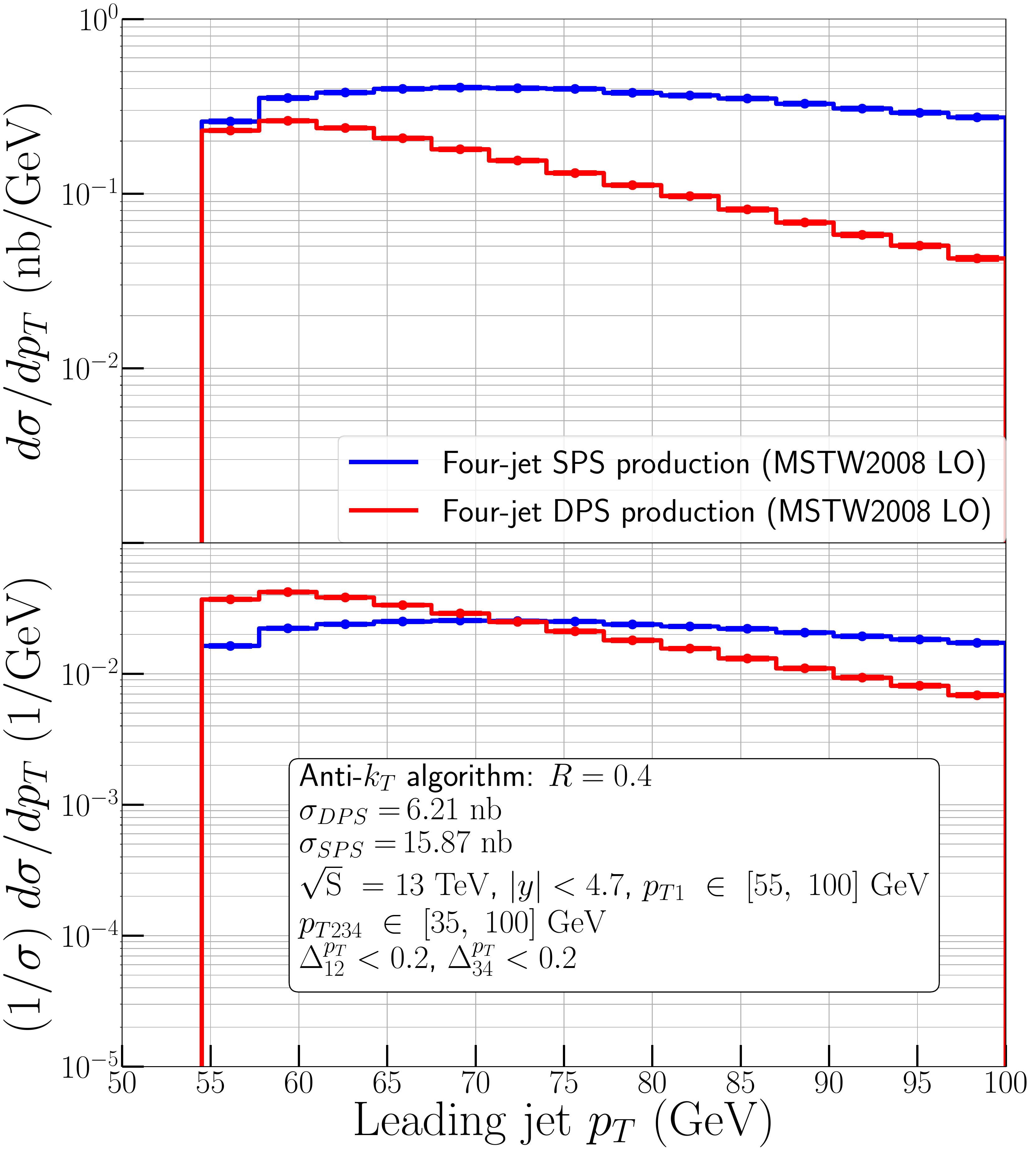}

\caption{Same as in Fig.~\ref{fig:sps_vs_dps_ps1} but after imposing additional cuts $\Delta_{12}^{\pperp} < 0.2$ and  $\Delta_{34}^{\pperp} < 0.2$.}
\label{fig:sps_vs_dps_ps11}
\end{figure}

\begin{figure}
\includegraphics[width=0.49\linewidth]{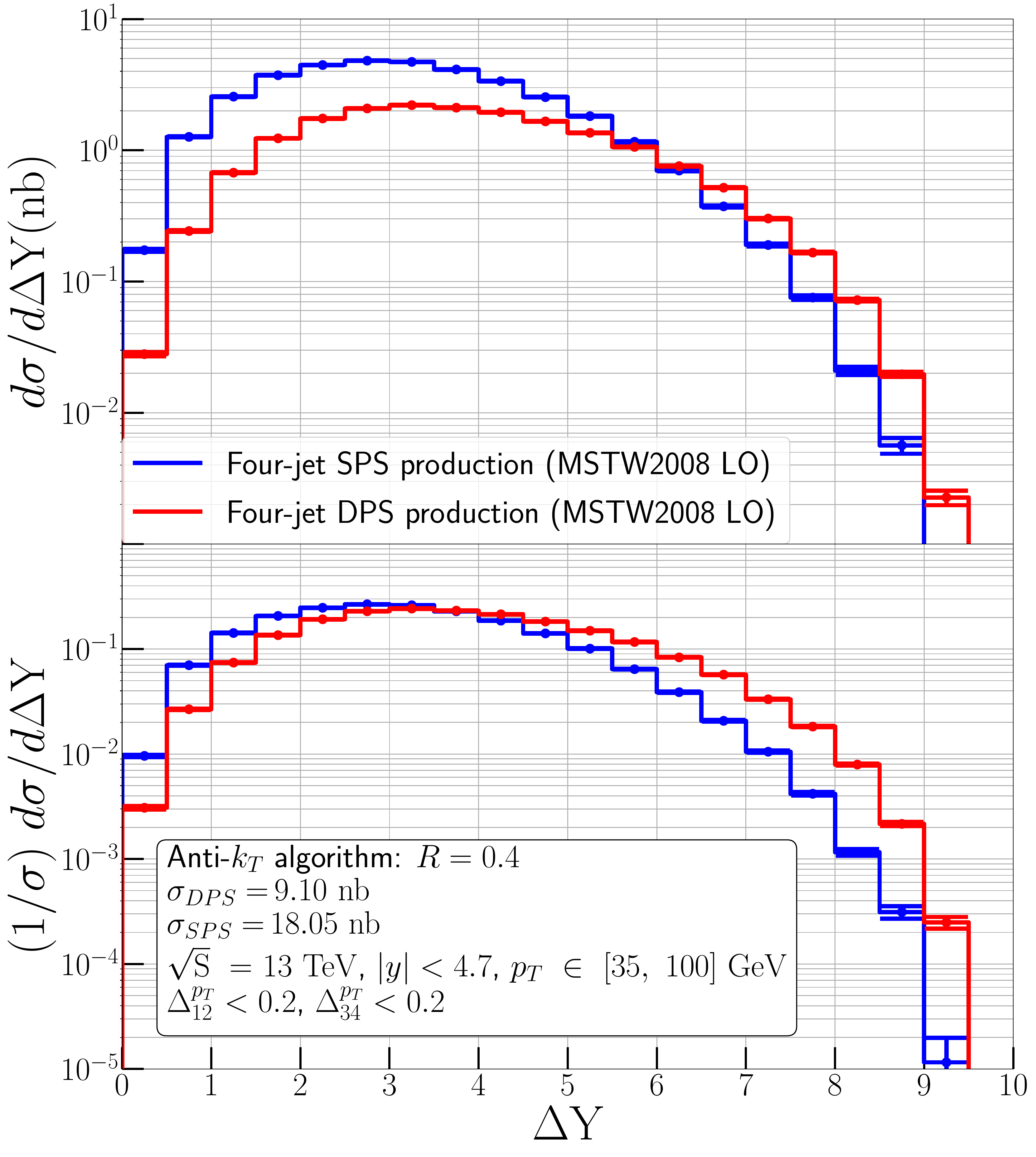}
\includegraphics[width=0.49\linewidth]{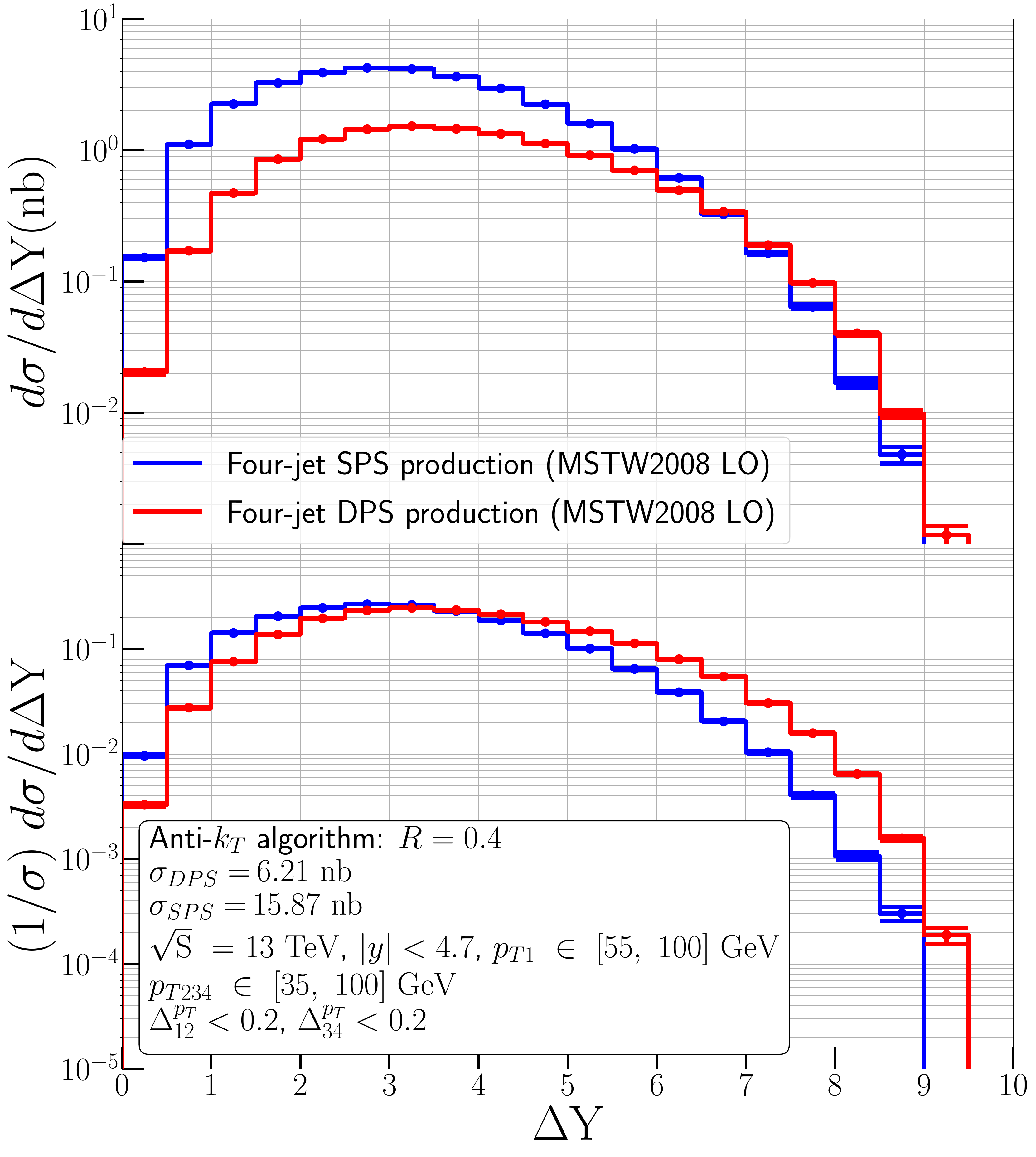}

\caption{Same as in Fig.~\ref{fig:sps_vs_dps_ps2} but after imposing additional cuts $\Delta_{12}^{\pperp} < 0.2$ and  $\Delta_{34}^{\pperp} < 0.2$.}
\label{fig:sps_vs_dps_ps12}
\end{figure}

\begin{figure}
\includegraphics[width=0.49\linewidth]{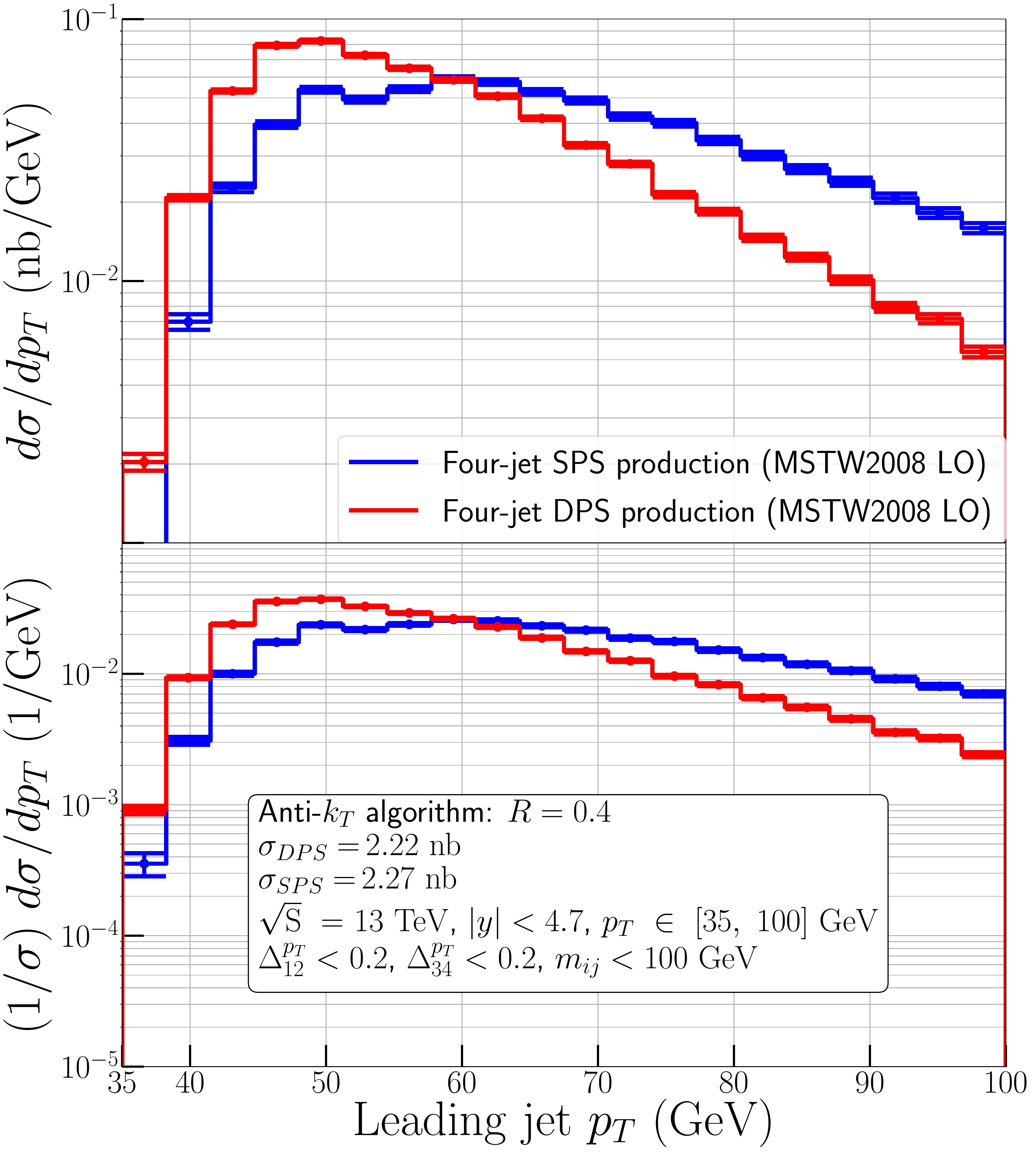}
\includegraphics[width=0.49\linewidth]{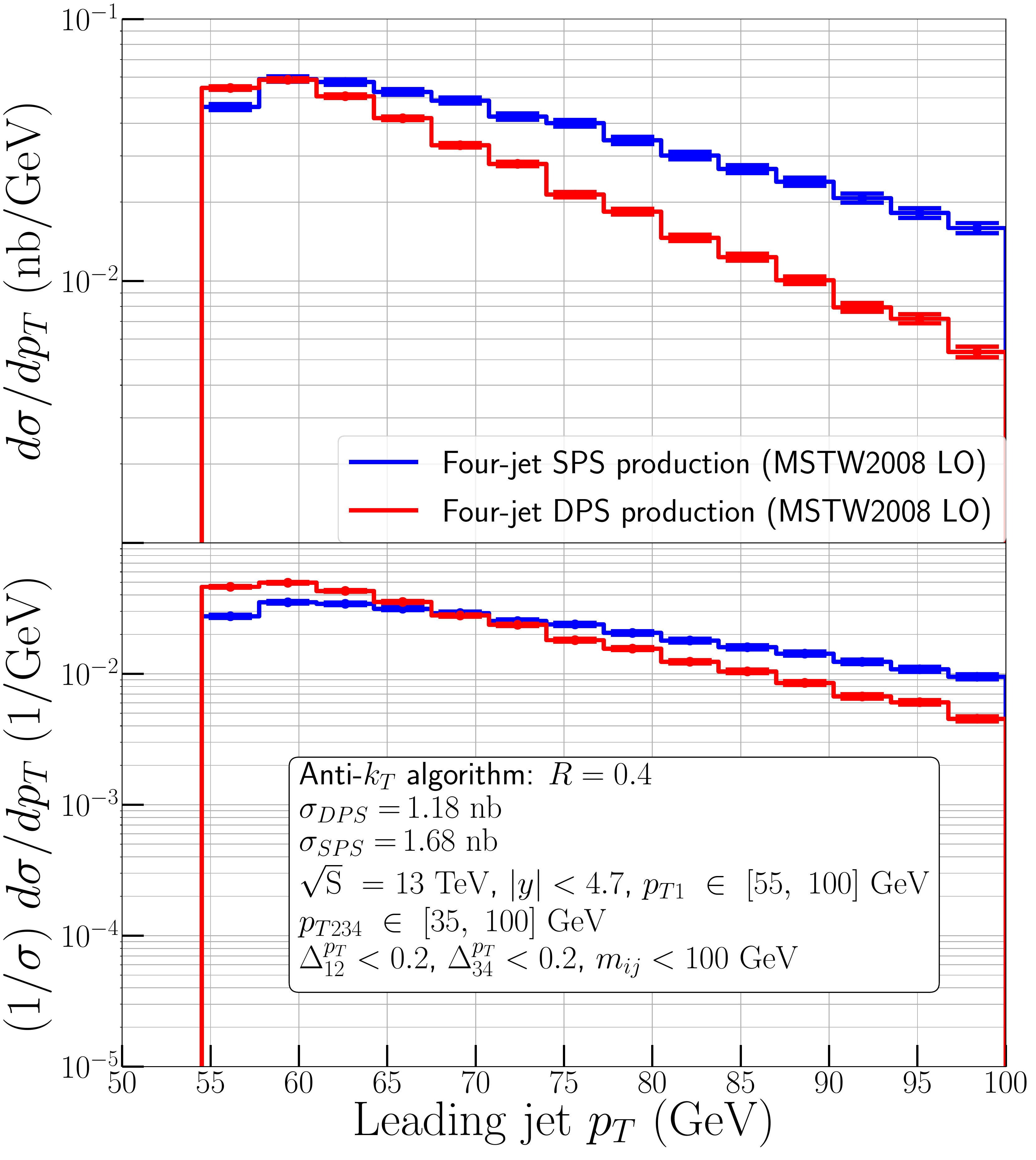}

\caption{Same as in Fig.~\ref{fig:sps_vs_dps_ps1} but after imposing additional cuts $\Delta_{12}^{\pperp} < 0.2$, $\Delta_{34}^{\pperp} < 0.2$ and $m_{ij}<100$ GeV.}
\label{fig:sps_vs_dps_ps13}
\end{figure}

\begin{figure}
\includegraphics[width=0.49\linewidth]{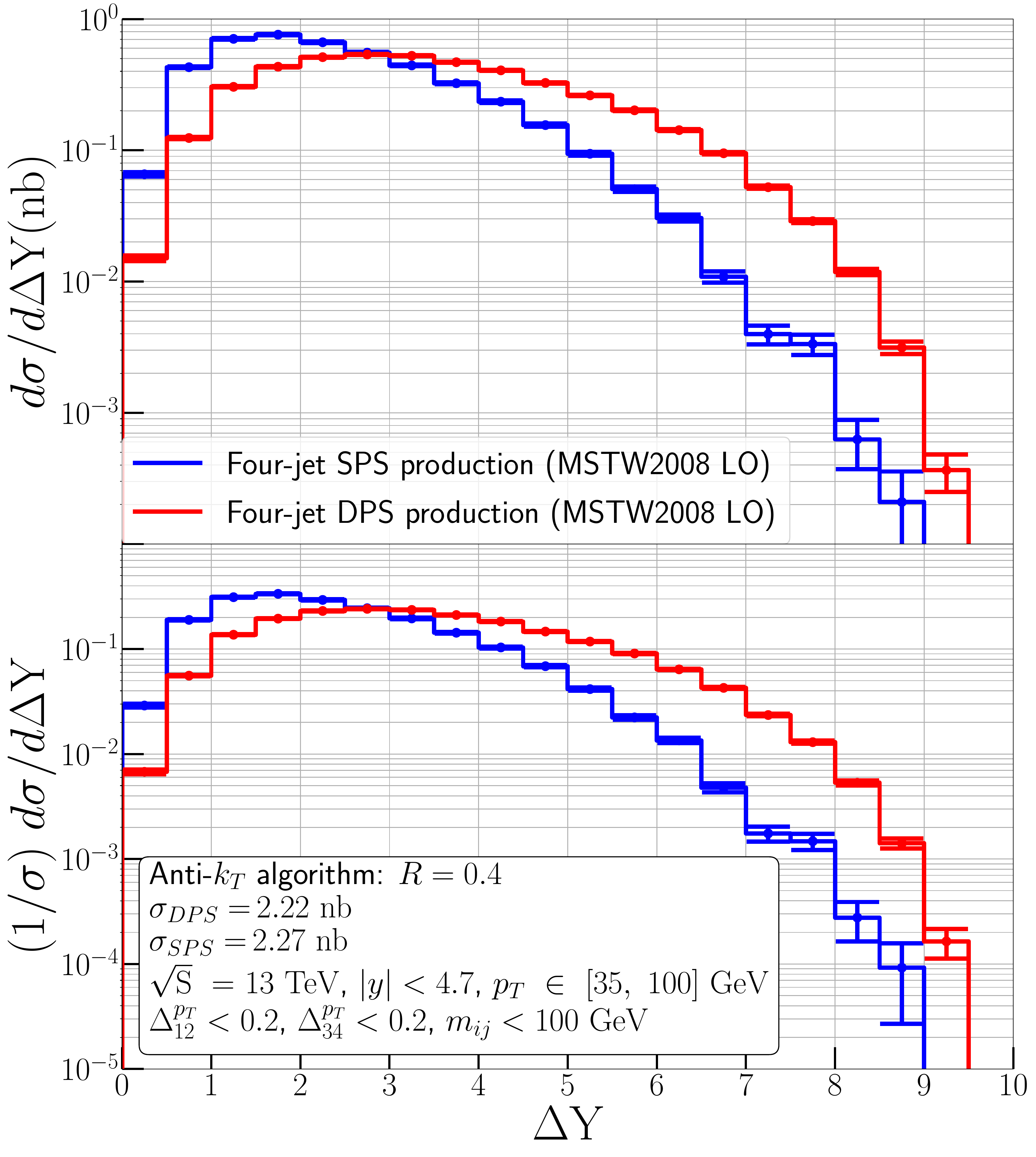}
\includegraphics[width=0.49\linewidth]{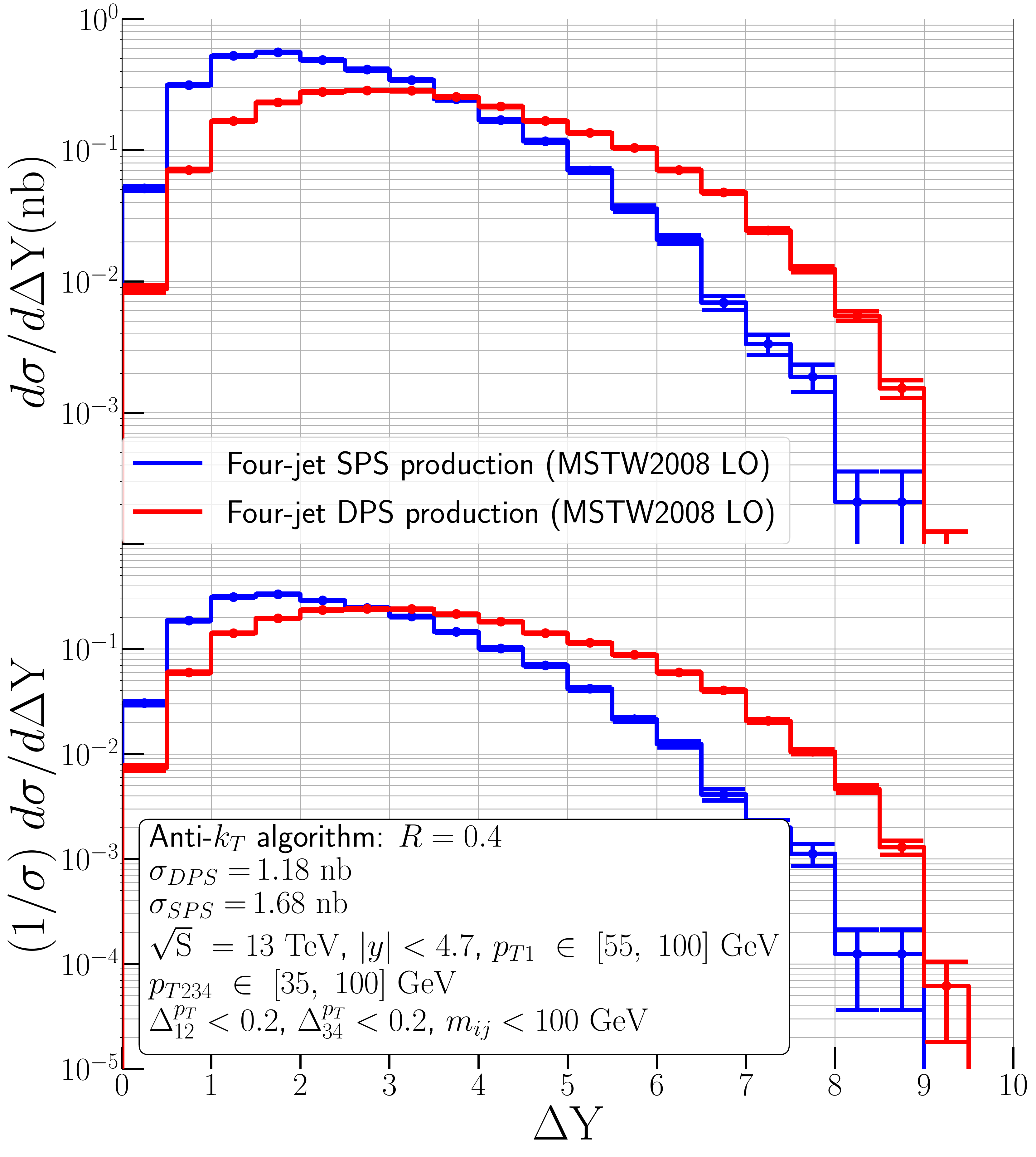}

\caption{Same as in Fig.~\ref{fig:sps_vs_dps_ps2} but after imposing additional cuts $\Delta_{12}^{\pperp} < 0.2$, $\Delta_{34}^{\pperp} < 0.2$ and 
$m_{ij}<100$ GeV.}
\label{fig:sps_vs_dps_ps14}
\end{figure}

\begin{figure}
\includegraphics[width=0.49\linewidth]{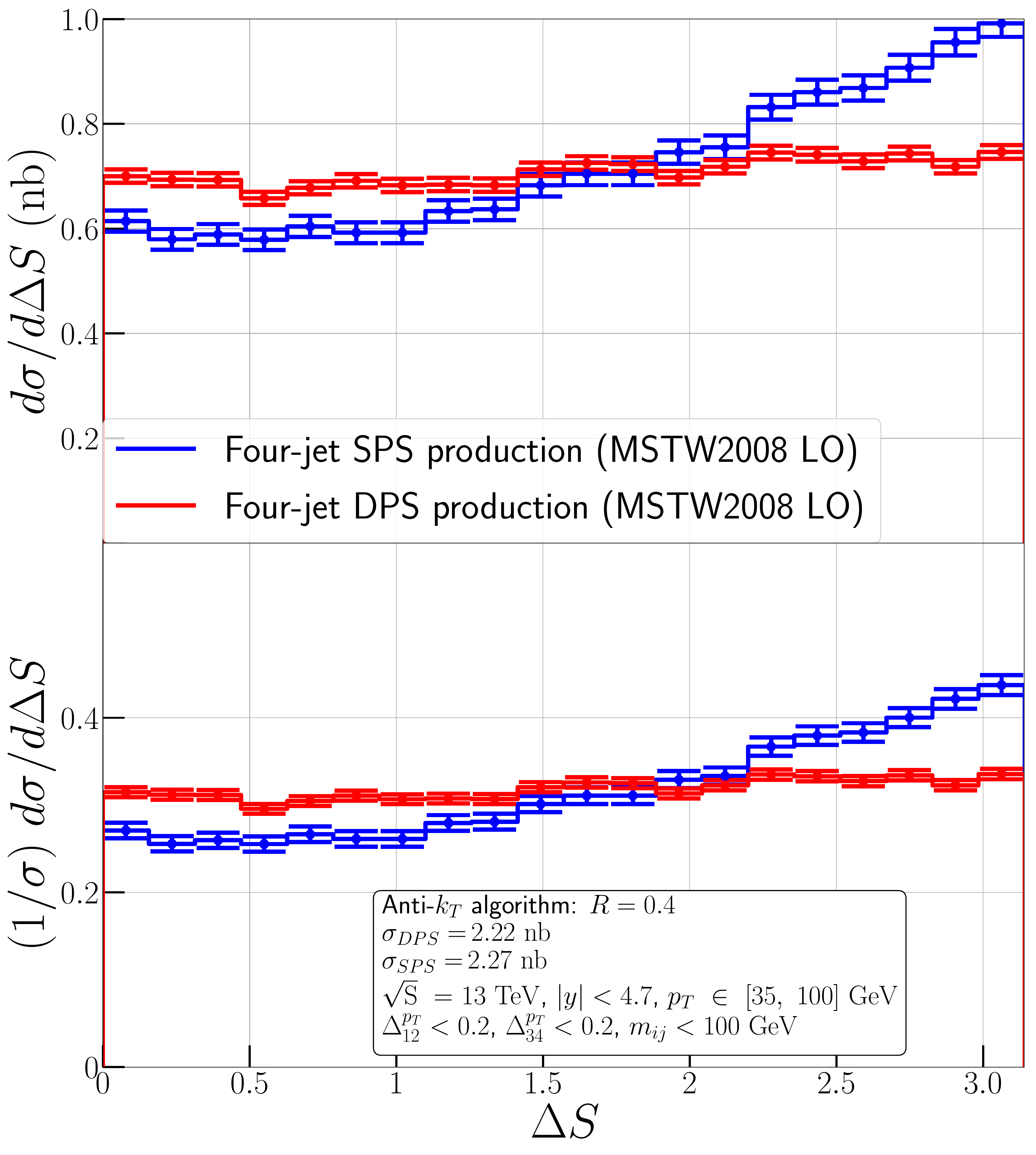}
\includegraphics[width=0.49\linewidth]{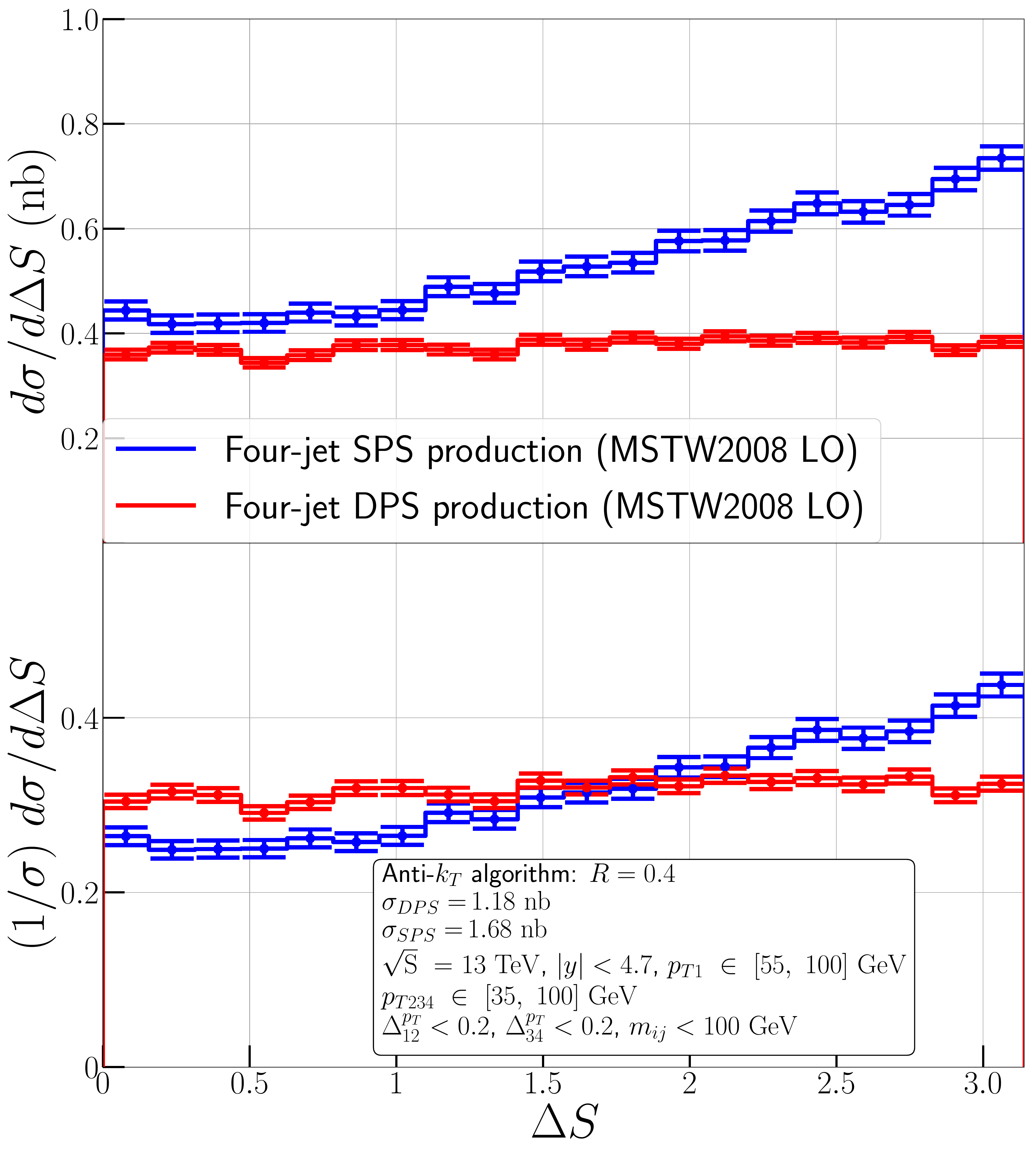}

\caption{Same as in Fig.~\ref{fig:sps_vs_dps_ps6} but after imposing additional cuts $\Delta_{12}^{\pperp} < 0.2$, $\Delta_{34}^{\pperp} < 0.2$, 
 and $m_{ij}<100$ GeV. }
\label{fig:sps_vs_dps_ps15}
\end{figure}

The information on shapes of the distributions can be harvested to improve discrimination power between DPS and SPS by imposing additional cuts. The choice of the cut parameters is driven by the shape of the distributions. In Table~\ref{tab:dps_4j_impact_of_ps_1e7}, we list values of the DPS 
and SPS total cross sections for various sets of cuts before and after radiation is included. Additionally, we provide percentage of DPS contributions to the total cross sections. We observe that, for our choice of the cut parameters, cuts on $\Delta_{12}^{\pperp}$ and $\Delta_{34}^{\pperp}$ 
increase the DPS signal in the most efficient way, even yielding some regions of phase-space where the DPS signal dominates, 
see Table~\ref{tab:dps_4j_impact_of_ps_1e7} and Figs.~\ref{fig:sps_vs_dps_ps11}, ~\ref{fig:sps_vs_dps_ps12}. 
 A further significant improvement can be achieved by combining these cuts with cuts on the invariant mass $m_{ij}$ of a di-jet pair with the smallest value of transverse momentum imbalance $\Delta_{ij}^{\pperp}$,see Figs.~\ref{fig:sps_vs_dps_ps13}, ~\ref{fig:sps_vs_dps_ps14}, ~\ref{fig:sps_vs_dps_ps15}.   At this level, the difference in the number of events, \textit{i.e.} lower statistics due to more stringent asymmetric set of cuts 
compared to symmetric set becomes visible. In this case, it might be advisable to decrease the cuts on the value of $p_T$ for all jets in the asymmetric set.

\begin{center}
\resizebox{\textwidth}{!}{
\begin{tabular}{|c|c|c|c|c|}
\hline
\multicolumn{2}{|c|}{\thead{Cuts and collision energy}} 		& \thead{$\sigma_{\rm SPS}$} 
          									  				& \thead{$\sigma_{\rm DPS}$} 
          													&  \thead{$\frac{\sigma_{\rm DPS} }{ \left(\sigma_{\rm DPS} + \sigma_{\rm SPS}\right)}$}\\ \hline
          														
\multirow{2}{*}{\thead{${p_T \in [35, 100]}$ GeV}}  		& \thead{No PS} 	& 316.78		&  333.83	& 51	 $\%$ \\ \cline{2-5} 
                   										& \thead{PS}		& 262.87 	&   47.67 	& 15	 $\%$  \\ \hline

\multirow{2}{*}{\thead{${p_{T1} \in [55, 100]}$ GeV,
					  ${p_{T2, 3,4} \in [35, 100]}$ GeV}}& \thead{No PS} 	& 263.59		& 95.69 	 	& 27	 $\%$ \\ \cline{2-5} 
                   										& \thead{PS}		& 240.75 	& 34.43	 	& 13	 $\%$ 
 \\ \hline

\multirow{2}{*}{\thead{${p_T \in [35, 100]}$ GeV,
       				  $\Delta^{p_T}_{12} < 0.2$, 
       				  $\Delta^{p_T}_{34} < 0.2$}}		& \thead{No PS} 	& 	80.38	& 333.83		& 81	 $\%$ \\ \cline{2-5} 
                   										& \thead{PS}		& 	18.05	& 	9.10		& 34	 $\%$ \\ \hline                					
\multirow{2}{*}{\thead{${p_{T1} \in [55, 100]}$ GeV,
                      ${p_{T2, 3, 4} \in [35, 100]}$ GeV,
       				  $\Delta^{p_T}_{12} < 0.2$, 
       				  $\Delta^{p_T}_{34} < 0.2$}}		& \thead{No PS} 	&  57.36		& 95.69		& 63		$\%$ \\ \cline{2-5} 
                   										& \thead{PS}		&  15.87		& 	6.21		& 28		$\%$ \\ \hline                					
                   										
\multirow{2}{*}{\thead{${p_T \in [35, 100]}$ GeV,
       					$\Delta^{p_T}_{12} < 0.2$, 
       					$\Delta^{p_T}_{34} < 0.2$, 
      					$m_{ij} < 100$ GeV}}										
      					 								& \thead{No PS} & 3.90	 	& 170.75		&  98	$\%$ \\ \cline{2-5} 
                   										& \thead{PS}		& 2.27		& 	2.22		&  49	$\%$ \\ 
\hline

\multirow{2}{*}{\thead{${p_{T1} \in [55, 100]}$ GeV,
                       ${p_{T2, 3, 4} \in [35, 100]}$ GeV,
       				  $\Delta^{p_T}_{12} < 0.2$, 
       				  $\Delta^{p_T}_{34} < 0.2$, 
      				  $m_{ij} < 100$ GeV}}										
      					 								& \thead{No PS} 	& 0.021		&  31.27		& 100  $\%$ \\ \cline{2-5} 
                   										& \thead{PS}		& 1.68 		& 	1.18  	& 41 	$\%$ \\ \hline

              																										             							\end{tabular}
}
\captionof{table}{Impact of parton shower (PS) effects on the DPS and SPS 
cross sections, $\sqrt{S} = 13$ TeV, ${|y| < 4.7}$. The parton-level SPS cross sections are multiplied by ${\rm K}$-factor equal to  $0.5$. All cross sections are given in nb. The DPS fraction is given at the 1$\%$ precision level.}
\label{tab:dps_4j_impact_of_ps_1e7}
\end{center}

Before finishing this section we would like to note that the different DPS-sensitive observables considered in this paper were already used in earlier publications \cite{Aaboud:2016dea, Berger:2009cm,  Blok:2015rka, Maciula:2015vza,  Gaunt:2010pi}. However, to the best of our knowledge, a combination of cuts on the aforementioned observables listed in Table~\ref{tab:dps_4j_impact_of_ps_1e7}, and the corresponding distributions in Figs.~\ref{fig:sps_vs_dps_ps11} - \ref{fig:sps_vs_dps_ps15}, are new. We hope that the new combinations of cuts proposed in this work would allow to 
achieve better separation between DPS signal and SPS background in the four-jet production processes and, as a consequence, to improve the estimate of value of $\sigma_{\rm eff}$.

\section{Summary and discussion}

In this work, we have studied double parton scattering in four-jet events 
at the LHC, both at the partonic level and incorporating the effects of QCD radiation. To this end, we have developed a parton-level Monte Carlo simulation outputting modified LHE files event records, which then can be showered by the \pythia event generator to which additional modifications 
are applied. Apart from studying the impact of various cuts and collider energies on the DPS and SPS contributions to the cross sections and estimating their uncertainties at the partonic level, we have also investigated the effect of longitudinal correlations as implemented in the GS09 package. After adding QCD radiation, we find that it can affect DPS and SPS predictions significantly, with DPS contributions being modified more. We have also examined a number of observables in regard to their discriminating power between DPS and SPS.  Applying just a basic set of cuts, we observe that many of these observables substantially differ in shape in a specific range of values. This information can be then used to propose more 
elaborated sets of cuts which increase the efficiency of the discrimination.  In particular, we find that a combination of cuts on jets' $\pperp$ and transverse momentum imbalance $\Delta_{ij}^{\pperp}$ as well as invariant mass $m_{ij}$  of a di-jet pair with the smallest value of transverse momentum imbalance $\Delta_{ij}^{\pperp}$ provides a very promising way 
for selecting DPS contributions in four-jet events.

In this foray towards including higher order effects in DPS predictions, we have implemented a simple model of dPDFs which only partially accounts 
for correlations between partons in the same proton. In particular, this model neglects  contributions from a perturbative splitting of one parton 
into two.  We have, however, checked that at the parton level the predictions obtained with a publicly available dPDF package GS09, which under assumption of transverse-longitudinal factorization of double distributions 
accounts for longitudinal correlations resulting from $1\to 2$  splittings, differ only minimally from predictions obtained with our naive dPDF model. As discussed above, a consistent theoretical framework according to~\cite{Diehl:2017kgu} would require generalized parton distributions dependent on the impact parameter. Their modelling involves an ``intrinsic'' and a ``splitting'' part, which both evolve according to the homogenous double DGLAP equation. Provided such set of double parton distributions, our parton shower calculations can easily accommodate them. We note that a complementary approach, relying on altering the parton shower to include $1 \to 2$ splittings was reported recently in~\cite{Cabouat:2019gtm, Cabouat:2020ssr}.

Naturally, in this exploratory study, we could only explore a few particular set-ups for the calculations. It is conceivable that \textit{e.g.} other kinematical cuts could lead to a better discrimination between DPS and SPS. Additionally, lowering a minimal cut on jets' $\pperp$ would select proportionally more DPS events. On the other hand however, it would inevitably lead to bigger contamination from background. Studies of various set-ups are beyond the scope of this work, but we hope that our work and simulation tool, which is available on request, can be used for optimizing the measurement of DPS in four-jet events at the LHC in the future.

\section*{Acknowledgments}
We are grateful to Torbj\"{o}rn Sj\"{o}strand for introducing modifications to the \pytppp code necessary to add ISR and FSR effects to our parton-level DPS simulations and  J. Gaunt for providing grids and interpolation routines for the GS09 dPDFs. We  also thank Ch. Klein-B\"{o}sing, O. Mattelaer and S. Prestel  for useful and fruitful discussions. The  work  of OF has been supported by the Deutsche Forschungsgemeinschaft (DFG) through 
the Research Training Group ``GRK 2149: Strong and Weak Interactions - from Hadrons to Dark Matter''. OF also acknowledges funding  from the European Union's Horizon 2020 research and innovation programme  as part of the Marie Sk\l{}odowska-Curie Innovative Training Network MCnetITN3 (grant agreement no. 722104) and the financial support through the curiosity-driven grant ``Using jets to challenge the Standard Model of particle physics'' 
from Universit\`{a} di Genova.

All simulations for this paper were performed by means of the PALMAII cluster provided by the University of M\"unster, Germany.

\appendix

\section{Double LHE files}
\label{s:double_lhe}
\begin{figure}
\center{\includegraphics[width=1.0\linewidth]{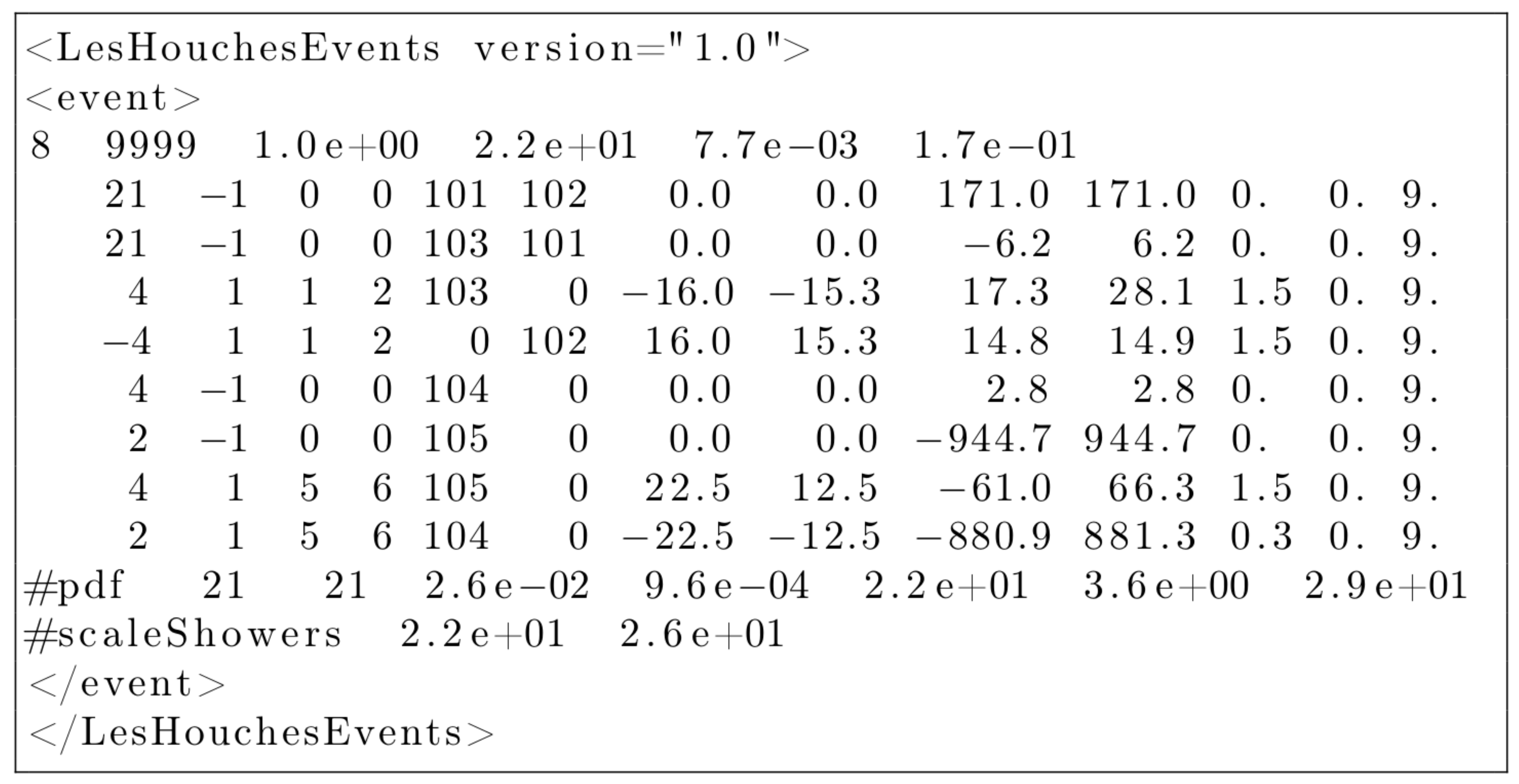}}
\caption{An extension of the Les-Houches version=``1.0'' standard to the DPS events. 
In order to ease the reading we keep only one digit after comma for the components of the four-momenta.}
\label{fig:double_lhe_code}
\end{figure}

Here we describe modifications to the \pythia code and the LHEF standard necessary to read and ``shower'' DPS events from LHE files. In \mbox{Fig.~\ref{fig:double_lhe_code}} we show an example of a modified LHEF standard for the $\left(g g \rightarrow c \bar{c}\right) \otimes \left(c u \rightarrow c u\right)$ DPS process. The two di-jet events which constitute a DPS event are stacked in the same event record. The extension of the LHEF 
standard to the DPS events also requires the correct mother-daughter information, \textit{c.f.} Fig. \ref{fig:double_lhe_code}. The  parent indices 1 and 2 of the  $c \, \bar{c}$ pair indicate that it originates from two initial-state gluons (first and second lines in the event record) and the parent indices 5 and 6 of the $c \, u$ pair tell us that it originates 
from the initial-state $c \, u$ pair
(fifth and sixth lines in the event record)\cprotect\footnote{Note that the numbering of lines between \verb|<event>| and \verb|</event>| tags starts from zero.}.
In addition to the aforementioned changes  a new line starting with the key-word \verb|#scaleShowers| was added. It contains factorization scales for the
first and second hard interactions correspondingly.

The \pythia event generator, starting from version ``8.240'' can generate 
and output DPS events into LHE files, as  shown 
in Fig. \ref{fig:double_lhe_code}. However, it cannot ``shower'' them correctly.
Several important modifications have to be added to the files \verb|Pythia.cc|, \verb|ProcessContainer.cc| and \verb|PartonLevel.cc|.
These  modifications are available starting from version ``8.243''. The  checks and description of these modifications are given in \cite{OFedkevych:PhD}. It also needs to be stressed that for the  aforementioned
modifications to \pythia version = ``8.240''  to work correctly, the LHE events have to be written as in \mbox{Fig. \ref{fig:double_lhe_code}} with a necessary tag
\verb|<LesHouchesEvents version="1.0">|.  If instead one uses \\
\verb|<LesHouchesEvents version="3.0">|  then \pythia will still read in and shower DPS events,
but in a wrong way. The reason is that \pythia reads in the LHE   version 
= ``1'' and the \mbox{LHE version = ``3''} files calling different routines from the \verb|LesHouches.cc| and \verb|LHEF3.cc| files, correspondingly. 
While reading in the LHE version = ``1'' files has been adopted for the 
DPS events, that's not yet the case for the LHE version = ``3'' files. Therefore, the usage of the tag \\
\verb|<LesHouchesEvents version="3.0">| will invoke calling routines from the file \\
\verb|LHEF3.cc|, leading then to a wrong assignment of the mother-daughter labels and MPI, ISR and FSR scales.

\bibliography{journal}

\end{document}